\documentclass[final,5p,times,twocolumn]{elsarticle}

\usepackage{subfig}
\usepackage[tableposition = bottom]{caption}
\makeatletter
\long\def\@makecaption#1#2{%
  \vskip\abovecaptionskip
  \sbox\@tempboxa{#1. #2}%
  \ifdim \wd\@tempboxa >\hsize
    #1. #2\par
  \else
    \global \@minipagefalse
    \hb@xt@\hsize{\hfil\box\@tempboxa\hfil}%
  \fi
  \vskip\belowcaptionskip}
\makeatother

\usepackage{lineno}
\usepackage{amssymb} %for some symbols
\usepackage{amsmath}

\usepackage{makecell} %for table

\usepackage{chngcntr}

\usepackage{float}
\usepackage[colorlinks=true, allcolors=blue]{hyperref}

\begin{document}
\begin{frontmatter}
\title{The BM@N spectrometer at the NICA accelerator complex.}
\author[jinr]{S.\,Afanasiev}
\author[jinr]{G.\,Agakishiev}
\author[jinr]{E.\,Aleksandrov}
\author[jinr]{I.\,Aleksandrov}
\author[jinr,kurchat]{P.\,Alekseev}
\author[jinr]{K.\,Alishina}
\author[jinr]{V.\,Astakhov}
\author[mephi]{E.\,Atkin}
\author[mipt]{T.\,Aushev}
\author[jinr]{V.\,Azorskiy}
% ...
\address[jinr]{Joint Institute for Nuclear Research (JINR), Dubna, Russia}
\address[inrras]{Institute for Nuclear Research of the RAS (INR RAS), Moscow, Russia}
\address[kurchat]{Kurchatov Institute, NRC, Moscow, Russia}
\address[lpiras]{Lebedev Physical Institute of the Russian Academy of Sciences (LPI RAS), Moscow, Russia}
\address[mipt]{Moscow Institute of Physics and Technology (MIPT), Moscow, Russia}
\address[mephi]{National Research Nuclear University MEPhI, Moscow, Russia}
\address[econom]{National Research University Higher School of Economics (HSE University), Moscow, Russia}
\address[sinp]{Skobeltsyn Institute of Nuclear Physics, Moscow State University (SINP MSU), Moscow, Russia}
\address[spbu]{St Petersburg University (SPbU), St Petersburg, Russia}
 \address[PlUni]{Plovdiv University ``Paisii Hilendarski'', Plovdiv, Bulgaria}
\address[moechin]{Key Laboratory of Nuclear Physics and Ion-Beam Application (MOE), Institute of Modern Physics, Fudan University, Shanghai, China}
\address[iptalmaty]{Institute of Physics and Technology, Satbayev University, Almaty, Kazakhstan}
\address[phtiuzas]{Physical-Technical Institute of Uzbekistan Academy of Sciences (PhTI of UzAS), Tashkent, Uzbekistan}
\address[IMechBAS]{Institute of Mechanics at the Bulgarian Academy of Sciences (IMech-BAS), Sofia, Bulgaria}
\address[IHEP]{State Research Center - Institute for High Energy Physics (IHEP), Protvino, Russia}
\address[Weizmann]{Weizmann Institute of Science, Rehovot, Israel}
\cortext[cor1]{Email Address: zavertyaevmv@lebedev.ru}

\begin{abstract}
BM@N (Baryonic  
Matter at Nuclotron) is the first experiment operating and taking data at
the Nuclotron/NICA ion-accelerating complex.The aim of the BM@N experiment 
is to study interactions of relativistic heavy-ion beams with fixed targets.
We present a technical description of the BM@N spectrometer including all its subsystems. 
\end{abstract}


\end{frontmatter}

\clearpage
\section{Introduction}

\begin{figure*}[tbp]
  \centering
\caption{Schematic view of the BM@N setup in the 2023 Xe run. Main components: 0) SP-41 analyzing magnet, 1) vacuum beam pipe, 2) BC1 beam counter, 3) Veto counter (VC), 4) BC2 beam counter, 5) Silicon Beam Tracker (SiBT), 6) Silicon beam profilometers, 7) Barrel Detector (BD) and Target station, 8) Forward Silicon Detector (FSD), 9) Gaseous Electron Multiplier (GEM) detectors, 10) Small cathode strip chambers (Small CSC), 11) TOF400 system, 12) drift chambers (DCH), 13) TOF700 system, 14) Scintillation Wall (ScWall), 15) Fragment Detector (FD), 16) Small GEM detector, 17) Large cathode strip chamber (Large CSC), 18) gas ionization chamber as beam profilometer, 19) Forward Quartz Hodoscope (FQH), 20) Forward Hadron Calorimeter (FHCal). }
  \includegraphics[width=0.8\textwidth]{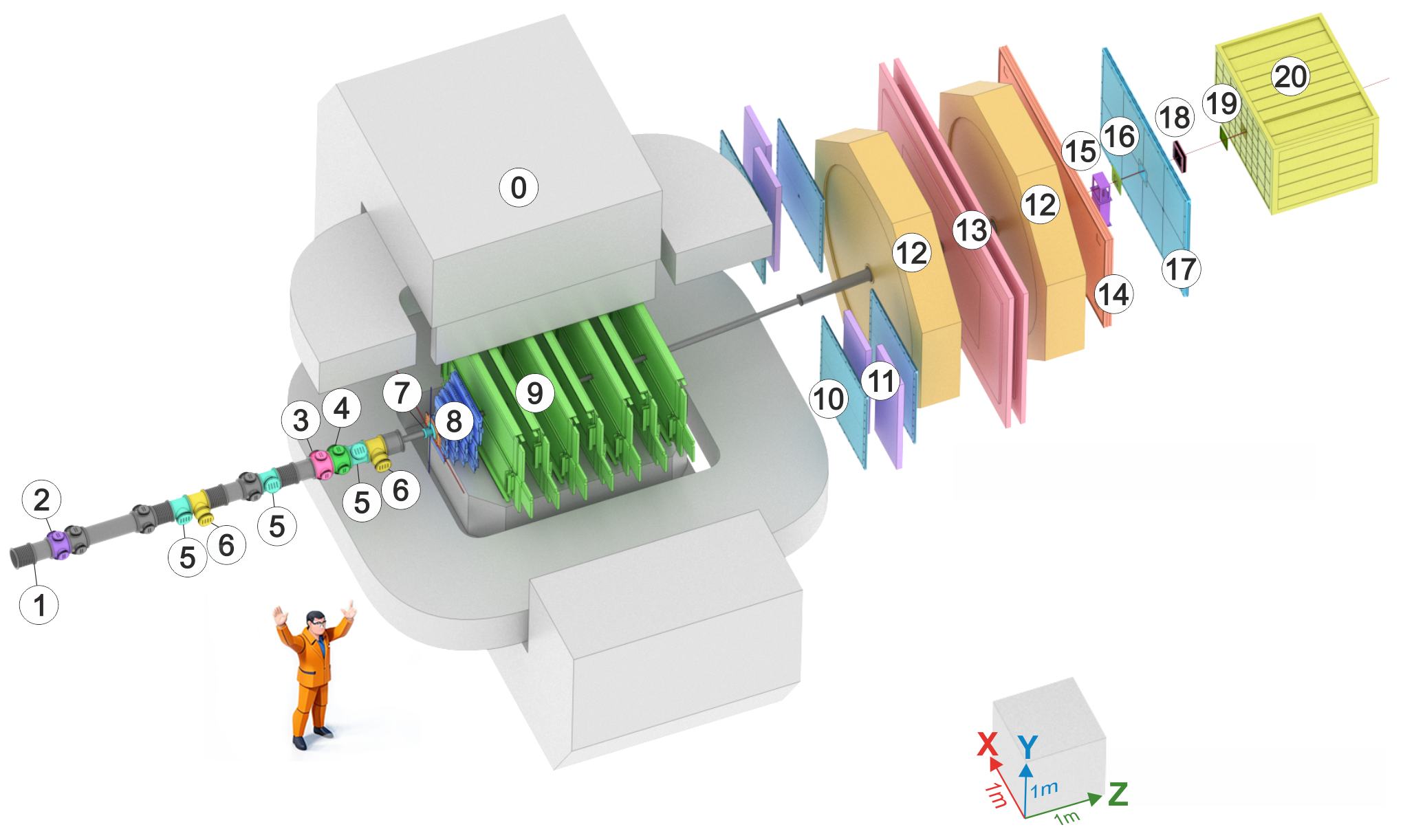}
  \label{fig:BMN_setup_RUN8}
\end{figure*}

BM@N (Baryonic Matter at Nuclotron) is the first experiment operational at
the Nuclotron/NICA ion-accelerating complex, dedicated to studying interactions of 
relativistic beams of heavy ions with fixed targets~\cite{BMN_EPJ} in the energy range that allows reaching high densities of baryonic matter~\cite{Cleymans}. The Nuclotron will provide the experiment with beams of a variety of particles, from protons to gold ions, with kinetic energy in the  range from 1 to $6\,{\rm AG}e{\rm V}$ for light ions with a $Z/A$ ratio of $\sim 0.5$ and up to $4.5\,{\rm AG}e{\rm V}$ for heavy ions with a $Z/A$ ratio of $\sim 0.4$.
At such energies, the density of nucleons in the fireball created by two colliding heavy nuclei is 3\,$-$\,4 times higher than the nuclear saturation density~\cite{Friman}. The primary goal of the experiment is to explore the QCD phase diagram in the region of high baryonic chemical potential, to search for the onset of critical phenomena, in particular, the conjectured critical end point, and to constrain the parameters of the equation of state (EoS) of high-density nuclear matter. In addition, the Nuclotron energies are high enough to study strange mesons and (multi)-strange hyperons produced in nucleus-nucleus collisions close to the kinematic threshold~\cite{NICAWhitePaper,BMN_CDR}. Studies of the excitation function of strange particle production below and near to the kinematical threshold make it possible to distinguish the hard behavior of the EoS from the soft one~\cite{Fuchs}.

The BM@N detector is a forward spectrometer that covers the
pseudorapidity range $1.6 \leq \eta \leq 4.4$. A schematic view of the BM@N setup, as used in the first physics run in 2023 with a Xe beam, is shown in Fig.~\ref{fig:BMN_setup_RUN8}. The setup comprises a dipole magnet along with several detector systems to monitor the beam, to identify produced charged particles, to measure their momentum, and to determine the
geometry of nucleus-nucleus collisions. 
Fig.~\ref{fig:BMN_setup_RUN8} lists the names of the detector systems and the abbreviations used throughout this paper.   
A detailed description of all subsystems is given in the sections below.

%\clearpage

\section{Beamline}

\subsection{Beam transport}

\begin{figure*}[tbp]
  \centering
  \caption{ NICA Complex.}
  \includegraphics[width=0.8\textwidth]{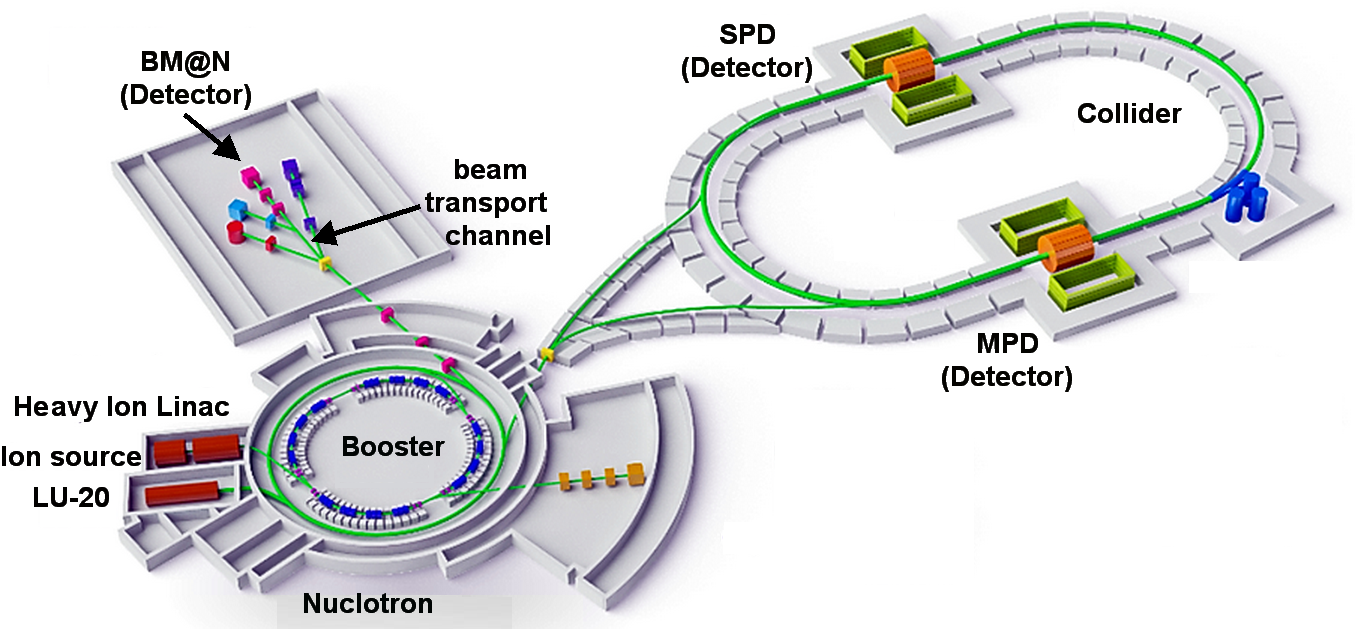}
  \label{fig:NICA}
\end{figure*}

The BM@N experiment operates as part of the NICA complex (see Fig.~\ref{fig:NICA}). For light and heavy ions, the accelerator facility provides different ways of injection into the Nuclotron. Light ions (d-Mg) are accelerated up to an energy of 5 AMeV by a linear accelerator LU-20, after which the extracted beam is directly injected into the Nuclotron ring. The heavy ion linear accelerator Linac injects the beam first into the Booster ring. After acceleration up to $500\, {\rm AM}e{\rm V}$, the beam is injected into the Nuclotron for further acceleration up to {4.5\,\rm AG}e{\rm V}. Ion beams extracted from the Nuclotron can be delivered either to the BM@N fixed target experiment or to the NICA collider.
At the collider, two experimental areas are reserved for two major experiments \,$-$\, the Multipurpose Detector (MPD) and the Spin Physics Detector (SPD) \cite{NICAWhitePaper}.

The beam extracted from the Nuclotron is transported to the BM@N experimental area over a distance of about $150\,m$ by a set of dipole magnets and quadrupole lenses. This transport line is enclosed in a vacuum stainless steel pipe. 
The final steering of the beam on the BM@N target is performed by a pair of VKM and SP-57 dipole magnets, which allow a small beam bending in vertical and horizontal planes, correspondingly. They are placed respectively at a distance of 
approximately 5.7 and $7.7\,m$ from the target (Fig.~\ref{fig:Magnets}). In addition, a doublet of quadrupole lenses is used for optimal focusing of the beam on the target. They are placed at about 10.0 and $12.5\,m$ upstream of the target, respectively. 

The target is located at the pole edge of the SP-41 analyzing magnet close to its entrance. After the target, the beam ions are deflected by the SP-41 magnetic field ($\int Bdl = 3.15\,\rm{Tm}$ at the maximum current of $2000\,A$) and the deflected beam passes through the entire setup also in the vacuum beam pipe. The position of the vacuum beam pipe after the target is fixed. For experiments at different beam energies, the magnetic field of the analyzing magnet is adjusted correspondingly. For example, studies of the Xe + CsI collisions during the 2023 Xe run were performed at Xe beam energies of {3.0\,\rm AG}e{\rm V} and {3.8\,\rm AG}e{\rm V} with the SP-41 current set to 1395 and 1720\,A, respectively.
 
\subsection{Vacuum beam pipe}

\begin{figure*}[tbp]
  \centering
  \caption{Technical design of the carbon beam pipe.}
  \includegraphics[width=1.0\textwidth]{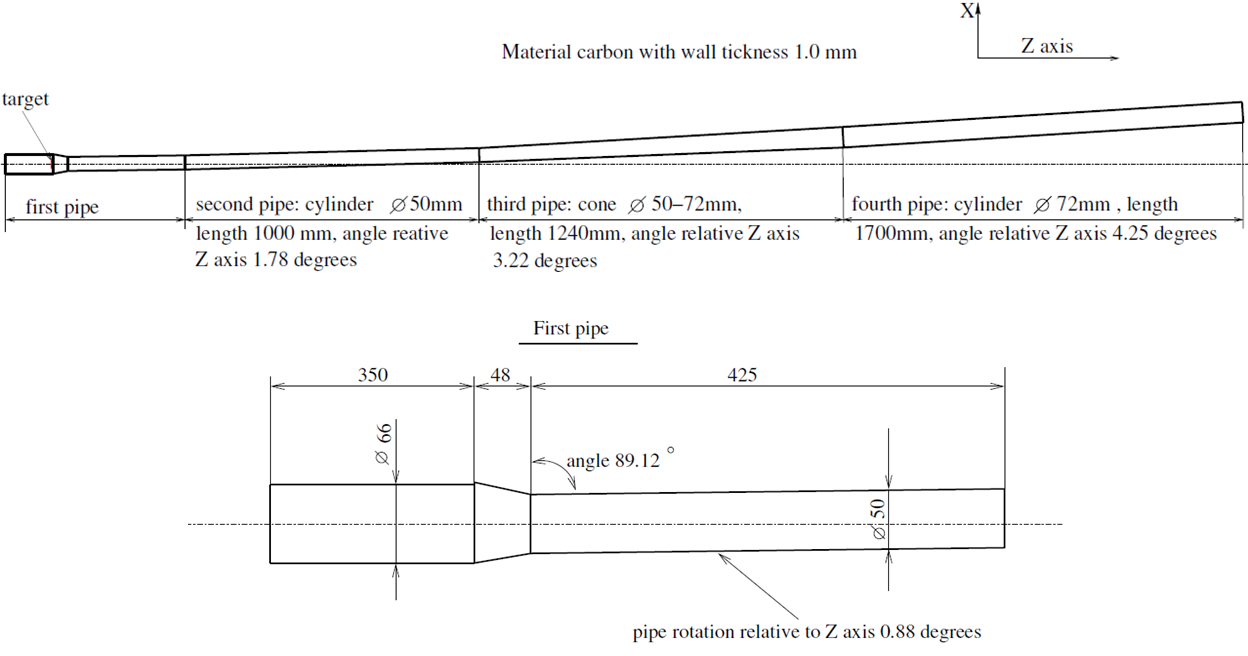}
  \label{fig:carbon_draw}
\end{figure*}

 \begin{figure*}[tbp]
  \centering
  \caption{Beam pipe: a) 3D models of the dismountable flangeless connection; b) the support scheme of the carbon beam pipe in the GEM detector notch. }
  \includegraphics[width=1.0\textwidth]{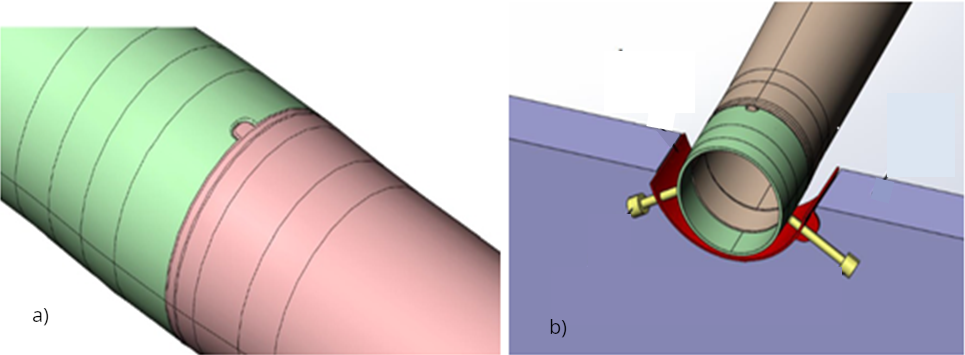}
  \label{fig:3d_connect}
 \end{figure*}

A vacuum beam pipe was integrated into the experimental setup in order to minimize the amount of scattering material in the path of the heavy ion beam. Vacuum in the entire beam pipe at the BM@N setup is achieved by a single roots pump installed upstream of the 1K200 quadrupole lens. The pressure maintained during the experiment is at the level of $10^{-4}\,Torr$. The ISO-K vacuum standard is adopted for flange connections. The stainless beam pipe with a diameter of $200\,mm$ is used in the beam transport line through the 1K200 and 2K200 quadrupole lenses and through the VKM and SP-57 corrective magnets up to the target. The vacuum level is monitored by two vacuum gauges, the data from which are recorded in the Slow Control System. The last $5\,m$ long part of the beam pipe before the target includes vacuum boxes containing the beam detectors described in the next section: two 3-way boxes for profilometers, three 3-way boxes for the Silicon Beam Tracker detectors and three 6-way boxes for the BC1, BC2, and VC trigger counters. All boxes located outside the magnetic field of the SP-41 analyzing magnet are made of stainless steel, while the vacuum pipe components, which are close to the target and placed in the magnet, are made of aluminum.

\begin{figure*}[tbp]
  \centering
  \caption{ Magnetic elements of the  BM@N setup. See text for details.}
\vspace*{0.3cm}
  \includegraphics[width=1.0\textwidth]{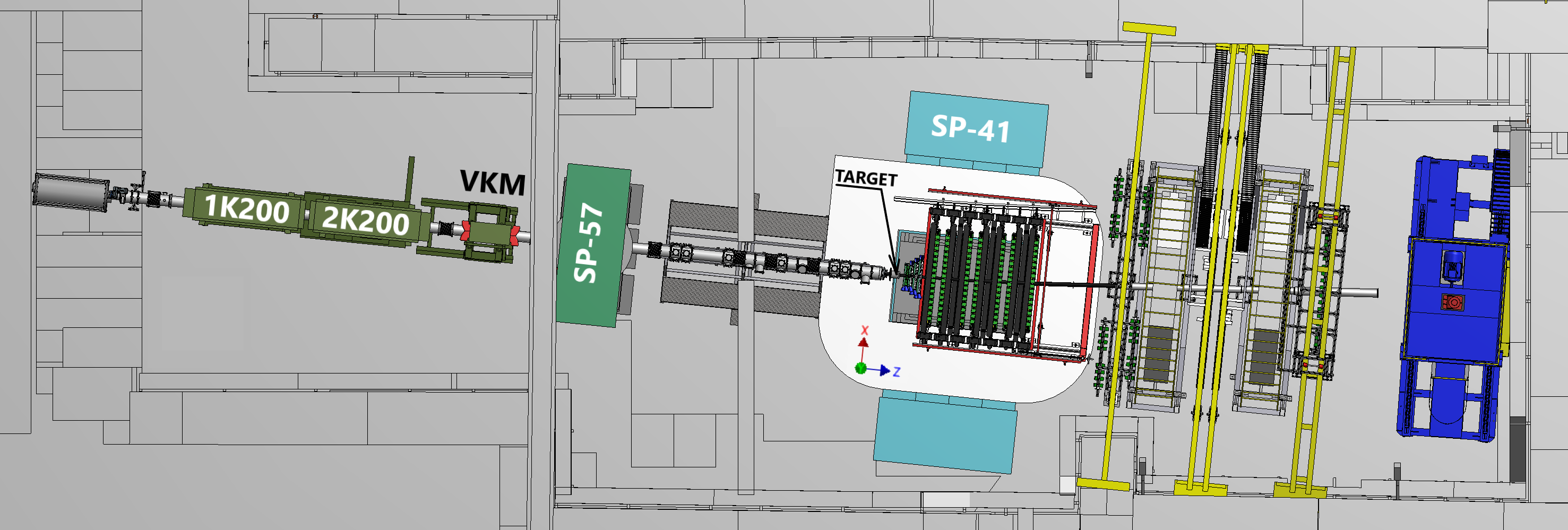}
  \label{fig:Magnets}
\end{figure*}

The bending of the beam ion trajectories by the magnetic field leads to a deflection from a straight line resulting in a few $mm$ displacement in the \textit{X} direction at the target location. During the assembly of the beam pipe vacuum elements, an adjustment is carried out in order to compensate for this deflection. For that purpose the corresponding grooves for the vacuum box O-rings are made slightly wider than dictated by the ISO standard and allow for slight off-center shifts of the vacuum pipe components. The target flange assembly is also made of aluminum as well as an ISO 240 to $66\,mm$ vacuum adapter  that provides a connection to the $4.5\,m$ long vacuum pipe that follows the target. This pipe is made of carbon fiber and consists of four straight sections of different lengths connected to each other by flangeless carbon fiber connections, which provide the possibility to align sections at slight angles with respect to each other as shown in Figs.~\ref{fig:carbon_draw} and ~\ref{fig:3d_connect}. The carbon beam pipe is suspended on two supports also made of carbon fiber and installed on two GEM detectors, the one closest to the target and the other most downstream. The supports have adjustment units for precise positioning of the carbon beam pipe on the beam axis (Fig.~\ref{fig:3d_connect}). The carbon beam pipe is designed to sustain vacuum up to $10^{-4}\,Torr$. In the straight segments its thickness is about $1\,mm$, while in flangeless connections it reaches $2\,mm$.

The last section of the beam pipe has a length of approximately $3.2\,m$ and provides vacuum along the beam trajectory through the Outer Tracker system. 
It consists of three cylindrical segments with lengths of $1.2$, $1.0$ and $1.0\,m$, made of aluminum tube with an outer diameter of $125\,mm$ and a wall thickness of $1.5\,mm$. At the end of this section, the vacuum line is closed by a $100\,\mu m$ thick titanium membrane installed in a frame after the adapter ($d = 125/150\,mm$).

\subsection{Target station}

\begin{figure}[tbp]
  \centering
  \caption{3D model of the target station:\ \ 1) Aluminum flange target station.\ \  2) Barrel Detector.\ \  3) Four targets.
  \mbox{4) Carbon beam pipe}.\ \  5) Pneumatic cylinders.}
  \includegraphics[width=0.5\textwidth]{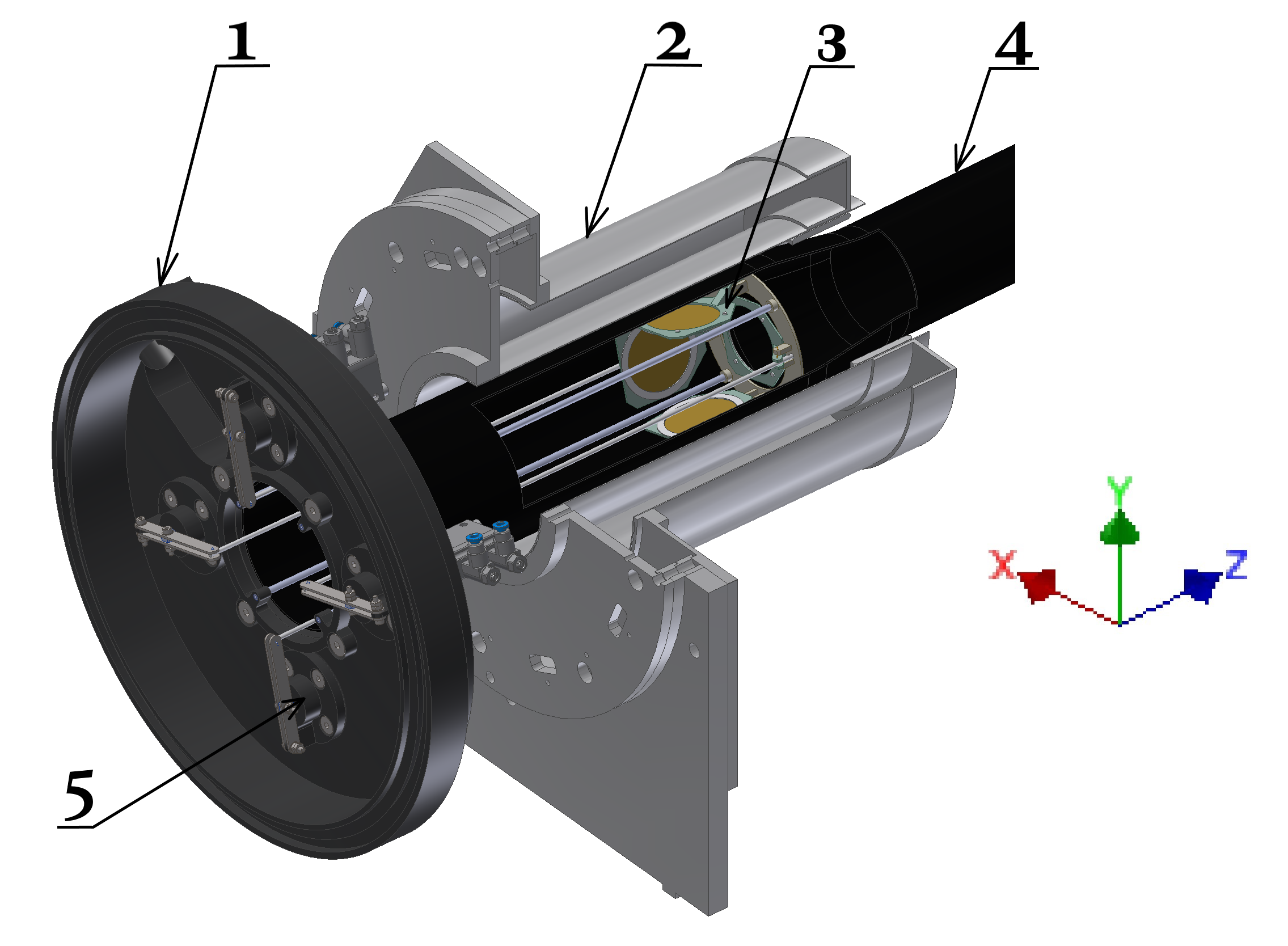}
  \label{fig:target_station}
\end{figure}

The target station is located at the end of the second beam pipe section. It is designed to provide the possibility to insert a target in the beam line inside the vacuum volume and to interchange several targets without breaking the vacuum. A 3d model of the target station is presented in Fig.~\ref{fig:target_station}. An aluminum flange of $240\,mm$ in diameter serves as a holder of the target assembly elements and as an adapter between the beam pipe upstream of the target station and the first section of the carbon beam pipe. On the outer part of this flange, four pneumatic cylinders are installed allowing four target frames to be alternately moved in and out of the beam. The pneumatic cylinders can be remotely operated, and an optocoupler sensor is used to control the position of the target frames via a dedicated electronic module.

The part of the target assembly placed inside the vacuum has a centering frame, which fits into the inner part of the first section of the carbon beam pipe.
The targets are installed in four petals. In the normal state, all the petals are leaning along the axis direction of the beam pipe. The retaining pins, made of carbon fiber, are $300\,mm$ long and $3\,mm$ in diameter.

In the 2023 Xe run, three disk targets with a diameter of $3.2\,cm$ were used: $1.75\,mm$ thick CsI, $0.85\,mm$ thick CsI, and $1.02\,mm$ thick Ge. One frame of the target assembly was left empty and was used to evaluate the background level caused by the interaction of beam particles with the structural elements of the target station.

\subsection{Magnetic field of the analyzing magnet}

\begin{figure*}[tbp]
  \caption{Analyzing magnet and magnetic field maps: 
           a) SP-41 analyzing magnet; b) $B_Y$ componentt at middle plane; 
	   c) $B_Y$ componentt at bottom plane; d) $B_Y$ componentt at upper plane.}  
  \centering \includegraphics[width=1.0\textwidth]{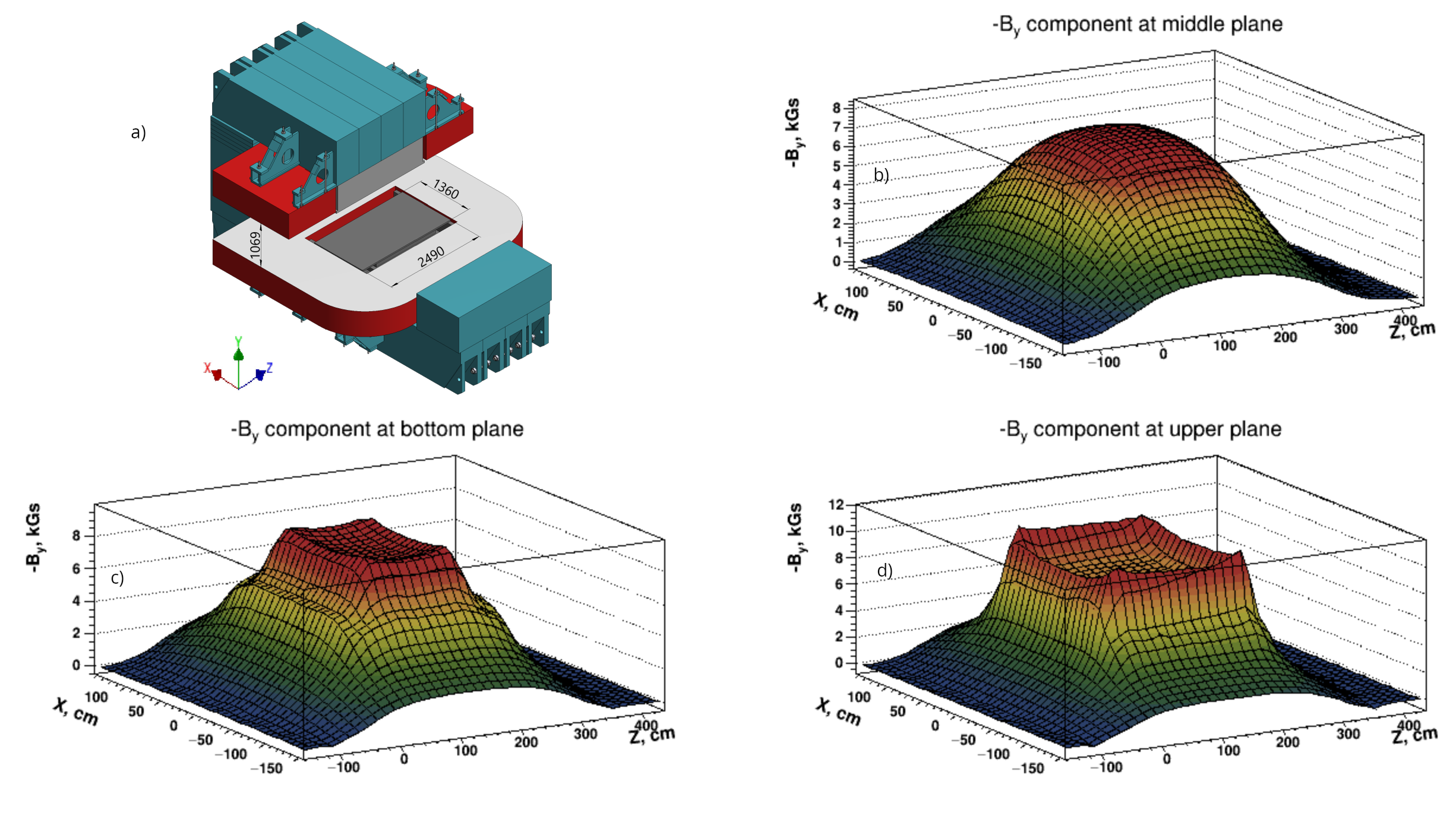}
  \label{fig:mag_field}
\end{figure*}

The SP-41 dipole magnet with large acceptance is used in the spectrometer as an analyzing magnet to measure the momenta of produced particles and beam fragments. During the preparation of the magnet for the BM@N experiment, the original configuration of the SP-41, used in previous experiments with a streamer chamber, was significantly upgraded. In particular, the camera hole in the upper pole was filled with steel to improve the uniformity of the magnetic field, and the distance between the poles was increased by approximately $30\,cm$ to provide the space required by the BM@N GEM chambers.  
The dimensions of the SP-41 pole in \textit{X} and \textit{Z} directions are about 1.4 and $2.5\,m$, respectively, while the vertical distance between the upper and lower poles after the upgrade is $1.07\,m$ (Fig.~\ref{fig:mag_field}). In the BM@N setup, the magnet is centered on the beam line. In the \textit{X} coordinate the beam axis goes through the magnet close to the center of the poles, while vertically the beam axis is shifted closer to the lower pole by approximately $40\,mm$. The leading edge of the pole defines the origin of the \textit{Z} axis, and the target is installed inside the SP-41 magnet at this position. 

Determination of the momentum of the produced particles requires a detailed knowledge of the value and orientation of the magnetic field.
After the upgrade of the SP-41, field measurements were performed by means of planar and 3D Hall probes \cite{SP41_2015}. In addition, the shape of the field was calculated by the TOSCA code using the known configuration of the yoke and coil materials.   
In order to obtain the field map for a wider \textit{X,Y,Z} range and with smaller steps, the magnetic field measurement was repeated prior to the 2023 Xe run. The measurements with 3D Hall probes covered $(-156, +145\,cm)$, $(-38, +54\,cm)$, $(-162, +439\,cm)$ and were performed in $(126\times47\times241)$ points in \textit{X, Y, Z} coordinates, respectively, allowing one to construct the field map on a $2.4\times2.0\times2.5\,cm^3$ three-dimensional grid (Fig.~\ref{fig:mag_field}). During simulation and event reconstruction, the magnetic field components at a given (\textit{x,y,z}) point are calculated by linear interpolation over eight neighboring measured nodes. The measurements of the field map were performed for four values of the current: 900, 1300, 1600, and $1900\,A$. 

%\clearpage

\section{Beam and trigger detectors} \label{BeamAndTriggerDetectors}

\begin{figure*}[tbp]
\centering
\caption{Beam, trigger, and fragment detector layout.}
\includegraphics[width=15.cm]{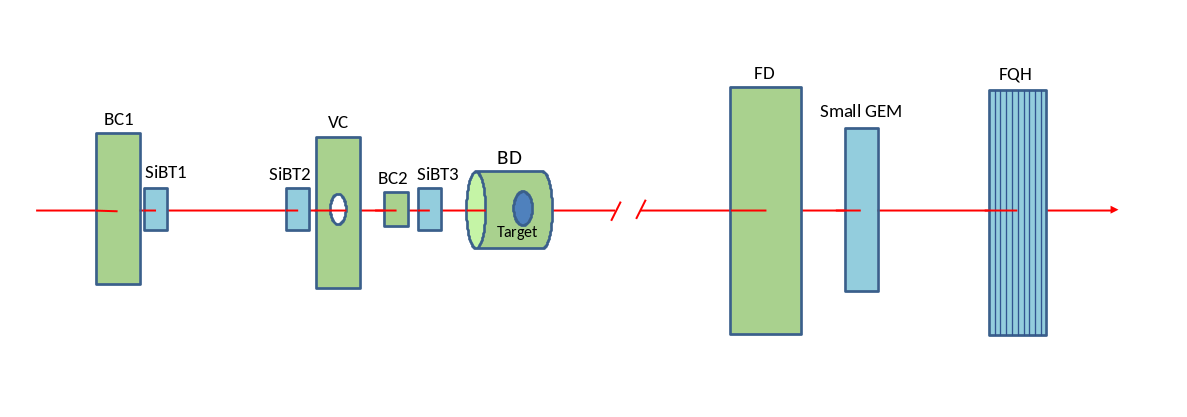}
\label{fig:trigger_fig_1}
\end{figure*}

Fig.~\ref{fig:trigger_fig_1} shows a schematic layout of the trigger detectors, placed on the beam line. In the target area the multiplicity Barrel Detector (BD) is also shown as part of the trigger system. 
Some physical parameters of the beam line detectors are summarized in Table ~\ref{table:BeamDetectors}.

The beam aperture is limited by the $25\,mm$ diameter hole in the scintillation Veto counter (VC), which rejects the beam halo. The diameter of the hole in the VC is chosen to be large enough to accept most of the beam ions, but smaller than the target diameter of $32\,mm$. Typically, in the 2023 Xe run, $80\,\%$ of the beam was accepted by the VC. 
In order to minimize interactions upstream of the target, the scintillators and active parts of the silicon detectors are located in vacuum, while the photomultiplier tubes (PMTs) of the scintillation counters and the front-end electronics of the silicon detectors are kept in the air with their housings mounted to the flanges of the beam pipe. 

\begin{table*}[tbp]
\caption{Beam line detectors.}
\label{table:BeamDetectors}
\begin{center}
\begin{tabular}{ |c|c|c|c|c|c| } 
 \hline
 Detector & \textit{Z} position, $cm$ & Active area, $mm\times mm$ & Material & Thickness, $mm$\\ 
 \hline
 BC1 & -422 & $100 \times 100$ & Scint. BC400B & 0.25\\ 
 \hline
 SiBT1 & -283 & $61 \times 61$ & Silicon & 0.175\\ 
 \hline
 SiBT2 & -183 & $61 \times 61$ & Silicon & 0.175\\ 
 \hline
  VC & -124 & $113 \times 113$ (hole $\oslash\,25$) & Plastic Scint. & 4\\ 
 \hline
 BC2 & -104 & $34 \times 34$ & Scint. BC400B & 0.15\\ 
 \hline
 SiBT3 & -84 & $61 \times 61$ & Silicon & 0.175\\ 
 \hline
 FD & +784 & $150 \times 150$ & Scint. BC408 & 0.5\\ 
 \hline
 Small GEM & +793 & $100 \times 100$ & & \\  
 \hline
 FQH & +970 & $160 \times 160$ & Quartz & 4\\
 \hline
\end{tabular}
\end{center}
\end{table*}

In all beam scintillation counters \,$-$\, BC1, BC2 and VC \,$-$\, the light from the scintillator is collected by Al-mylar light guides to a pair of PMTs, placed above and below the scintillator. This orientation of the PMTs in the BC2 and VC detectors is dictated by the requirement that they should operate in the magnetic field of the analyzing magnet, since they are located close to the target. Hamamatsu R2490-07 mesh dynode PMTs are used in the BC1 and VC detectors, whereas the BC2 has Photonis XPM85112/A1 Q400 microchannel plate PMTs.

The detectors BC1 and BC2 define the start time for the time-of-flight system. The requirement to obtain a precise time measurement favored the design of the BC1 and BC2 with light collection by two PMTs, whereas the input in the trigger logic is configured to accept one pulse from each of the beam counters, the BC1, BC2, and VC. Individual signals from the top or bottom PMT are affected by light collection non-uniformity to a larger degree than the summed signal from the individual PMTs. Therefore, the signals from individual PMTs are split by fast fan-out modules,  one output of which is used to form the summed pulse for the trigger logic, while the signals from the other output are fed to a TQDC module for offline processing of the individual pulses. Both types of PMTs used in the beam counters, Hamamatsu R2490-07 and Photonis XPM85112/A1 Q400, have excellent timing characteristics. The fan-outs have a time jitter of about $10\,ps$ and preserve a high quality of the time response. After offline correction for time walk (slewing), the time resolution obtained in the 2023 Xe run using pulses from top and bottom PMTs was found to be $\sigma_t \approx 40\,ps$ for the BC1 and BC2 individually, and $\sigma_t \approx 30\,ps$ for the combined response of the system of two counters. 

Upstream of the target the beam position is traced by three double-sided silicon strip detectors. These detectors are kept permanently in the beam and provide information about the beam ion trajectory for each event. A detailed description of the Silicon Beam Tracker (SiBT) is given in the next chapter. In addition to the SiBT, the beam position and profile can also be measured by a pair of beam profilometers, which are similar in design and parameters to the SiBT stations, but have a course pitch of $1.8\,mm$ in \textit{X} and \textit{Y}. The readout of the profilometers is organized independently of the main BM@N DAQ in order to facilitate beam tuning at the early stages of the run. The detectors of the beam profilometers can be moved in and out of the beam by remotely controlled drivers without breaking the vacuum. During data taking, the detectors of the beam profilometers are positioned outside of the beamline. 

The trigger signal based on the multiplicity of particles produced in the interaction is provided by the Barrel Detector (BD), which consists of 40 scintillator strips covering 
a cylindrical surface $\sim90\,mm$ in diameter oriented along the beam line. Each BD strip has a size of $150\times7\times7\, mm^3$, is viewed from one side by a $6\times 6\, mm^2$ silicon photomultiplier (SensL, J-ser.), and is coated with aluminized mylar.  The target is situated inside the BD, centered in the XY plane and longitudinally placed at a distance of 35 mm from the downstream edge of the BD strips. This position of the BD is dictated by the requirements for the detector to cover a sufficiently large solid angle while leaving free the acceptance of the tracking detectors of the spectrometer.

The $\delta$-electrons generated by beam ions in the target and curved by the magnetic field can contribute significantly to the number of fired strips in the BD. In order to reduce this background, the scintillator strips are protected by Pb-shielding: a $3\,mm$ thick cylinder inside the BD and a $10\,mm$ thick outer plate.   

Downstream of the analyzing magnet the beam goes through the Fragment Detector (FD), the Small GEM detector and the Forward Quartz Hodoscope (FQH). These detectors are placed in the air, the FD is positioned directly behind the $100\,\mu m$ titanium window of the vacuum beam pipe. The amplitude of the pulse in the FD reflects the charge squared of the ion passing through the counter. This amplitude is used in the trigger system in order to distinguish events with and without interactions in the target. To minimize the background from interactions within the FD itself, its radiator has to be thin, while in the \textit{X} and \textit{Y} directions the radiator should be wide enough to cover all the beam ions going through the target without interaction. The \textit{X} and \textit{Y} dimensions of the FD radiator  $15\,cm\times15\,cm$ were chosen to be large enough for a potential configuration, in which the vacuum beam pipe is extended further downstream and the FD is moved close to the FHCal. In the 2023 Xe run, the radiator made of a $0.5\,mm$ thick BC408 scintillator  was viewed by a single Hamamatsu R2490-07 PMT  placed about $50\,cm$ below the beam line. Light collection was done by an air light guide made of aluminized mylar. The pulse height resolution for the Xe peak was found to be $\sigma \approx 5.2\,\%$.    

In addition to the FD, the beam ions or spectator fragments can be detected by a $4\,mm$ thick quartz hodoscope FQH located in front of the beam hole in the FHCal. Information from this hodoscope is used in the offline analysis for event selection and determination of event centrality. The FQH amplitude resolution for Xe ions is about $2\,\%$. The detailed description of the hodoscope is given in section~\ref{FQH}. 

The small GEM detector is placed between the FD and FQH and used to monitor the position, shape and spot size of the beam downstream of the analyzing magnet. Its active area covers $10\,cm\times10\,cm$ in \textit{X} and \textit{Y}. The detector has three GEM foils and a multilayered readout board with two planes of parallel strips oriented along the \textit{X} and \textit{Y} axes with 256 strips in each coordinate.

%\clearpage

\section{Silicon Beam Tracker}

 \begin{figure}[tbp]
  \centering
  \caption{Three stations SiBT1, SiBT2, SiBT3 with detectors and FEE electronics, view along the beam.}
\vspace*{0.3cm}
  \includegraphics[width=0.45\textwidth]{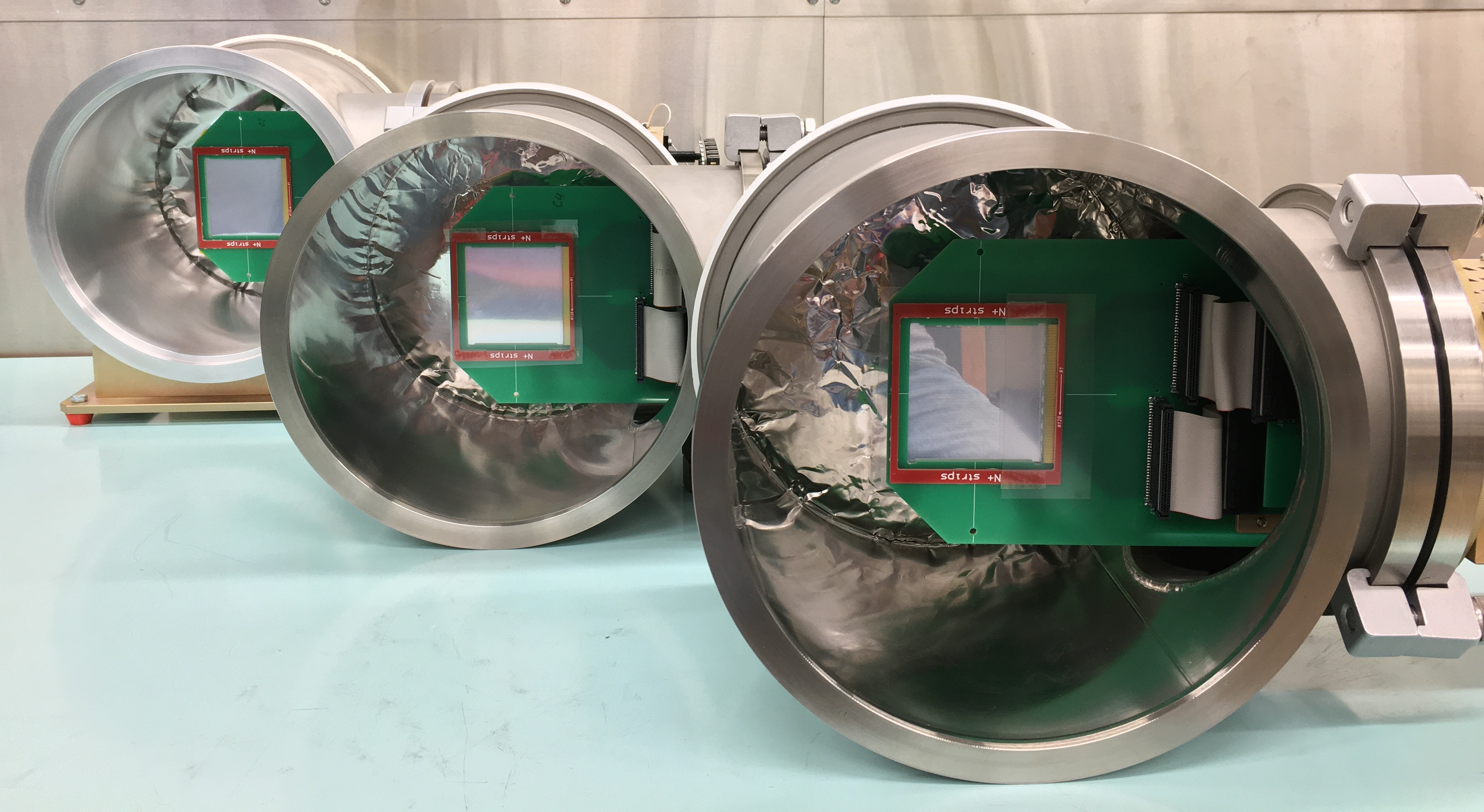}
  \label{fig:SiBT-Fig1}
\end{figure}

The main task of the Silicon Beam Tracker (SiBT) is to measure the beam ion trajectory in each event and determine the primary vertex coordinates as well as the impact angle of the beam projectile. The SiBT consists of three stations, each of which utilizes a double-sided silicon strip detector (DSSD) with dimensions of $63\times63\times0.175\,mm^3$. The DSSDs are made of high-resistivity silicon wafers obtained by the Float Zone method. The detector thickness of $175\,\mu m$ was chosen as small as possible, taking into account the limitations of the planar technology applied to 4" ($100\,mm$) wafers. 
The minimum thickness of the detectors allows not only reducing the amount of material in the beam, but also decreasing the volume of the space charge region of the detector. This reduces the noise caused by the radiation defects per strip, which is very important considering that the detectors are exposed to heavy ion beams with intensity up to $\sim1\,MHz$.

\begin{figure}[tbp]
  \centering
  \caption{SiBT stations installed in the vacuum beam pipe.}
\vspace*{0.3cm}
  \includegraphics[width=0.45\textwidth]{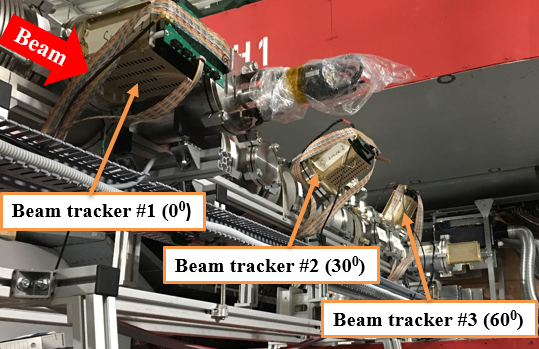}
  \label{fig:SiBT-Fig41}
\end{figure}

\begin{figure*}[tbp]
  \centering
  \caption{Two-dimensional beam profiles measured in the 2023 Xe run: a) without the Veto counter in the trigger; b) with the Veto counter in the trigger.}
\vspace*{0.3cm}
  \includegraphics[width=0.7\textwidth]{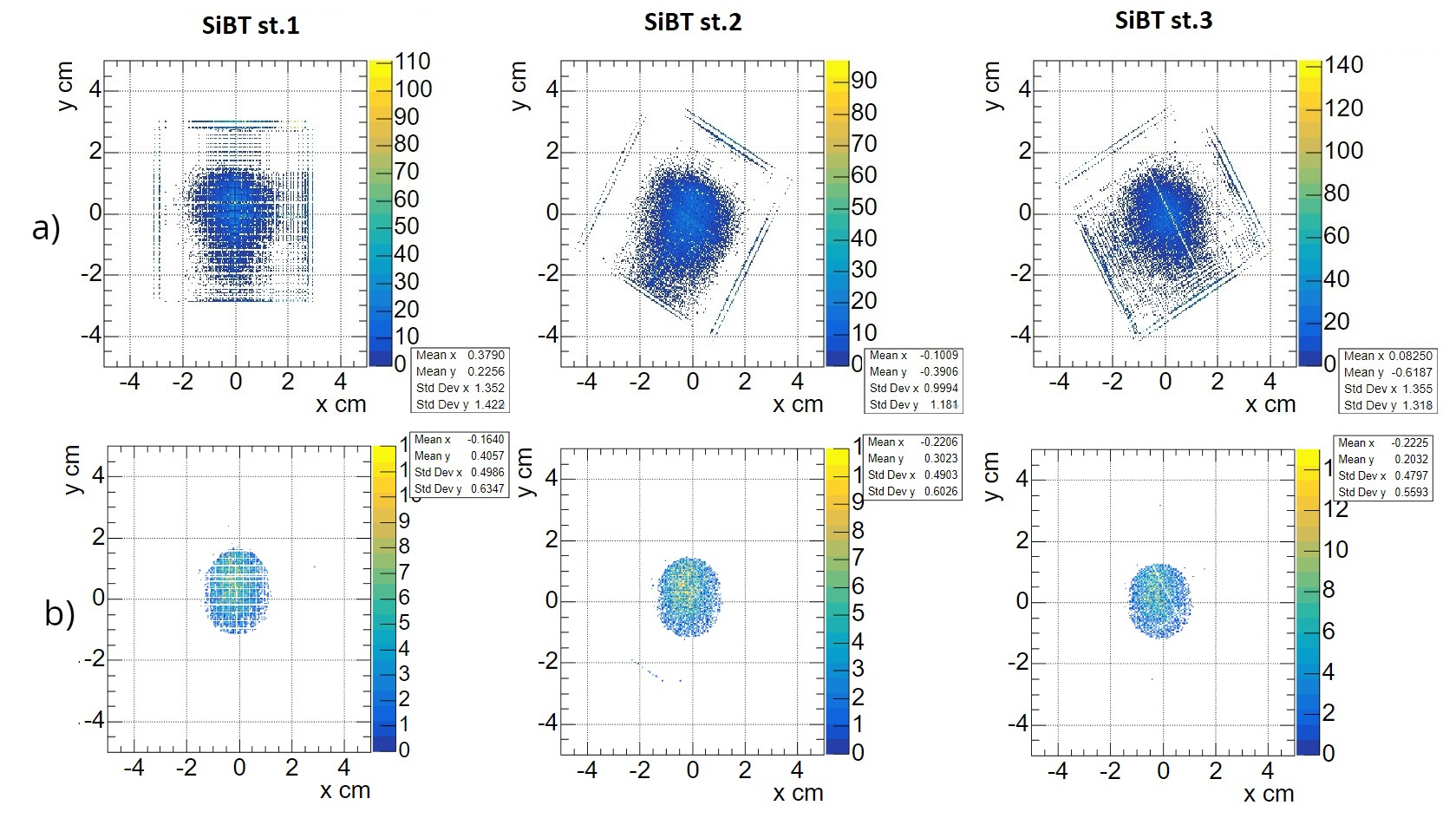}
  \label{fig:SiBT-Fig42}
\end{figure*}

Each detector has an active area of $61\times61\,mm^2$, 128 strips on both the $p^+$ and the $n^+$ sides with a pitch of $470\,\mu m $ resulting in a total of  $2\times128$ readout channels. The strips on the two sides are oriented orthogonally with respect to each other. The silicon plate in the SiBT1 detector is positioned inside the beam pipe such that the strips are aligned along the \textit{X} and \textit{Y} axes, whereas the plates of the SiBT2 and SiBT3 detectors are rotated azimuthally by $30^{\circ}$ and $60^{\circ}$, respectively.

Fig. \ref{fig:SiBT-Fig1} shows the three vacuum stations with the DSSD installed inside. The 3d coordinate positions of each DSSD relative to the geometrical axis of the beam pipe were measured using a NORGAU NVM II-5040D video meter with an accuracy of $\pm5\,\mu m$. Structurally, the detectors are assembled on printed circuit boards with gold contact pads, which are connected by ultrasonic bonding (US-bonding) with Al-plated strips on the DSSD. The signals from the detector strips, grouped in four bundles of 64 channels each, are sent via flat cables to 4 vacuum connectors fixed on the vacuum flange. The front-end electronic (FEE) plates for 128 $p^+$ and 128 $n^+$ strips are mounted on the flange outside the vacuum volume. In this case, the detector electronics is located outside of the high radiation zone. Moreover, the FEE is available for testing and tuning, and, if necessary, can be replaced without breaking the vacuum in the beam pipe.

The chip VATA64HDR16.2 (IDEAS, Norway)~\cite{FEE_IDEAS} was chosen for the FEE because of its large dynamic range \((-20\,pC\,-\,+50\,pC)\) suitable for operation with highly ionizing heavy ion beams. For example, the charge in the input signal caused by 3--$4\,AGeV$ Xe ion going through a $175\,\mu m$ layer of silicon is $11\,pC$. The ASIC VATA64HDR16.2 accepts up to 64 input channels. Therefore, four chips are used in each SiBT stations. After passing through the pulse shapers, at the time defined by an external trigger, the values of signal amplitudes from 64 strips are stored in memory capacitors. After that, in sequential reading mode using an analog multiplexer, the 64 signals are transmitted for digitization into a single ADC channel. The same read-out scheme, but with different types of ASICs (all made by IDEAS), is used for other tracking systems, namely, FSD, GEM and CSC. Main parameters of the ASICs chosen for the FEE of these tracking systems are given in Table~\ref{table:Central Tracking System}.

Fig.~\ref{fig:SiBT-Fig41} shows the three SiBT stations mounted in the vacuum beam pipe. The histograms in Fig.~\ref{fig:SiBT-Fig42}
 represent the online monitoring of the 2D distribution of beam ion hits in the SiBT. The typical RMS value of the beam profile in 
 2023 year Xe run, measured for trigger selected events, i.e., for ions passing through the $2.5\,cm$ diameter hole of the 
 Veto counter, was $0.5\,cm$ and $0.6\,cm$ in the \textit{X} and \textit{Y} coordinates, respectively.

%\clearpage

\section{Central Tracking System}

The Central Tracking System (CTS) is based on two large tracking detector systems placed inside the SP-41 analyzing magnet. These systems are the Forward Silicon Detector (FSD) located right behind the target area and a set of Gaseous Electron Multiplier (GEM) detectors installed downstream, inside the interpole volume. The FSD has four tracking planes, while the GEM system consists of seven tracking planes. In order to accommodate the beam vacuum pipe going through the setup, each tracking plane in both systems is divided into two half-plane detectors, an upper and a lower one.

The detector position and configuration of the CTS in the 2023 Xe run is shown in Figs.\,\ref{fig:CTS_side_view} and~\ref{fig:CTS_top_view}.
A detailed description of each tracking subsystem is given below.

\begin{figure*}[tbp]
  \centering
  \caption{Side view of the subsystems inside the SP-41 analyzing magnet. 1) Target station.
  2) Barrel Detector. \mbox{3) Forward Silicon Detector}. 4) GEM detectors. 5) Beam pipe.}
\vspace*{0.3cm}
  \includegraphics[width=1.0\textwidth]{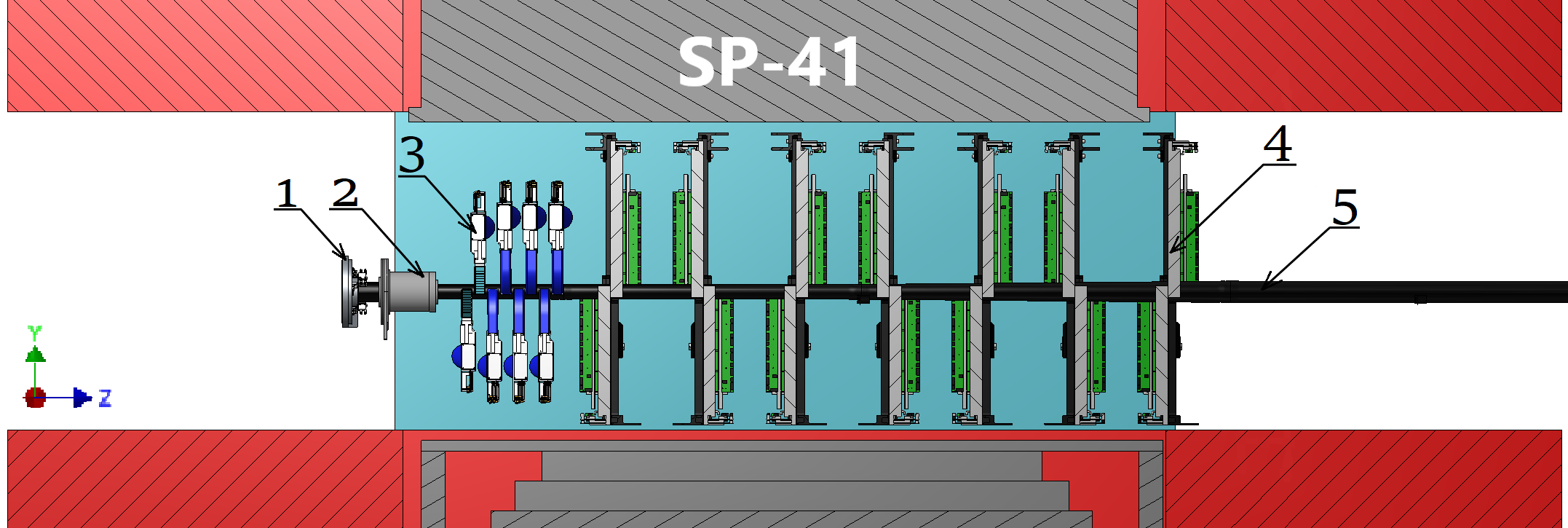}
  \label{fig:CTS_side_view}
\end{figure*}

\begin{figure*}[tbp]
  \centering
  \caption{Top view of the subsystems inside the analyzing magnet. 1) Target station.
  2) Barrel Detector. 3) Forward Silicon Detector. 4) GEM detectors. 5) Beam pipe.}
   \includegraphics[width=1.0\textwidth]{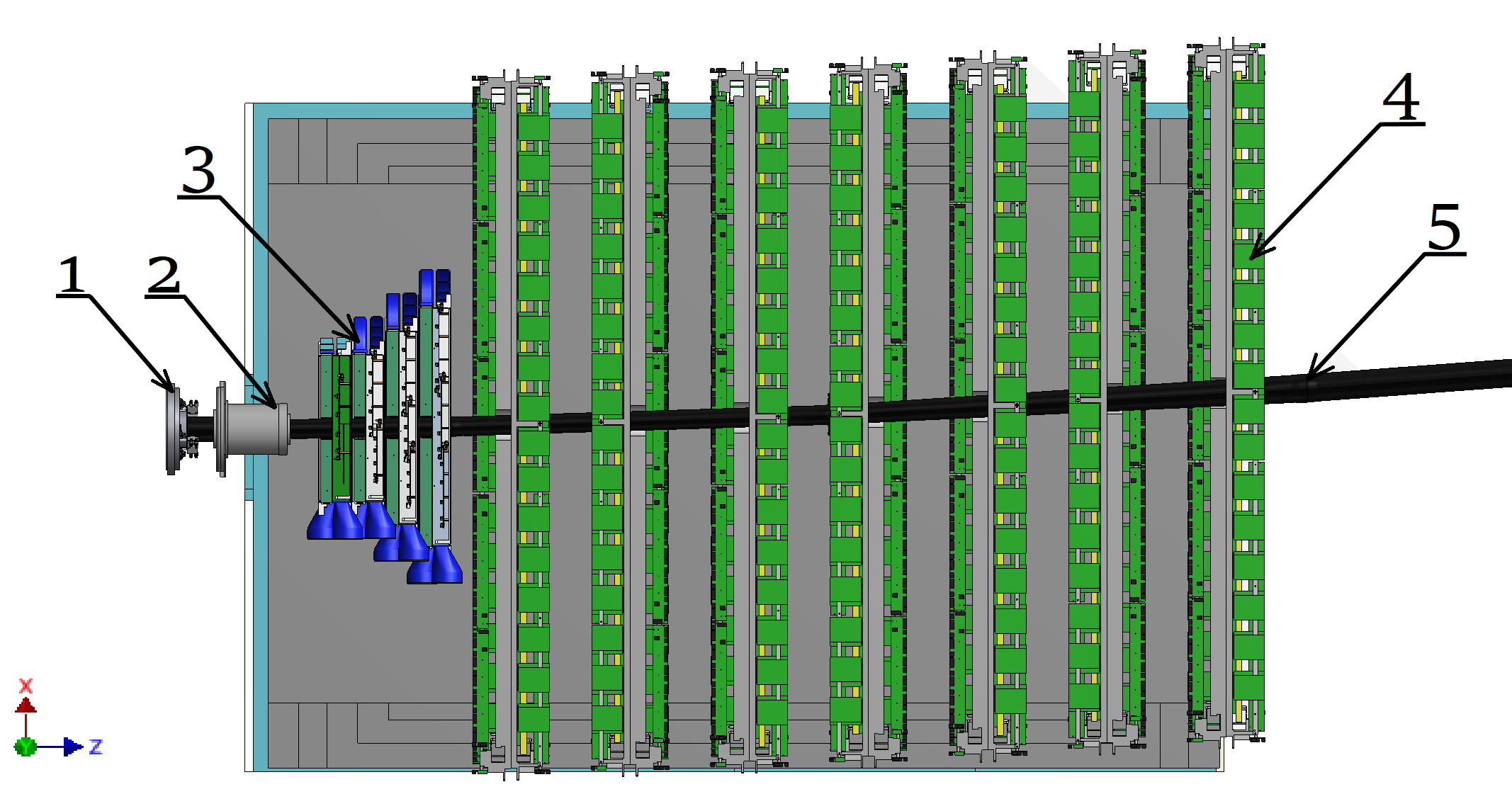}
   \label{fig:CTS_top_view}
\end{figure*}

\begin{table*}[tbp]
\begin{center}
\caption{Main FEE ASIC parameters of the SiBT, FSD, GEM and CSC tracking systems.}
\label{table:Central Tracking System}
\vspace*{0.3cm}
\begin{tabular}{ |c|c|c|c| }   
\hline
Detector &SiBT & FSD & GEM, CSC\\
\hline
Type of chip & VATAG64HDR16 & VATAGP 7.1 & VA163 \\
\hline
No of input channels in chip & 64 & 128 & 32\\
\hline
Dynamic range (AC) & $-20\,pC \div +50\,pC$ & $\pm 30\,fC$ & $\pm 750\,fC$\\
\hline
Detector signal range & $\pm 15\,pC$ & $\pm (0.5\div20)\,fC$ & $(20\div100)\,fC$\\
\hline
Noise level, $\sigma_{0} (C_{in}=0)$ & $1\,fC$ & $70\,e$ & $1069\,e$\\
\hline
Shaping time, $ns$ & 300 & 500 & 500\\
\hline
Multiplexer frequency, $MHz$ & 2.5 & 3.5 & 2.6 \\
\hline
Power consumption, $mW$ & 960 & 280 & 77\\
\hline
\end{tabular}
\end{center}
\end{table*}

%------------------------------------------------------------------------------------

\subsection{Forward Silicon Detector}

\begin{figure}[tbp]
  \centering
  \caption{The photo of the 2nd plane of the FSD taken with the light and electromagnetic shield removed to show the arrangement of the silicon modules and electronic boards.}
\vspace*{0.3cm}
  \includegraphics[width=0.48\textwidth]{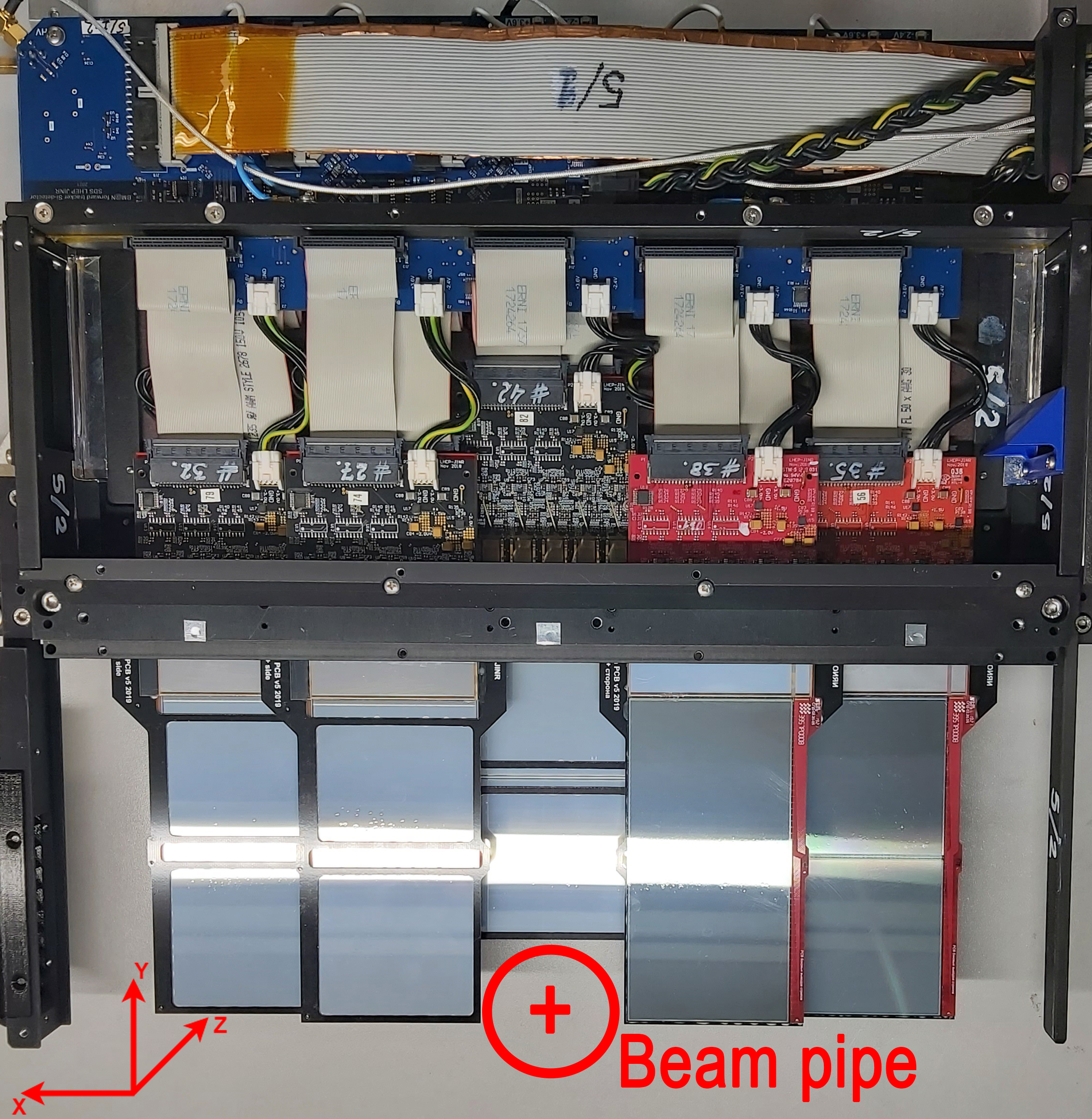}
  \label{fig:FSD-Fig.1}
\end{figure}

Each half-plane of the FSD forms an independent detector incorporating the following systems: coordinate modules based on DSSD, electronics cross-board, suspension and precise positioning mechanics, cable patch panel, air cooling, temperature monitors, light and EM shield.  The top and bottom halves of each plane are made structurally identical and interchangeable. In addition, the design allows vertical shift of the half-planes during assembly in order to provide the possibility to mount/dismount the planes regardless of the installed beam pipe and to minimize the chances of its mechanical damage. In the working position, the upper and lower half-planes form a single coordinate system with active regions overlapping along the \textit{Y} coordinate. In the center of each plane there is an insensitive $57\times57\,mm^2$ zone which makes room for the beam pipe.  A top view of the assembled eight half-planes around the beam pipe inside the SP-41 magnet is shown in Fig. \ref{fig:FSD-Fig.1}.

The first plane consists of 6 modules, each of which uses one DSSD with dimensions of $93\times63\times0.32\, mm^3$ positioned in such a way that the long side is aligned with the \textit{Y} coordinate. The detector modules of the remaining three coordinate planes use two $63\times63\times0.32\,mm^3$ DSSDs mounted on a common frame with an accuracy of $\pm20\,\mu m$. There, the strips of the same type of one DSSD are connected to the strips of another DSSD by US-bonding with an aluminum wire of $25\,\mu m$ in diameter. Table ~\ref{table:FSD-T1} provides information about the number of modules and electronic components in each FSD plane.

\begin{table*}
\begin{center}
\caption{Main parameters of the Forward Silicon Detector.}
\label{table:FSD-T1}
\vspace*{0.3cm}
\begin{tabular}{ | w{l}{4.5cm} | w{c}{1.5cm} | w{c}{1.5cm} | w{c}{1.5cm} | w{c}{1.5cm} | w{c}{1cm}| }  
\hline
Parameters & 1\textsuperscript{st} plane & 2\textsuperscript{nd} plane & 3\textsuperscript{rd} plane & 4\textsuperscript{th} plane & Total\\
\hline
Number of Si- modules & 6 & 10 & 14 & 18 & 48\\
\hline
Number of DSSDs & 6 & 20 & 28 & 36 & 90\\
\hline
DSSD size, $mm^2$ & $93 \times 63$ & $63 \times 63$ & $63 \times 63$ & $63 \times 63$ & \\
\hline
Number of ASICs & 60 & 100 & 140 & 180 & 480\\
\hline
Number of PAs & 12 & 20 & 28 & 36 & 96\\
\hline
Number of FEE PCBs & 12 & 20 & 28 & 36 & 96\\
\hline
Number of channels & 7680 & 12800 & 17920 & 23040 & 53760\\
\hline
Area, $m^2$ & 0.035 & 0.073 & 0.102 & 0.132 & 0.307\\
\hline
\end{tabular}
\end{center}
\end{table*}

The DSSDs with dimensions $63\times63\times0.32\,mm^3$ and $93\times63\times0.32\, mm^3$ were manufactured at RIMST (Zelenograd, Russia) and ZNTC (Zelenograd, Russia), respectively. The detector material is high-resistivity silicon wafers with diameters of 4" and 6", produced by the Float Zone method ($\rho>5\,k\Omega\times cm$). Each side, $p^+$ and $n^+$, contains 640 strips. The strip spacing is 95 and $103\,\mu m$, respectively, and the relative angle between the strips on the two sides is $2.5^{\circ}$. The detectors are positioned in such a way that the strips of the $p^+$ side are aligned with the \textit{Y} axis.

 \begin{figure}[tbp]
   \centering
   \caption{Example of the FSD module:
   1) Readout electronics; 2) Pitch Adapter; 3) DSSD1; 4) DSSD2;
   5) Example of US-bonding PA + DSSD1; 6) Example of US-bonding DSSD1 + DSSD2; 7) Positioning frame.}
\vspace*{0.4cm}
   \includegraphics[width=0.48\textwidth]{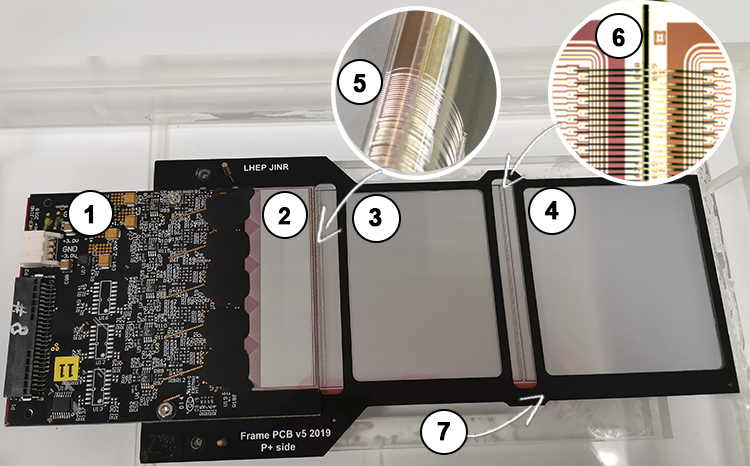}
   \label{fig:FSD-Fig.3}
\end{figure}

 \begin{figure}[tbp]
  \centering
  \caption{Functional diagram of the signal readout from a silicon detector module.}
%\vspace*{0.3cm}
  \includegraphics[width=0.48\textwidth]{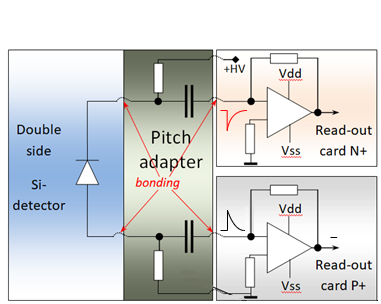}
  \label{fig:FSD-Fig.4}
\end{figure}

Fig. \ref{fig:FSD-Fig.3} shows a module with two DSSDs and a demonstration of US-bonding. A diagram of the FEE is presented in Fig. \ref{fig:FSD-Fig.4}. The detector topology (DC) does not contain integrated bias resistors and capacitors for DC decoupling of the strips from the inputs of the readout electronics. The role of the RC-bias element in the DC circuit is performed by the integrated Pitch-Adapter (PA), which also performs the matching of the strip pitch with the pad topology of inputs in the FEE ASIC. The PAs, also manufactured at ZNTC, are based on a silicon-on-sapphire structure. Each PA has 640 RC channels with $1\,M\Omega$ polysilicon bias resistors and $120\,pF/100\,V$ integral capacitors. The PA-640 integrated circuits have low leakage currents (less than $10\,pA/capacitor/100\,V$) and an electrical breakdown value of $150\,V$, which corresponds to an electric field strength in the capacitor of more than $3\,MV/cm$.

After passing the PA, the signals from the $p^+$ and $n^+$ strips of the detector are fed to the inputs of a 128-channel specialized integrated circuit VATAGP7.1 (IDEAS, Norway)~\cite{FEE_IDEAS}. Each electronic registration channel has a charging amplifier ($\sigma$ - 200 $e$), a pulse shaper (peaking time $t_s = 500\,ns$), and a memory capacitor,
which stores the pulse amplitude at the trigger time. The ASIC also uses an analog multiplexer that channels 128 inputs into 1 output sent to the readout in the DAQ by ADC. Two printed circuit boards are used in each FSD module in order to accommodate its input signals, one for the 640 negative polarity signals from the $n^+$ strips, the other for the 640 positive polarity signals from the $p^+$ strips. Correspondingly, 5 ASICs are mounted on each PCB, bonding into the pitch adapters and sealed with a compound. 

After the assembly of the modules into a half-plane, the position and rotation angles of every DSSD with respect to geodetic markers on the half-plane housing is measured using the NORGAU NVM II-5040D video meter with an accuracy of $\pm5\,\mu m$. The markers are subsequently used during the installation in order to bind the position of each detector to the common coordinate system of the experimental setup.

\begin{figure*}[tbp]
  \centering
  \caption{\textit{XY} distribution of hits in the 3\textsuperscript{rd} plane of the FSD: 
  a) in cosmic rays tests; b) in the 2023 Xe run. }
  \includegraphics[width=1.0\textwidth]{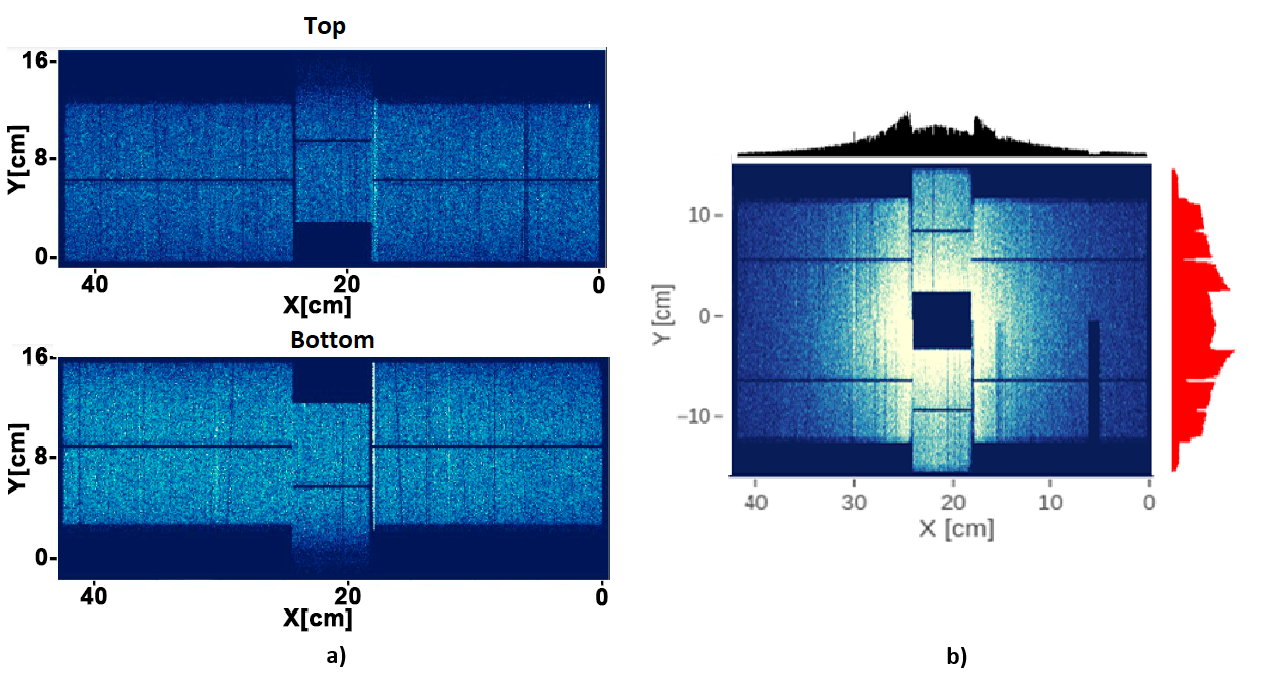}
  \label{fig:FSD-Fig.prof}
\end{figure*}

Fig. \ref{fig:FSD-Fig.prof} illustrates the distribution of hits in the 3\textsuperscript{rd} plane of the FSD observed in tests with cosmic rays and in the 2023 Xe run. Dark bands in the distribution indicate insensitive groups of 128 channels (1 chip). The number of such faulty chips at the end of the run was equal to 0, 1, 1, 8 for the 1\textsuperscript{st}, 2\textsuperscript{nd}, 3\textsuperscript{rd} and 4\textsuperscript{th} planes, respectively, which corresponds to 0.0, 1.0, 0.7 and $4.4\,\%$ of the channels. This malfunction can be due to the following reasons: 1) broken electrical contact in the transmission circuit from the chip (FEE buffer, cross-board connector, patch panel cable, long ADC-64 cable) 2) failure of the chip (no programming of the operating mode) or breakage of the US-bonding. Defects of the first type can be repaired, while the failures in the second group are of a permanent nature.  

\subsection{GEM detectors} 

Triple-GEM detectors are located inside the SP-41 analyzing magnet  downstream of the FSD.
The full GEM tracking system configuration implemented in the 2023 Xe run consists
of 14 detectors forming 7 tracking planes: 7 top detectors above the vacuum beam pipe
and 7 bottom detectors below the pipe. Since the beam line inside the SP-41 is
closer to the bottom pole of the magnet, in order to cover the maximum possible 
acceptance, the top and bottom detectors have been designed with different active 
area sizes, $163\times45\,cm^2$ and $163\times39\,cm^2$, respectively. 

\subsubsection{Design of GEM detectors} 

\begin{figure*}[tbp]
  \caption{Design of the GEM detectors: a) Microscopic picture of the GEM foil; 
  b) Cross-section of the triple GEM detector; c) Microscopic picture of the multilayer 
     readout board with inclined strips on the top; d) Schematic view of the detector readout board.}
\vspace*{0.3cm}
\begin{minipage}[h]{0.32\linewidth}
\centering
    \includegraphics[width=1\textwidth]{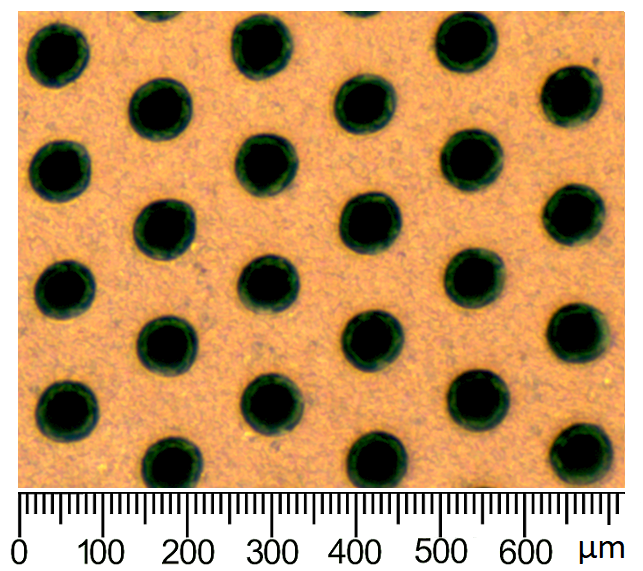}%
    \\ a) \\
\end{minipage}
\hfill
\begin{minipage}[h]{0.32\linewidth}
\centering
    \includegraphics[width=1\textwidth]{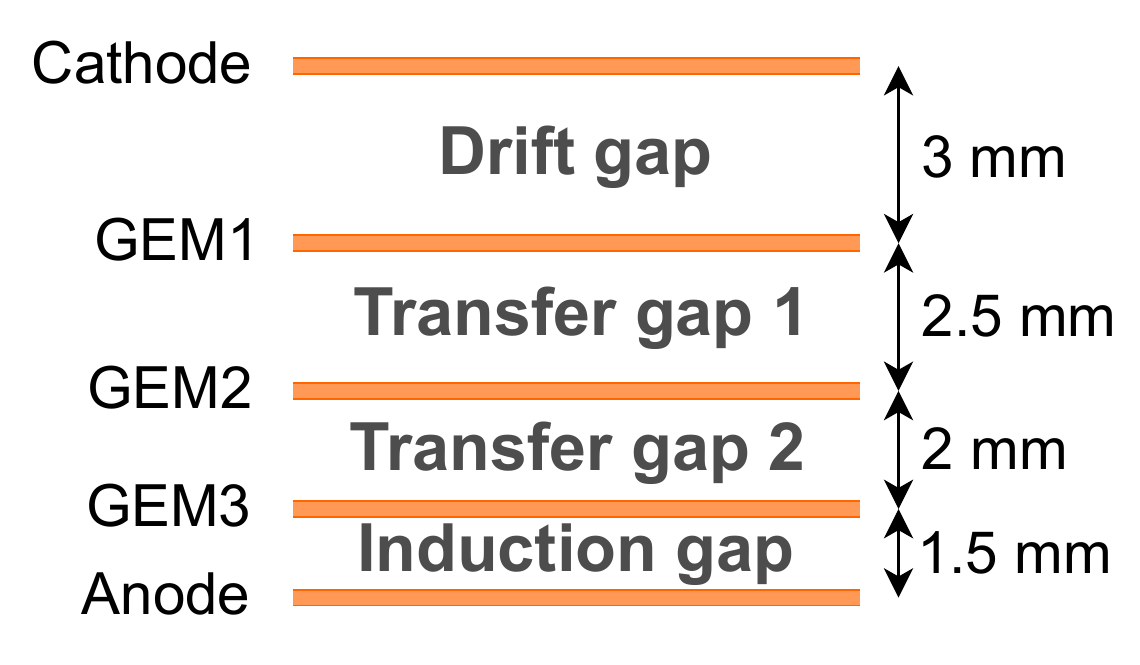}%
    \\ b) \\
\end{minipage}
\hfill
\begin{minipage}[h]{0.32\linewidth}
\centering
    \includegraphics[width=1\textwidth]{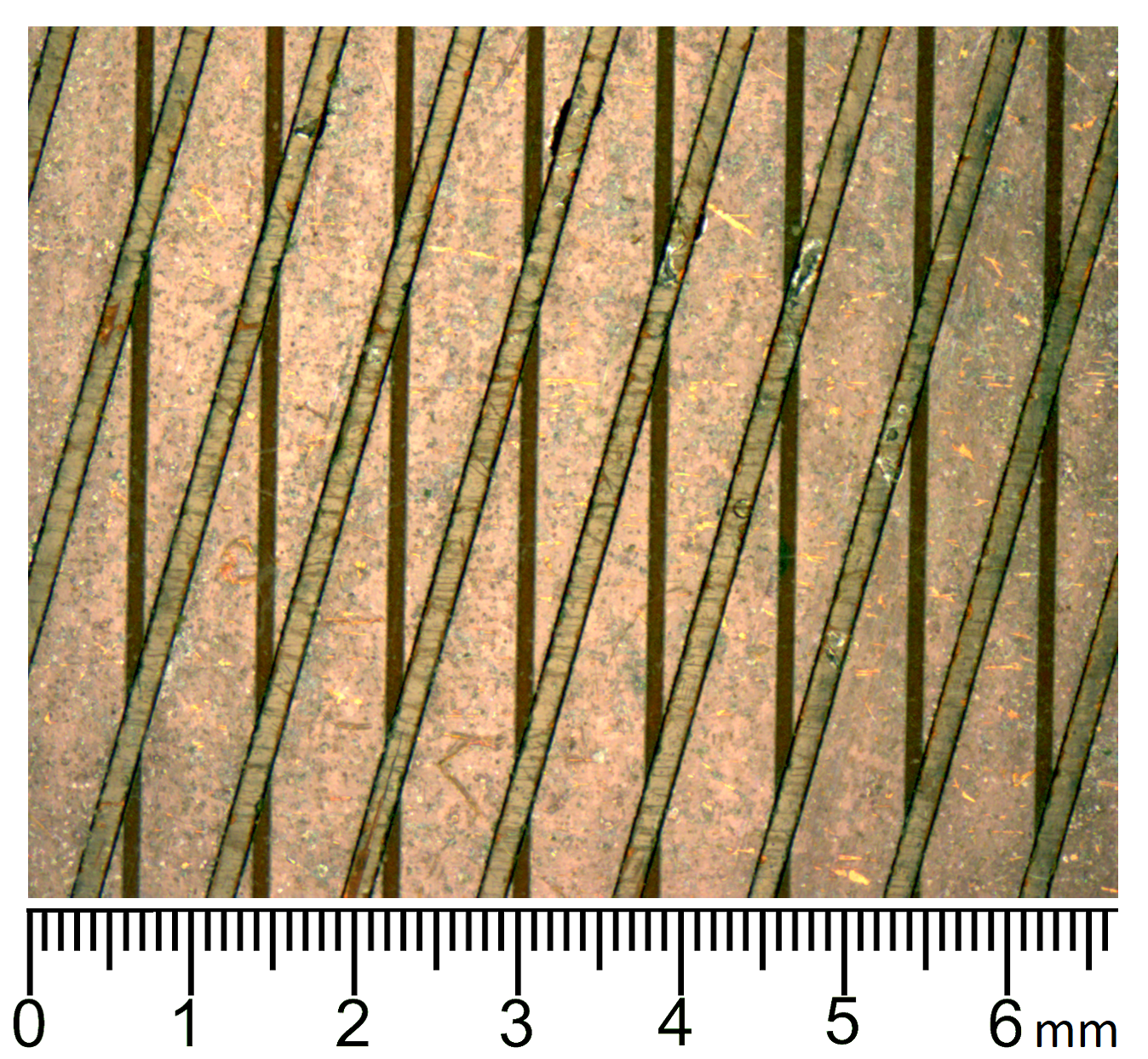}%
    \\ c) \\
\end{minipage}
\hfill

\begin{minipage}[h]{1\linewidth}
\centering
    \includegraphics[width=1\textwidth]{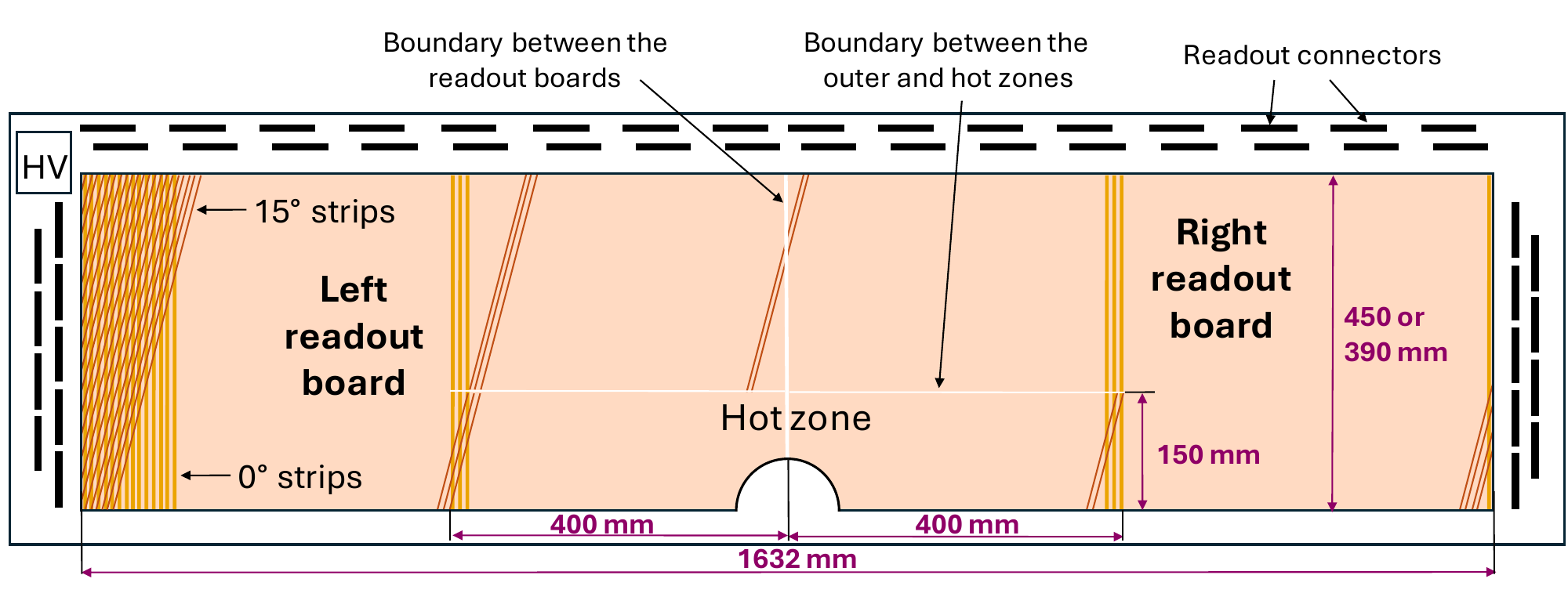}%
    \\ d) \\
\end{minipage}
  \label{fig:GEM_details}
\end{figure*}

The BM@N GEM detectors were produced using non-glue ``foil-stretching" technology and assembled at CERN in the PH Detector Technologies and Micro-Pattern Technologies workshop. All three GEM foils in a detector are identical and made of a $50\,\mu m$ thick Kapton foil covered on both sides with $5\,\mu m$ copper electrodes. The foils are perforated by double-conical holes with an outer diameter of $70\,\mu m$ and an inner diameter of $50\,\mu m$. The hole pitch is  $140\, \mu m$. Microscopic picture of the GEM foil is shown in Fig.~\ref{fig:GEM_details}\,a. The gaps between the electrodes are shown in Fig.~\ref{fig:GEM_details}\,b. 

The cathode is made of $50\,\mu m$ Kapton covered by $5\,\mu m$ copper on one side. The anode plane is used for the readout and organized as a multilayer board with two types of parallel strips: aligned with the vertical axis and inclined by 15 degrees with respect to it,
as shown in Fig.~\ref{fig:GEM_details}\,c. The width of vertical and inclined strips is
$680\,\mu m$ and $160\,\mu m$, respectively, while the pitch for both types of strips is $800\,\mu m$.
The readout plane is subdivided by two halves and, in addition, a separate readout is organized for the region close to the beam pipe where higher density of hits is expected (Fig.~\ref{fig:GEM_details}\,d). The size of this ``hot zone" is approximately $80\times15\,cm^2$. The readout FEE boards are mounted on the frames of the detectors outside of the acceptance. Both the cathode and anode are glued to honeycomb planes to provide rigidity of the detector.

A ceramic HV divider is used to apply voltages to the GEM foils and the cathode. The operating high voltage is $-3390\,V$, which corresponds to a divider current of $435\,\mu A$. The electric fields in the gaps are $E_{drift} = 1.8\,kV/cm$, $E_{trans1} = 2.3\,kV/cm$, $E_{trans2} = 3.3\,kV/cm$, and $E_{Ind} = 3.8\,kV/cm$. The voltages applied to the GEM foils are $\Delta{V_{GEM1}} = 360\,V$, $\Delta{V_{GEM2}} = 340\,V$, and $\Delta{V_{GEM3}} = 325\,V$. More details on the design, tests and preparation of the detectors can be found in ~\cite{GEM_tracking_system_JINST_17, GEM_tracking_system_JINST_20}.

The main parameters of the GEM system are presented in Table ~\ref{table:GEM_Summary}.
\begin{table*}[tbp]
\begin{center}
\caption{Main parameters of the GEM system.}
\vspace*{0.3cm}
\label{table:GEM_Summary}
\begin{tabular}{ |c|c|c|c|c|c|c|c|c| } 
 \hline
  & \multicolumn{4}{c|}{Top GEM detector}
  & \multicolumn{4}{c|}{Bottom GEM detector}
  \\
  \cline{2-9}

  & \multicolumn{2}{c|}{Left}
  & \multicolumn{2}{c|}{Right}
  & \multicolumn{2}{c|}{Left}
  & \multicolumn{2}{c|}{Right}
  \\
  & \multicolumn{2}{c|}{readout}
  & \multicolumn{2}{c|}{readout}
  & \multicolumn{2}{c|}{readout}
  & \multicolumn{2}{c|}{readout}
  \\
  & \multicolumn{2}{c|}{board}
  & \multicolumn{2}{c|}{board}
  & \multicolumn{2}{c|}{board}
  & \multicolumn{2}{c|}{board}
  \\
  \cline{2-9}

  & Outer & Hot & Outer & Hot & Outer & Hot & Outer & Hot 
  \\
  & zone & zone & zone & zone & zone & zone & zone & zone 
  \\
  \hline
  
  Number of $0^{\circ}$ strips & 1019 & 500 & 1019 & 500 & 1021 & 501 & 1019 & 500
  \\
  \hline
  Number of $15^{\circ}$ strips & 1081 & 488 & 1130 & 506 & 1062 & 488 & 1111 & 506
  \\
  \hline
  Total number of detector strips &\multicolumn{4}{c|}{6243} &\multicolumn{4}{c|}{6208} 
  \\
  \hline
  Detector active area, ${cm^2}$ &\multicolumn{4}{c|}{$163\times45$} &\multicolumn{4}{c|}{$163\times39$} 
  \\
  \hline
  FEE on one detector &\multicolumn{4}{c|}{50} &\multicolumn{4}{c|}{50} 
  \\
  \hline
  Number of detectors &\multicolumn{4}{c|}{7} &\multicolumn{4}{c|}{7} 
  \\
  \hline
  Total active area, $m^2$ &\multicolumn{8}{c|}{9.58}  
  \\
  \hline
  Total number of FEE channels &\multicolumn{8}{c|}{87157}  
  \\
  \hline

\end{tabular}
\end{center}
\end{table*}

\subsubsection{Mechanical support} 

The mechanical support of the GEM detectors inside the SP-41 magnet was designed and manufactured by  LLC ``Pelcom Dubna Machine-Building Plant'' (Dubna, Russia). The support structure is made of non-magnetic material and satisfies strict requirements for precise positioning of the detectors. The weight of one GEM detector equipped with mechanics, front-end electronics and cables is about $19.5\,kg$.
\begin{figure*}[tbp]
  \centering
  \caption{Photo of GEM detectors installation viewed from the back side of the SP-41 magnet (in the direction opposite to the beam).}
\vspace{0.3cm}
  \includegraphics[width=0.8\textwidth]{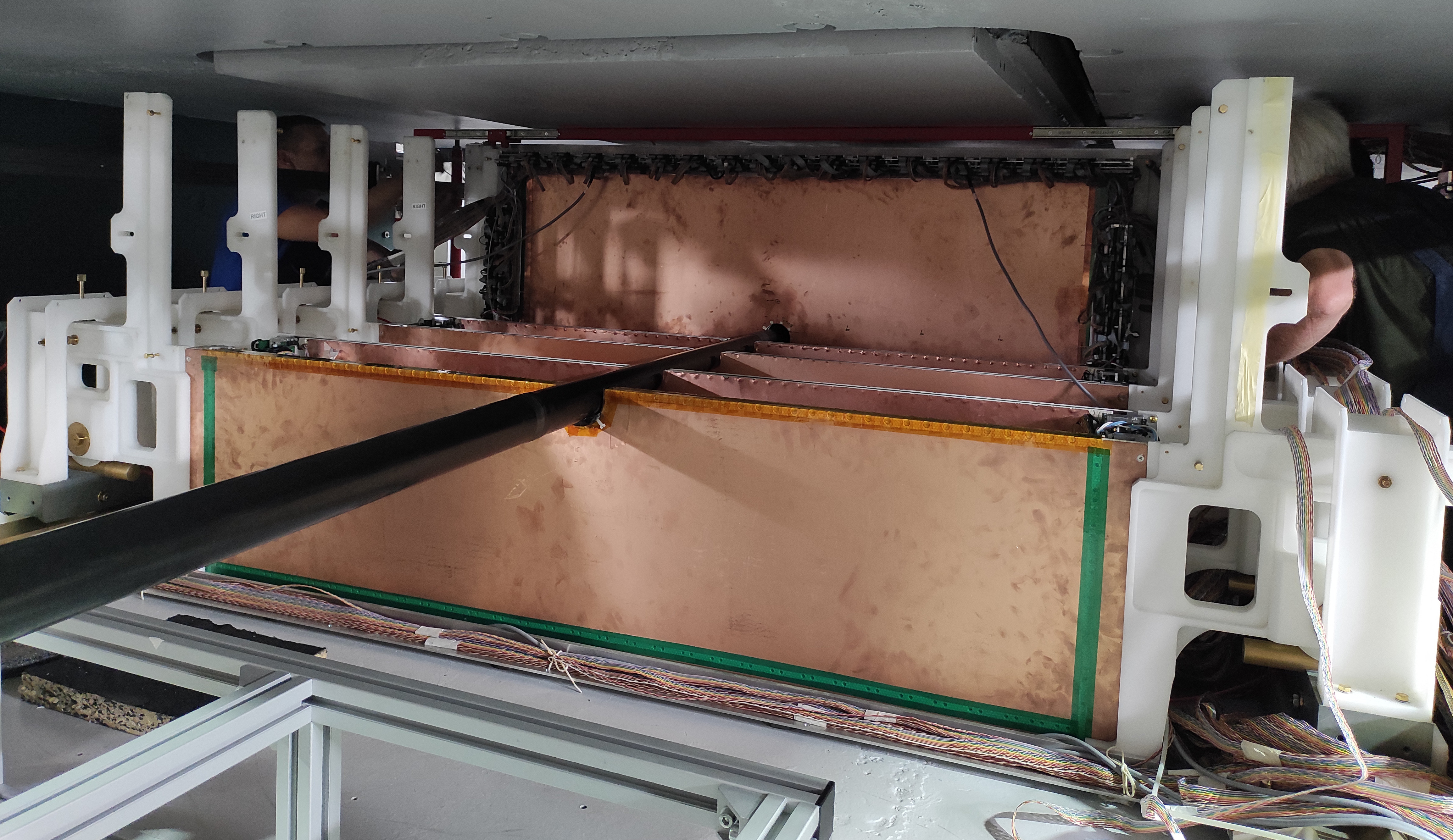}
  \label{fig:GEM_installation}
\end{figure*}
The whole assembly of 14 GEM detectors can be vertically adjusted by $\pm10\,mm$ relative to the surface of the magnet coil. In addition, the setup allows shifting each GEM detector vertically by $\pm5\,mm$ with respect to the mechanical support. The accuracy of positioning of each detector relative to the other (one half-plane relative to the other half-plane) does not exceed $0.2\,mm$. After the installation and alignment of each detector, the coordinates of the frame corners and the center of the semi-circular notch for the beam pipe are measured with an accuracy better than $0.5\,mm$.

The detector installation sequence is as follows: 1) the bottom detectors are installed sequentially, starting from the detector closest to the target; 2) the carbon beam pipe is installed 
on top of the bottom detectors; 3) the top detectors are installed sequentially, starting from the detector closest to the target. A photo of the installation process is shown in Fig.~\ref{fig:GEM_installation}.

Prior to the installation, detectors were tested with cosmic muons in order to determine the gain uniformity across the detector area. Lower amplification in the outer parts of the detectors was observed, with typical variations in the range of $\pm20\,\%$. The detectors, which due to this effect can have lower efficiency in the outer parts, were installed closer to the target, where the outer regions are less important for track reconstruction.

\subsubsection{Gas system} 
The $Ar(80)C_4H_{10}(20)$ gas mixture was chosen for the operation in the 2023 Xe run, while the overall GEM detector amplification was maintained at the level of $3\times10^4$. 
The $H_{2}O$ and $O_{2}$ removal filter was installed in the gas system after the gas mixer. The gas line was divided into two identical lines for independent connection of the top and bottom GEM detectors. Two rotameters were installed in the lines, allowing one to regulate the gas flow. Each line connected a group of top or bottom detectors in series, starting with the detector closest to the target. The small GEM detector (see the section ``Beam and trigger detectors") was connected last in the line for the bottom detectors. The gas flow rate in each line during the 2023 Xe run was at the level of $3\,l/h$.

\subsubsection{Front-end electronics} 

Front-end electronics are based on the 32-channel integrated circuit VA163 (IDEAS, Norway)~\cite{FEE_IDEAS}.
Each channel of the ASIC has a charge sensitive preamplifier, a shaper with $500\,ns$ peaking time, and a sample holder circuit. An analog multiplexer sends 32 sampled signals channel by channel into one serial readout. Four ASICs are joined in one front-end board. The multiplexed data from each board are transmitted through $13\,m$ of a twisted pair flat cable to a 12-bit analog-to-digital
converter. A more detailed description of the FEE can be found in ~\cite{GEM_performance_18}.

%\clearpage

\section{TOF systems}

 Two time-of-flight systems are used in BM@N for charged particle identification. The first system, TOF400, is placed at about 4 meters from the target and consists of two arms to the left and right of the beam axis. It is focused on identifying particles flying at high polar angles. The time-of-flight distance does not allow effective separation of charged particles near the beam axis. The second wall, TOF700, is located at a distance of about 7 meters from the target, sufficient for an effective separation of particles at small angles. The arrangement of both systems provides continuous geometric acceptance and overlap with the FSD, GEM and Outer Tracker subsystems. 
The choice of detectors and their parameters was dictated by the following requirements:
\begin{enumerate}
    \item[-] high granularity and rate capability to keep the overall system occupancy below $15\,\%$, while minimizing efficiency degradation due to double hits;
    \item[-] position resolution better than 1 cm in order to provide effective matching of TOF hits with tracks;
    \item[-] high combined geometrical and detection efficiency (better than $85\,\%$);
    \item[-] separation of pions and kaons in the momentum range $0.1 < p < 3\,{\rm G}e{\rm V}$;
    \item[-] separation of kaons and protons in the momentum range $0.3 < p < 5\,{\rm G}e{\rm V}$.
\end{enumerate}

To achieve necessary performance, a strip-readable Multigap Resistive Plate Chamber (MRPC) detector was chosen for both TOF subsystems. This type of detectors is widely used for time-of-flight measurements. It shows good efficiency, excellent time resolution and the ability to work with particle flux up to tens of $kHz/cm^2$.

\subsection{TOF400}

\begin{figure}[tbp]
  \centering
  \caption{Schematic view of the TOF400 system placed downstream the analyzing magnet. The green rectangles represent MRPCs.}
  \includegraphics[width=0.48\textwidth]{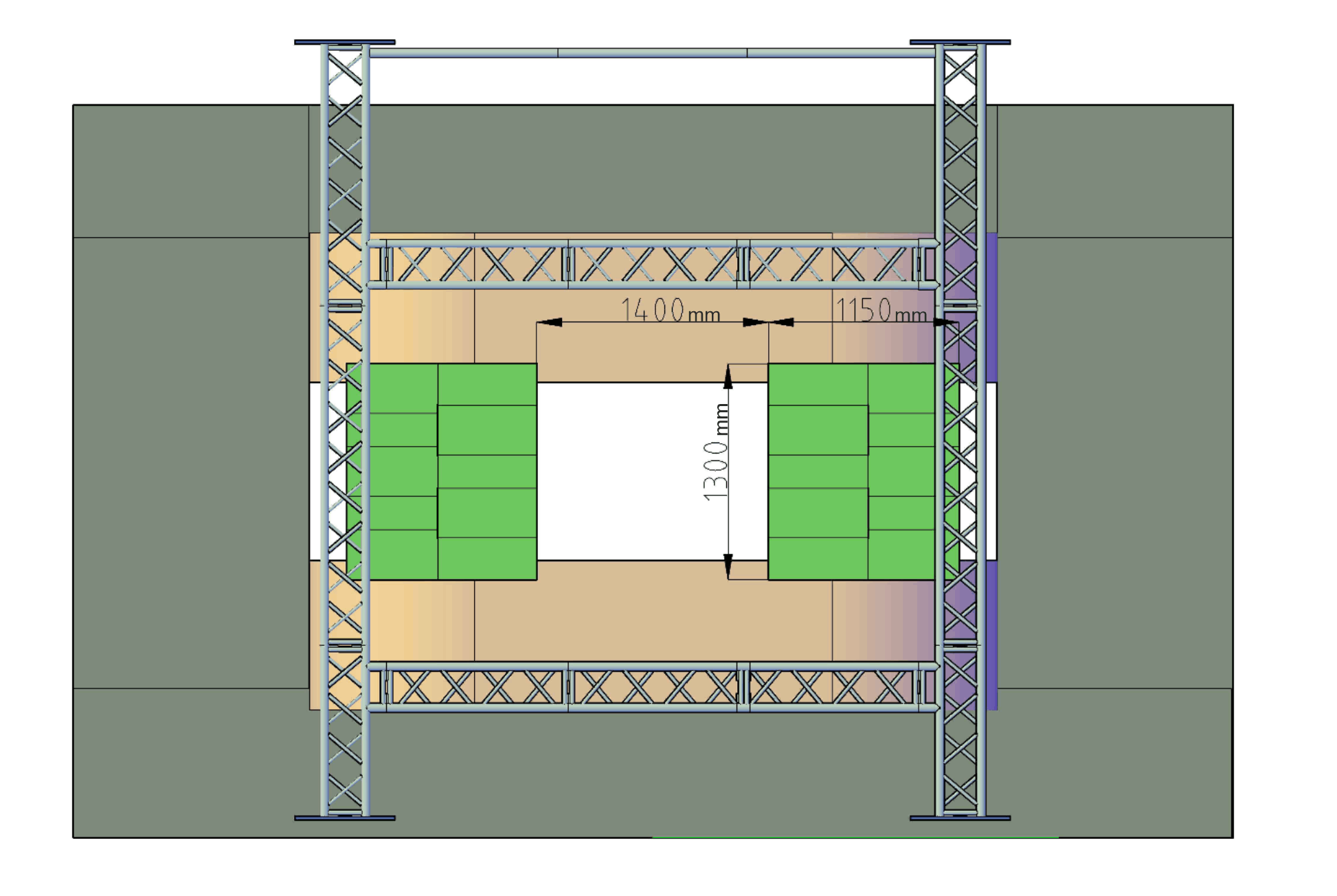}
  \label{fig:TOF400_Front_View}
\end{figure}

The left and right arms of the TOF400 system are placed at a distance of about 4 m from the target, symmetrically with respect to the beam. Each arm consists of two gas boxes (modules), each having 5 MRPC detectors (Fig.~\ref{fig:TOF400_Front_View}). The active area of one detector is $60\times30\,cm^2$. Inside the box, the active areas of adjacent detectors overlap vertically by $50\,mm$, while the horizontal overlap of the gas boxes ensures crossing of the detector active area by $50\,mm$ as well. This makes the total active area of each of the two arms to be equal to $1.1\times1.3\,m^2$, matching the geometrical acceptance of the $1.1\times1.1\,m^2$ CSCs and covering a significant fraction of the GEM system acceptance. Each gas box is formed by an aluminum frame closed from the front and back sides by aluminum honeycomb plates, which provide sufficient rigidity while having small thickness in radiation lengths.  

\begin{figure}[tbp]
  \centering
  \caption{Schematic cross-section of the TOF400 MRPC.}
  \includegraphics[width=0.48\textwidth]{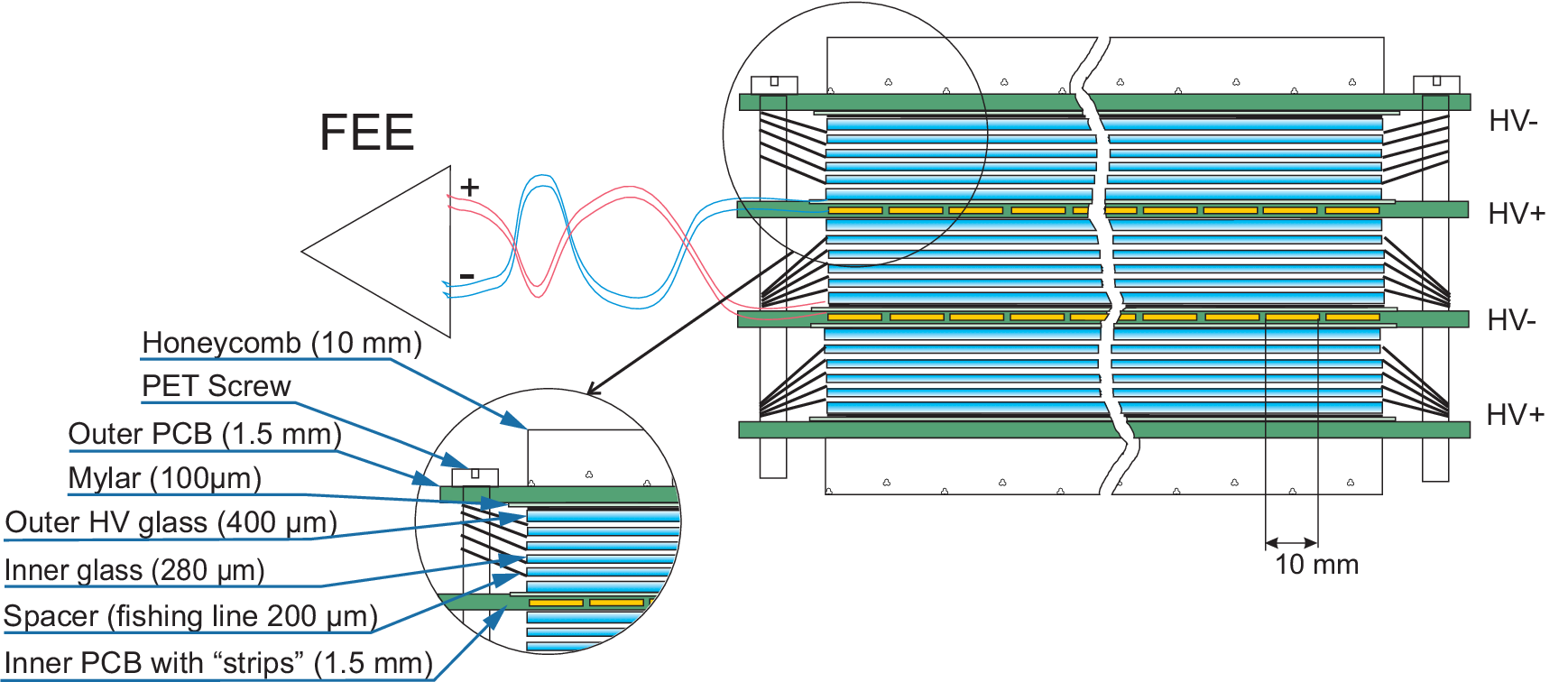}
  \label{fig:TOF400_Schematic}
\end{figure}

Fig.~\ref{fig:TOF400_Schematic} shows a schematic cross-section of the TOF400 MRPC. The detector consists of three stacks inserted between two outer $1.5\,mm$ thick PCBs and separated by two inner PCBs, also $1.5\,mm$ thick. In order to add stiffness to the structure, fiberglass honeycombs with a thickness of $10\,mm$ were glued on the outer sides of the external PCBs. Each stack has 5 gas gaps between glass sheets, which are used as resistive electrodes. The two external glass sheets in a stack have a thickness of $400\,\mu m$, while the four internal sheets are $280\,\mu m$ thick. Fishing line as a spacer defines the $200\,\mu m$ gap between all the glass plates. High voltage is applied to the outer part of the external glass electrodes covered by conductive paint with the surface resistivity of about $2-10\,M\Omega/sq$. All internal glass electrodes are left electrically floating. Two PCBs with pickup readout pads are placed on both sides of the inner stack, one serving as the cathode readout plane, the other as the anode one. 
Correspondingly, the HV is applied in an alternating sequence, as shown on the right side of Fig.~\ref{fig:TOF400_Schematic}, in order to form a symmetrical configuration for the cathode and anode readout planes. Such a configuration was chosen to ensure that propagation of signals to the FEE has equal speed on positive and negative lines, thus preventing dispersion of the differential signal. 

Readout pads are strips of $300\times10\,mm^2$ arranged vertically with a $12.5\,mm$ pitch, making 48 readout strips in each PCB. A $2.5\,mm$ spacing between adjacent strips is introduced in order to reduce crosstalk between them. Signals from the strips are transferred to the front-end electronics by twisted pair cables. In order to achieve a better time resolution and determination of hit coordinate along the strip, the signals are read out from both ends of the strip. 

The FEE of the TOF400 is based on the NINO amplifier/discriminator ASIC developed at CERN for the time-of-flight system of the ALICE experiment ~\cite{TOF400_NINO_chip}.  The chip 
is processed on $0.25\,\mu m$ technology and has 8 input channels. Each channel includes an ultra-fast preamplifier with peaking time less than $1\,ns$, a discriminator with a minimum detection threshold of $10\,fC$, and an output stage which provides an LVDS output signal. The duration of the LVDS signal is proportional to the charge of the input signal and can be used for the amplitude-time correction. The 24-channel FEE board, which combines signal processing for three NINO chips, was developed at LHEP JINR ~\cite{TOF400_FEE_board}. In order to ensure optimal operation of the FEE, the boards are placed as close to the MRPC as possible and mounted on the front cover of the gas box. Measurements with a test signal from a generator showed an intrinsic time resolution of the FEE chain of about $7\,ps$. Additional features of the FEE board include the ability to remotely control the threshold levels of the NINO discriminators and to measure the supply voltage and temperature on the board via the RS-485 interface.

LVDS signals from the FEE boards are transmitted over a distance of up to $10\,m$ via a special cable without loss of time resolution. The signals are digitized in 72-channel time-to-digital converters (TDC72VHL) based on the HPTDC chip ~\cite{TOF_HPTDC}. The TDC72VHL were developed at LHEP JINR and operate in ultra high resolution mode with a binning of $23.4\,ps$. 
Such fine binning allows determining the leading and trailing edges of the input LVDS signals with high accuracy. The TDCs exhibit significant integral non-linearity, which, if not corrected, causes significant degradation of the time resolution. The method of uniform filling the TDC time window with random events (code density test) is used for non-linearity calibration of every channel of the TDC module. After applying the non-linearity correction, the intrinsic time resolution of individual TDC72VHL channels is equal to $20\,ps$ on average.

\begin{figure}[tbp]
  \centering
  \caption{ TOF400 MRPC detector efficiency and time resolution as a function of applied HV for different NINO thresholds;}
    \includegraphics[width=0.48\textwidth]{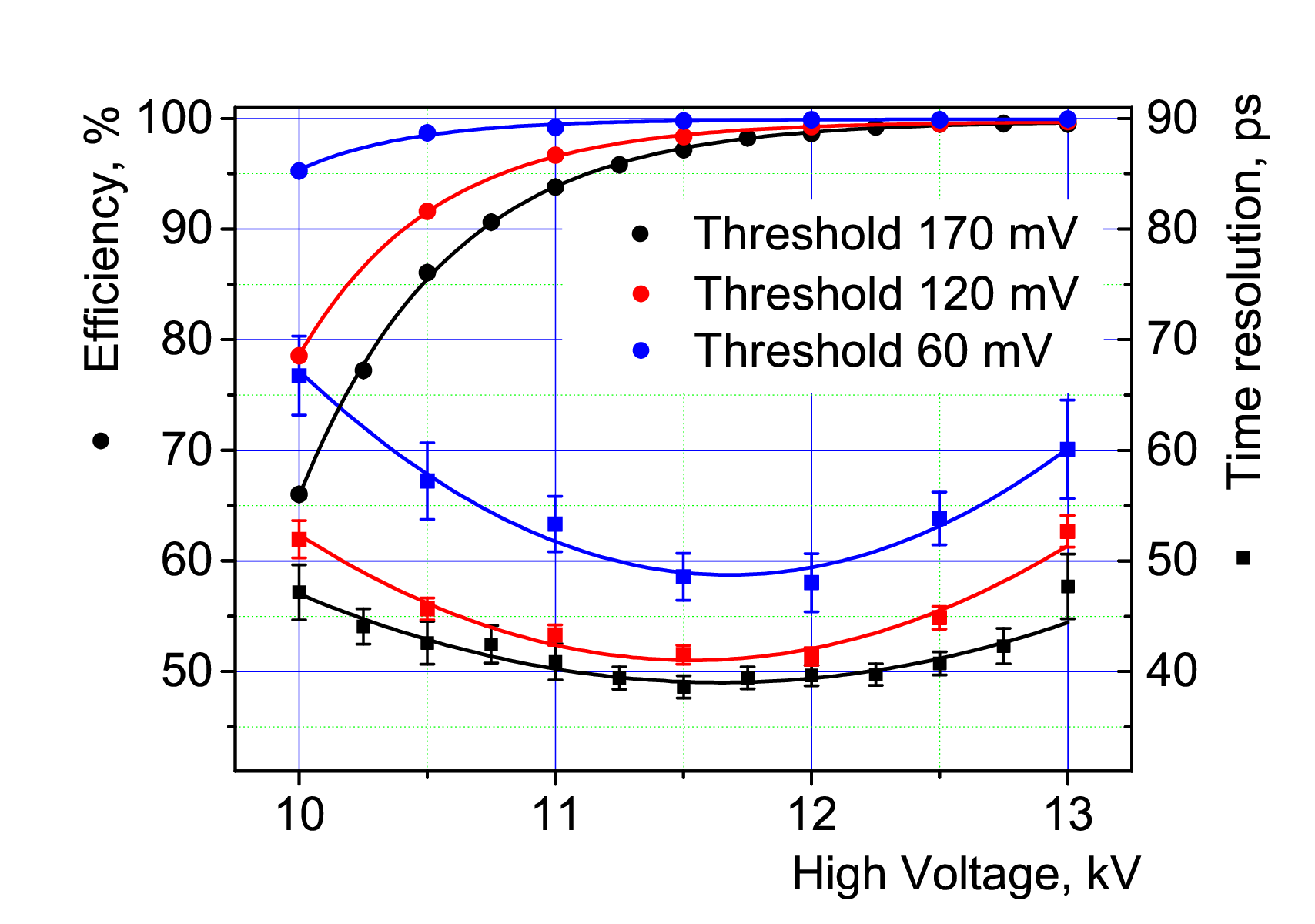}%
  \label{fig:MRPC_Eff_Res}
\end{figure}

\begin{figure}[tbp]
  \centering
  \caption{TOF400 MRPC  detector performance. Dependence on the particle flux for $11.5\,kV$ HV and $120\,mV$ threshold.}
    \includegraphics[width=0.48\textwidth]{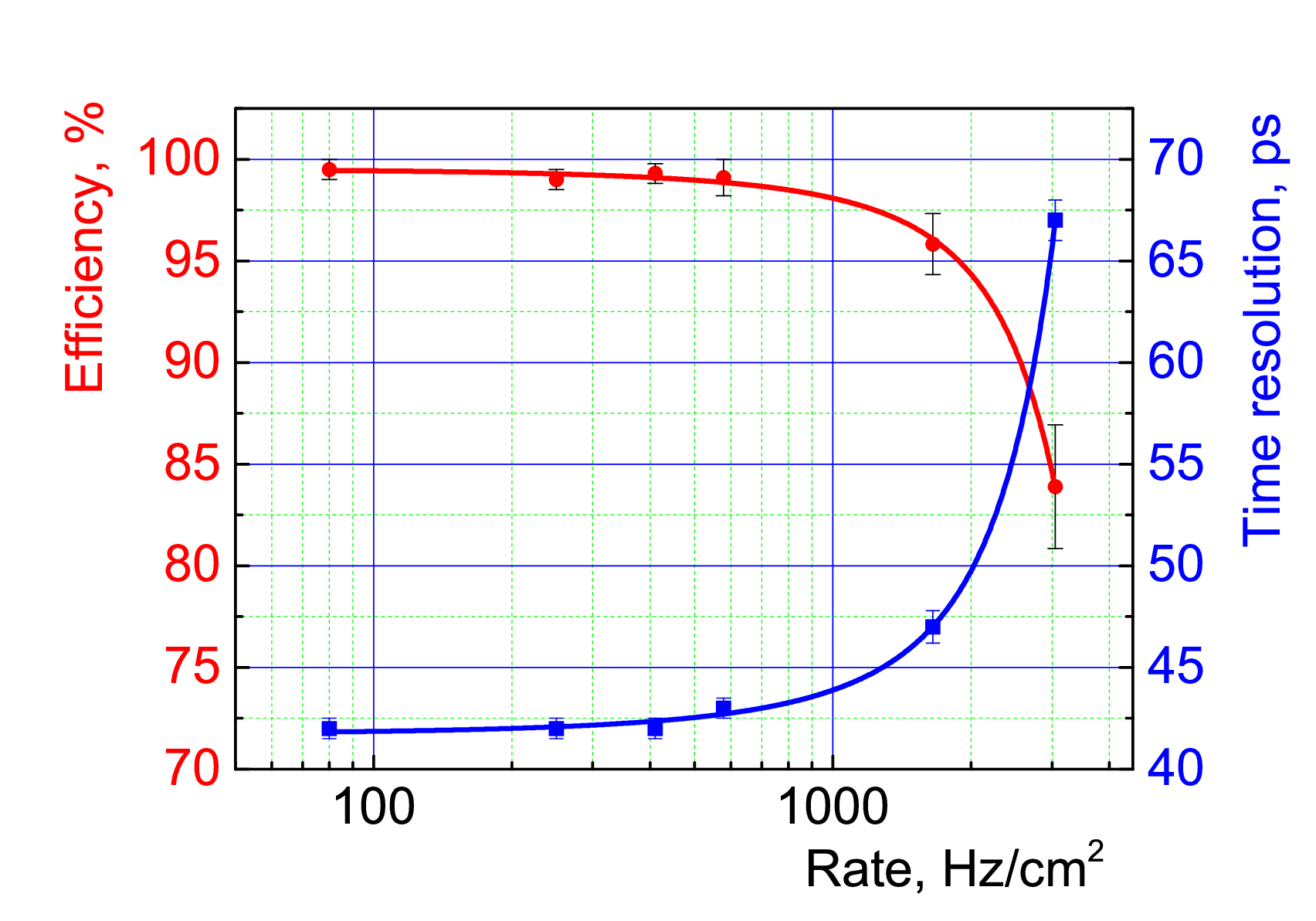}%
  \label{fig:MRPC_Eff_Res_vs_Rate}
\end{figure}

A full scale MRPC prototype with a complete readout chain and a $90\,\%\,C_2H_2F_4 + 5\,\%\,i-C_4H_{10} + 5\,\%\,SF_6$ gas mixture was tested in the Nuclotron deutron beam ~\cite{TOF400_TEST}. A fast Cherenkov counter with a time resolution of $37\,ps$ was used as a start detector. The measured efficiency and time resolution as a function of high voltage for different levels of the NINO discriminator threshold are presented in Fig. ~\ref{fig:MRPC_Eff_Res}. All the results include contributions from the front-end and data acquisition electronics. Based on the measured prototype performance, a high voltage of $11.5\,kV$ and a discriminator threshold of $120\,mV$ were chosen as the operating point of the TOF400 modules. With these settings, the dependence of efficiency and time resolution on the particle rate was studied with the prototype. The results of these tests are presented in Fig. ~\ref{fig:MRPC_Eff_Res_vs_Rate}. Monte Carlo simulations show that under the BM@N conditions, even at the highest heavy ion beam intensity, the particle flux in the TOF400 does not exceed $1\,kHz/cm^2$ for an Au+Au collision with the maximum energy and intensity of BM@N. Therefore, a time resolution better than $50\,ps$ and an efficiency higher than $95\,\%$ are expected. 

\subsection{TOF700}

\begin{figure}[tbp]
  \centering
  \caption{Arrangement of 40 ``warm" (red) and 32 ``cold" (blue) MRPCs in the TOF700 active area.}
  \includegraphics[width=0.48\textwidth]{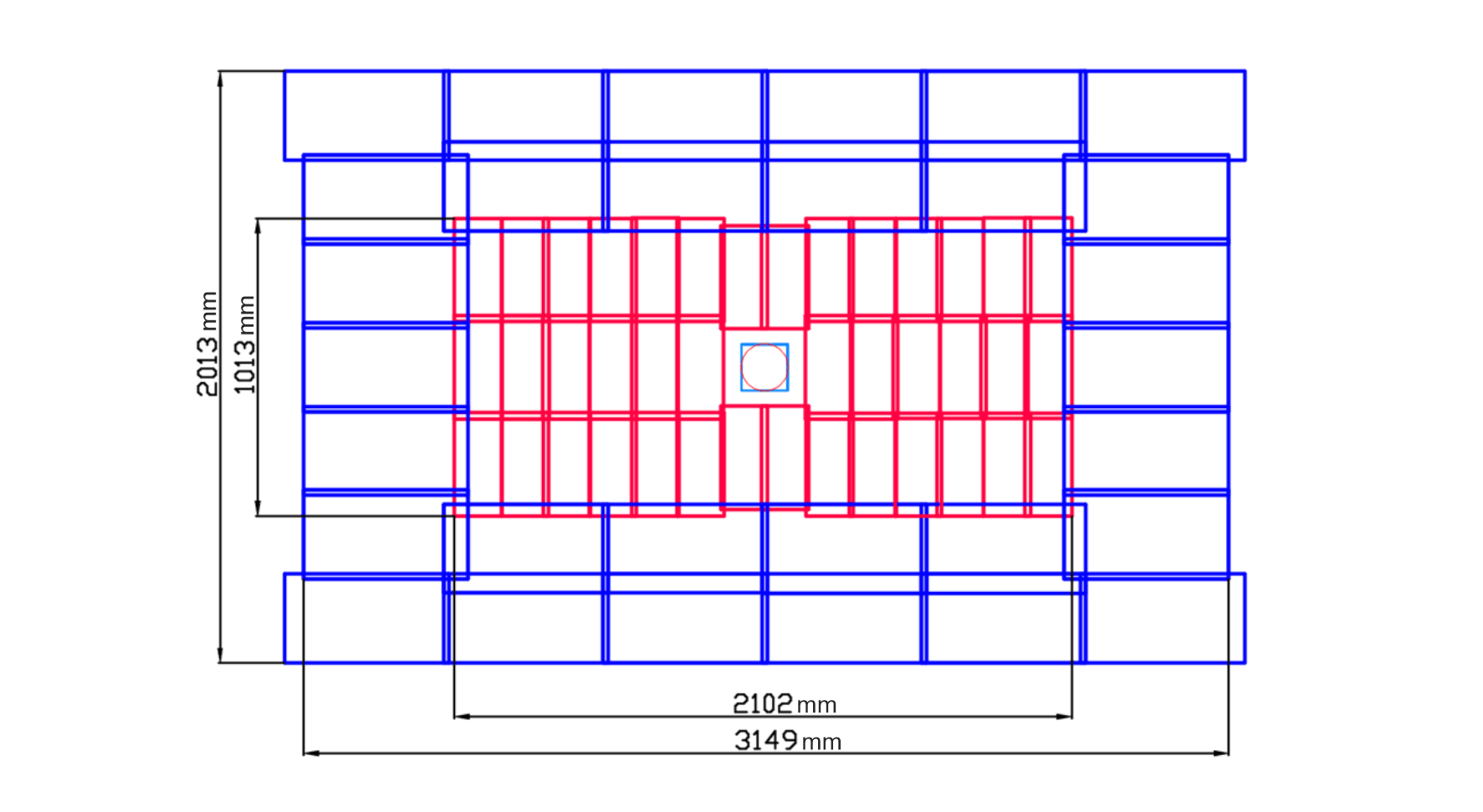}
  \label{fig:TOF700_Front_View}
\end{figure}

The TOF700 wall is placed at about 7 meters from the target and has an active \textit{XY} area of $3.2\times2.2\,m^2$ defined to overlap with the geometrical acceptance of the Outer Tracker detectors ($2.2\times1.5\,m^2$ CSC) as well as to provide substantial overlap with the GEM system acceptance. At the center of the TOF700 wall, there is an opening for the vacuum beam pipe. Since the density of hits from particles produced in heavy ion collisions is significantly higher in the region close to the beam, two types of MRPC detectors are used for the TOF700: ``cold" -- with an active area of $30\times56\,cm^2$ and 16 readout strips of $18\times560\,mm^2$ for the outer region with a low particle flux, and ``warm" -- with an active area of $16\times35\,cm^2$ and 32 strips of $10\times160\,mm^2$ for the region near the beam line. The horizontal orientation and size of the readout strips were dictated by the expected hit occupancy and the requirement of unambiguous matching of hits with particle tracks.   
The arrangement of the TOF700 MRPCs in the \textit{XY} plane is shown in Fig.~\ref{fig:TOF700_Front_View}.
The detectors are mounted on two sub-walls, which can slide relative to each other to facilitate access for installation and maintenance of the detectors. In addition, the MRPCs in each sub-wall are arranged in two layers in order to provide geometrical overlap between adjacent detectors. 

\begin{figure}[tbp]
  \centering
  \caption{Schematic view of the ``cold" MRPC detector for the TOF700 system.}
\vspace*{0.3cm}
  \includegraphics[width=0.48\textwidth]{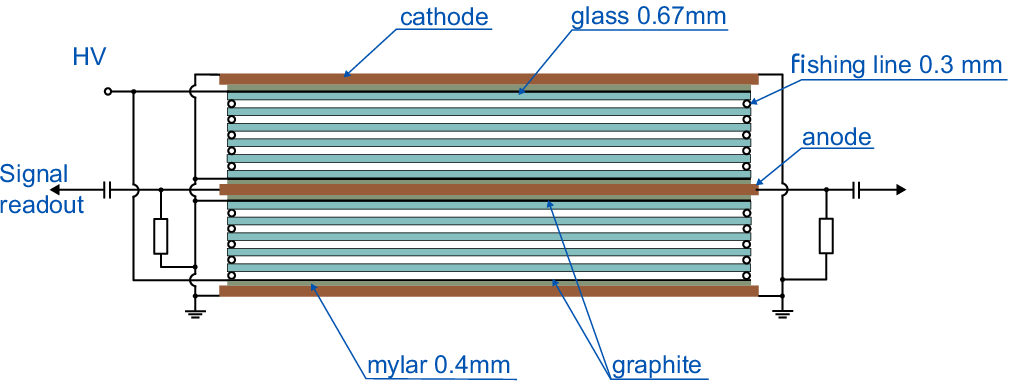}
  \label{fig:TOF700_RPC_scheme}
\end{figure}

``Cold" and ``warm" MRPCs have a similar two-stack design with a single anode readout plane placed between the stacks. A schematic cross-section of a ``cold" MRPC is shown in Fig. ~\ref{fig:TOF700_RPC_scheme}. Each stack is formed by six $0.67\,mm$ thick glass plates with the bulk resistivity of $2\times10^{12}\, \Omega\times cm$. Fishing line spacers define a $0.3\,mm$ gap between the glass sheets. A graphite conductive coating with the surface resistivity of $\sim1\,M\Omega/sq$ is painted on the outer surfaces of the external glass plates in order to apply both high voltage and ground connections. The anode readout plane is arranged on a $100\,\mu m$ one-sided PCB. Unipolar signals are taken from both ends of the strips, which makes it possible to determine the coordinate of the particle hit along the strip by measuring the time difference between the signals. Each detector is placed in an individual gas box, which is formed by a $2.5\,mm$ thick aluminum frame and two cover plates. One cover is made of a $2.5\,mm$ thick PCB and is designed to take out signal wires from the box volume to the readout electronics.  The other cover is made of a $1.5\,mm$ thick aluminum sheet.  

The design of ``warm" MRPCs has only minor modifications. In order to increase their rate capability, the gas gaps and thickness of the glass plates in warm MRPCs are reduced to $0.22\,mm$ and $0.55\,mm$, respectively. Such a reduction leads to lower signal amplitudes due to increased anode strip \textendash{} cathode capacity. To compensate for this signal weakening, the number of gaps in the chamber was increased from 10 to 12 (six gaps per stack).

The FEE boards are developed specially for BM@N.
Signals from the MRPC are sent to the FEE over $50\,\Omega$ coaxial cables with MMCX connectors. 
The boards are based on the NINO ASICs, which, as already mentioned, process the signals in such a way that the duration of the LVDS output signals is proportional to the amplitude of the input signals, suitable for implementation of the time-over-threshold method. The output signals are transmitted to the digitizing module using DHR-78F connectors. A 64-channel VME TDC64VHLE time-to-digital converter based on the HPTDC chip is used for digitization. Using a special module (PWR\&CTRL), it is possible to remotely control the power supply, discrimination threshold and hysteresis value in the FEE boards.

\begin{figure}[tbp]
  \centering
    \caption{Efficiency of ``warm" MRPC designed for the TOF700 system. }
    \includegraphics[width=0.42\textwidth]{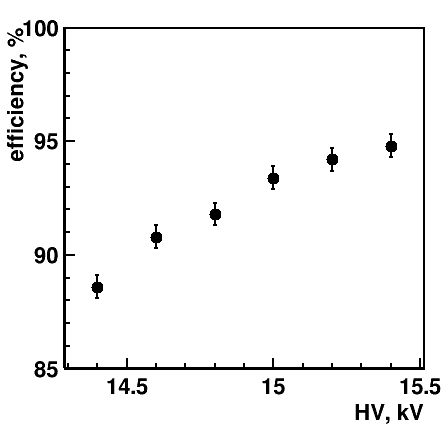}
    \label{fig:TOF700_Groop_1}
\end{figure}

\begin{figure}[tbp]
  \centering
    \caption{Time resolution of ``warm" MRPC designed for the TOF700 system.}
    \includegraphics[width=0.42\textwidth]{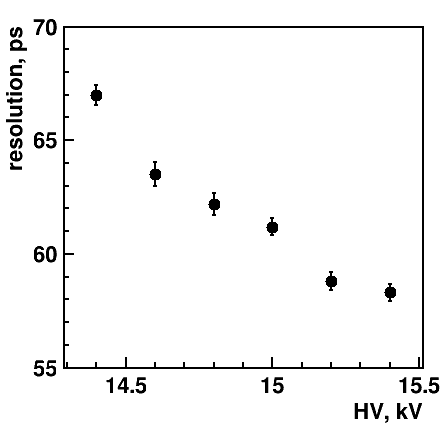}
    \label{fig:TOF700_Groop_2}
\end{figure}

Prototypes of the ``cold" and ``warm" TOF700 detectors were tested in the secondary muon beam of the U-70 accelerator at IHEP (Protvino, Russia). The test was carried out at the ``MUON" facility with a particle flux of about $1\,kHz/cm^2$ \cite{Kuzmin2019}. The test results of the ``warm" MRPC prototype are shown in Fig.~\ref{fig:TOF700_Groop_1} and Fig.~\ref{fig:TOF700_Groop_2}. The time resolution of the MRPCs with a complete chain of electronics (FEE and readout) is at the level of $60\,ps$. The efficiency is more than $95\,\%$.
\begin{table*}[!ht]
\begin{center}
\caption{Main parameters of the TOF system.}
\vspace*{0.3cm}
\begin{tabular}{ |c|c|c| } 
 \hline
  & TOF400 & TOF700 \\ 
 \hline
 MRPC active area & $30\times60\,cm^2$ & \makecell{$30\times56\,cm^2$ ``cold" \\ $16\times35\,cm^2$ ``warm"}\\ 
 \hline
  FEE on one MRPC & 96 & \makecell{32 for ``cold" \\ 64 for ``warm"}\\ 
 \hline
  Number of MRPCs & 20 & \makecell{30 ``cold" \\ 40 ``warm" }\\ 
 \hline
  Total active area & 2 Arms $\times1.1\times1.3\,m^2$ & $3.2\times2.2\,m^2$\\ 
 \hline
  Total number of FEE channels& 1920 & 3520\\ 
 \hline
  Time resolution of the MRPC& $50\,ps$ & $60\,ps$\\ 
 \hline
  Efficiency of the MRPC& $>95$ \% & $>95$ \%\\ 
 \hline
\end{tabular}
\label{table:TOF_Summary}
\end{center}
\end{table*}

Both TOF400 and TOF700 systems use the same non-flammable Freon rich gas mixture containing $90\,\% \,C_2H_2F_4$, $5\,\%\,i-C_4H_{10}$, and $5\,\%\,SF_6$. A simple open-loop gas system was designed for the BM@N experiment. This system is based on the MKS 1479A controllers produced by MKS Instruments, Inc. (Andover, Massachusetts, USA) for measuring and adjusting the absolute flow of components with an accuracy of $0.3\,\%$. The flow rate of the gas mixture can be adjusted in the range from $6\,l/h$ to $90\,l/h$.The typical operational flow is $21\,l/h$ which corresponds to the exchange of 2 volumes per day. The content of  $O_2$ and  $H_2O$ at the outlet of the system does not exceed $600\,ppm$ and $1000\,ppm$, respectively. The additional channel of the gas system is available to purge the system with nitrogen for cleaning and drying. A special PC program has been written to control the parameters of the gas system via the Ethernet interface.

The MRPC detector operates at very high voltages of $12\,kV$ and $15\,kV$ for the TOF400 and TOF700, respectively. On the other hand, the dark currents of the detector are quite small at the level of tens of $nA$. Also, the detector is very sensitive to voltage ripples due to the large capacitive coupling between the high voltage layer and the readout strips. Therefore, the high voltage system is subject to strict requirements for voltage stability and current measurement accuracy. The high voltage power supply systems for both TOF subsystems are based on commercially available Iseg modules (Iseg Spezialelektronik GmbH, Germany) and a system module specially designed for BM@N by HVSys (JINR, Dubna, Russia). Remote control of all elements of the system is organized via the Ethernet interface. 

The main parameters of the TOF400 and TOF700 subsystems are summarized in Table ~\ref{table:TOF_Summary}.

%\clearpage

\section{Outer Tracker}

The detectors of the Outer Tracker are situated downstream of the analyzing magnet. 
In the 2023 Xe run the Outer Tracker consisted of two large aperture drift chambers (DCHs) and five cathode strip chambers (CSCs), four of them $1.1\times1.1\,m^2$ (small) and one $2.2\times1.5\,m^2$  (large). Track localization in the Outer Tracker is used not only to improve particle momentum reconstruction, but also to facilitate matching of tracks reconstructed in the Central Tracking System with corresponding hits in the time-of-flight detectors. Therefore, the size and location of the small CSC were chosen to provide significant overlap with the TOF400 acceptance, while the DCH were placed to cover most of the TOF700 acceptance. The granularity of the DCH is sufficient to perform measurements with beams of light and medium nuclei. However, in experiments with Au or Bi beams, the DCH occupancy becomes too high to perform efficient track separation. Therefore, in the process of preparation for the experiments with heaviest ions, the two drift chambers will be replaced by two large cathode strip chambers. The first of them was tested during the 2023 Xe run. In the final configuration, relative position of the TOF700 and two large CSC will be optimized.

\subsection{Drift chambers}

\begin{figure}[tbp]
  \centering
  \caption{DCH  integrated into the BM@N experimental setup.}
\vspace*{0.3cm}
  \includegraphics[width=0.45\textwidth]{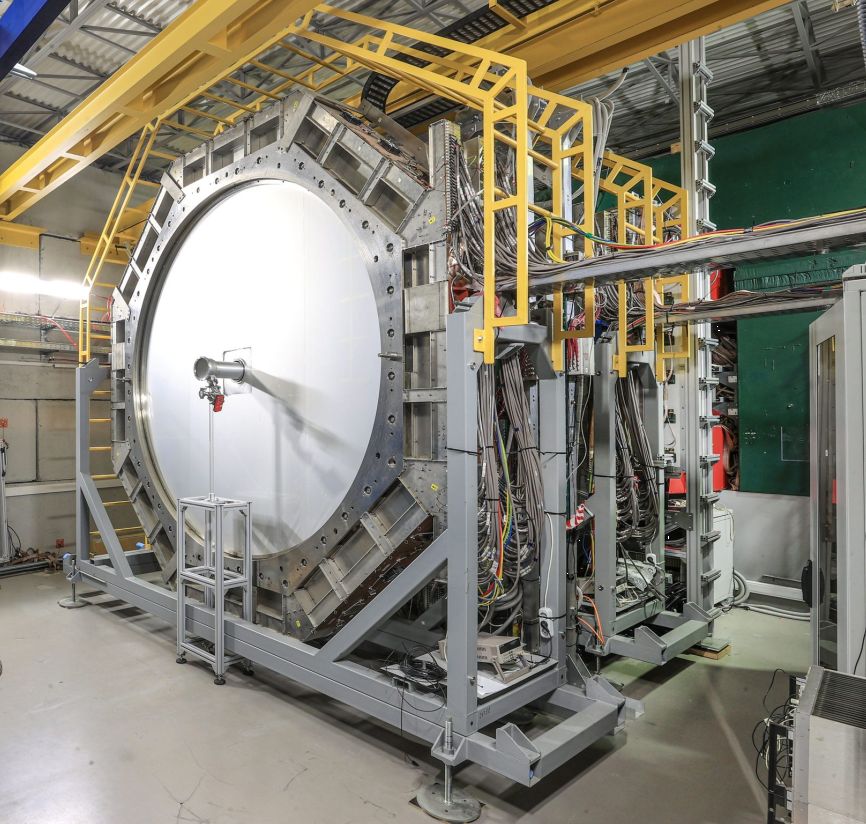}
  \label{fig:DCH_view}
\end{figure}

The drift chambers, formerly used in the NA48 experiment at CERN~\cite{DCH}, have an octagonal shape with a transverse width of $2.9\,m$ (Fig.~\ref{fig:DCH_view}). Their fiducial area is about $4.5\,m^2$. The hole for the beam pipe in the center of the chamber has a diameter of $160\,mm$. The chambers are constructed with minimal amount of material along the beam direction, thus reducing multiple scattering effects. 

\begin{figure}[tbp]
  \centering
  \caption{Drift cell geometry of the DCH.}
\vspace*{0.35cm}
  \includegraphics[width=0.35\textwidth]{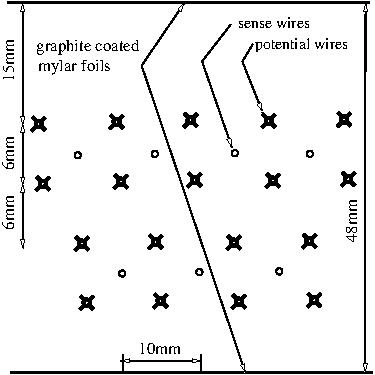}
  \label{fig:geom_dch}
\end{figure}

Each chamber contains four \textit{X}, \textit{Y}, \textit{U}, \textit{V} coordinate planes with wire inclination angles with respect to the \textit{Y} axis of $0^\circ$, $90^\circ$, $-45^\circ$ and $+45^\circ$, respectively. A schematic view of the drift cell geometry is shown in Fig.~\ref{fig:geom_dch}. In order to resolve left-right ambiguity of the hit position relative to the signal wire, every coordinate plane has two staggered rows of wires. Each coordinate plane has $2\times256$ signal wires with a wire pitch of $10\,mm$. The central wires in the region of the beam pipe are split in two. The sense wires are grounded. The electric field is created by the negative voltage applied to two planes of field wires located on each side of the sense wire plane at a distance of $3\,mm$. 
The sense wires made of gold-plated tungsten have a diameter of $20\,\mu m$, while the gold-plated Ti-Cu field wires have a diameter of $120\,\mu m$. Thin Mylar foils ($22\,\mu m$) coated with graphite are used to shape the electric field in the drift cell. In addition, they serve as walls separating adjacent \textit{X}, \textit{Y}, \textit{U}, \textit{V} coordinate planes. 

The chambers normally operate at a high voltage of about 2--$2.5\,kV$ between the field and sense wires,
whereas the Mylar foils are kept at a negative voltage of 1--$1.5\,kV$. Typical gas amplification for such operating conditions is at the level of $2\times10^4$. A drift distance of $5\,mm$ corresponds to a $\sim100\,ns$ drift time, which ensures high rate capability of the detector.   

The front-end amplifiers and discriminators developed for the readout of the DCH in the NA48 experiment~\cite{DCH_el} are used in the BM@N without modifications. The FEE cards are mounted on the frame of the drift chambers. The amplifiers are designed to provide accurate timing with minimal cross-talk between neighbouring channels. For pulses with a $10\,ns$ rise time, the cross-talk is suppressed with $\gtrsim46\,dB$. The outputs of the amplifiers are AC-coupled to high-speed discriminators in LeCroy MVL407 IC units. The discriminator thresholds are set based on an external DC voltage and can be remotely controlled. The discriminator output is a differential ECL pulse with a $50\,ns$ width followed by $50\,ns$ of dead time. The output pulses are transmitted via $12\,m$ long twisted pair cables to TDC64VL VME modules, developed by AFI Electronics (Dubna, Russia).
These 64-channel $100\,ps$ multihit TDCs provide a timestamp of the pulse front. The overall accuracy of the entire readout chain is about $1\,ns$.

\subsection{Cathode strip chambers}

\begin{figure*}[tbp]
   \centering
   \caption{Schematic view of the small CSC.}
   \includegraphics[width=0.9\textwidth]{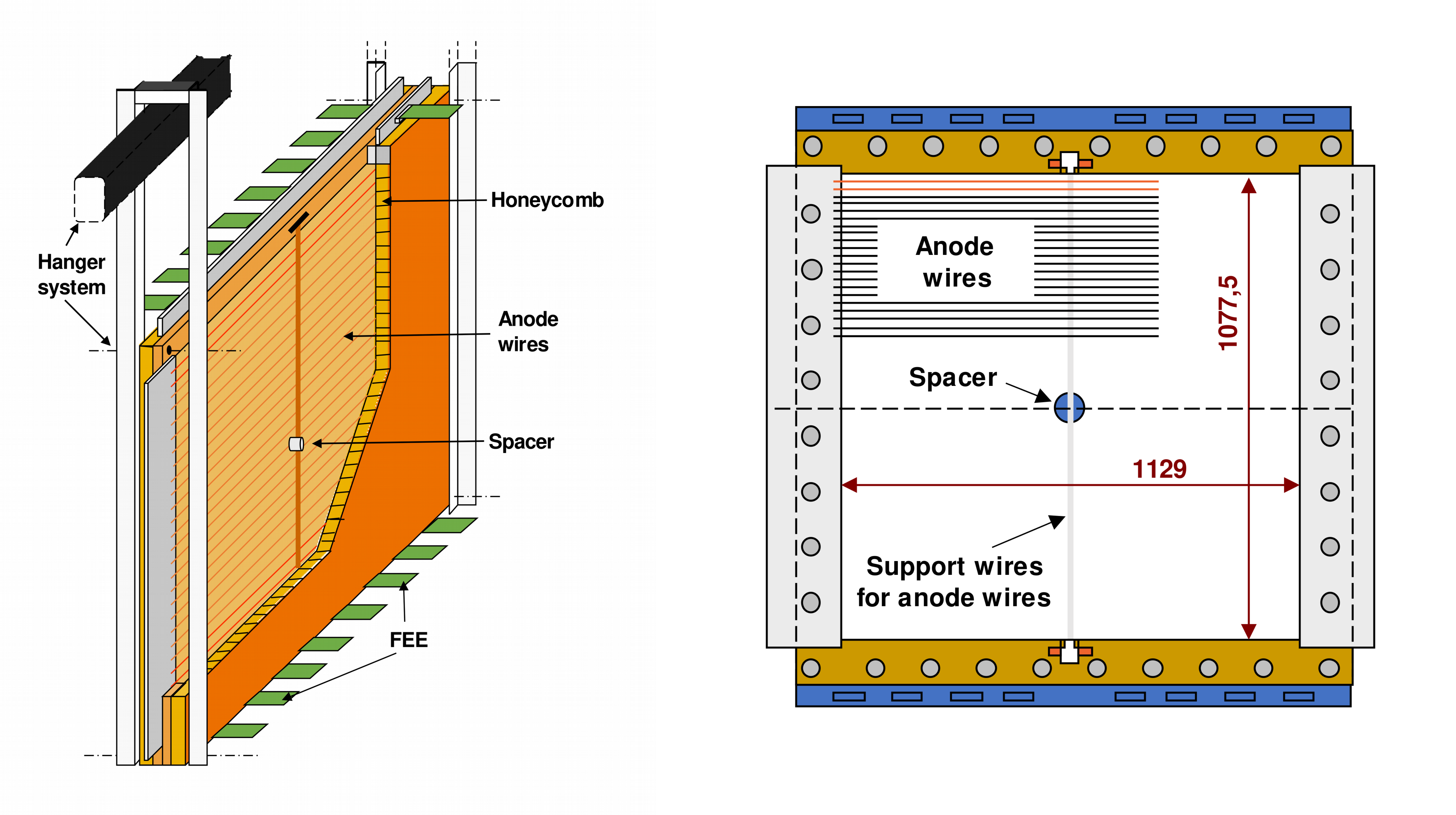}
   \label{fig:CSC_schematic_view}
\end{figure*}

\begin{figure*}[tbp]
  \centering
  \caption{Technical drawing of the large cathode strip chamber (the 15-degree strips are not shown).}
  \includegraphics[width=0.9\textwidth]{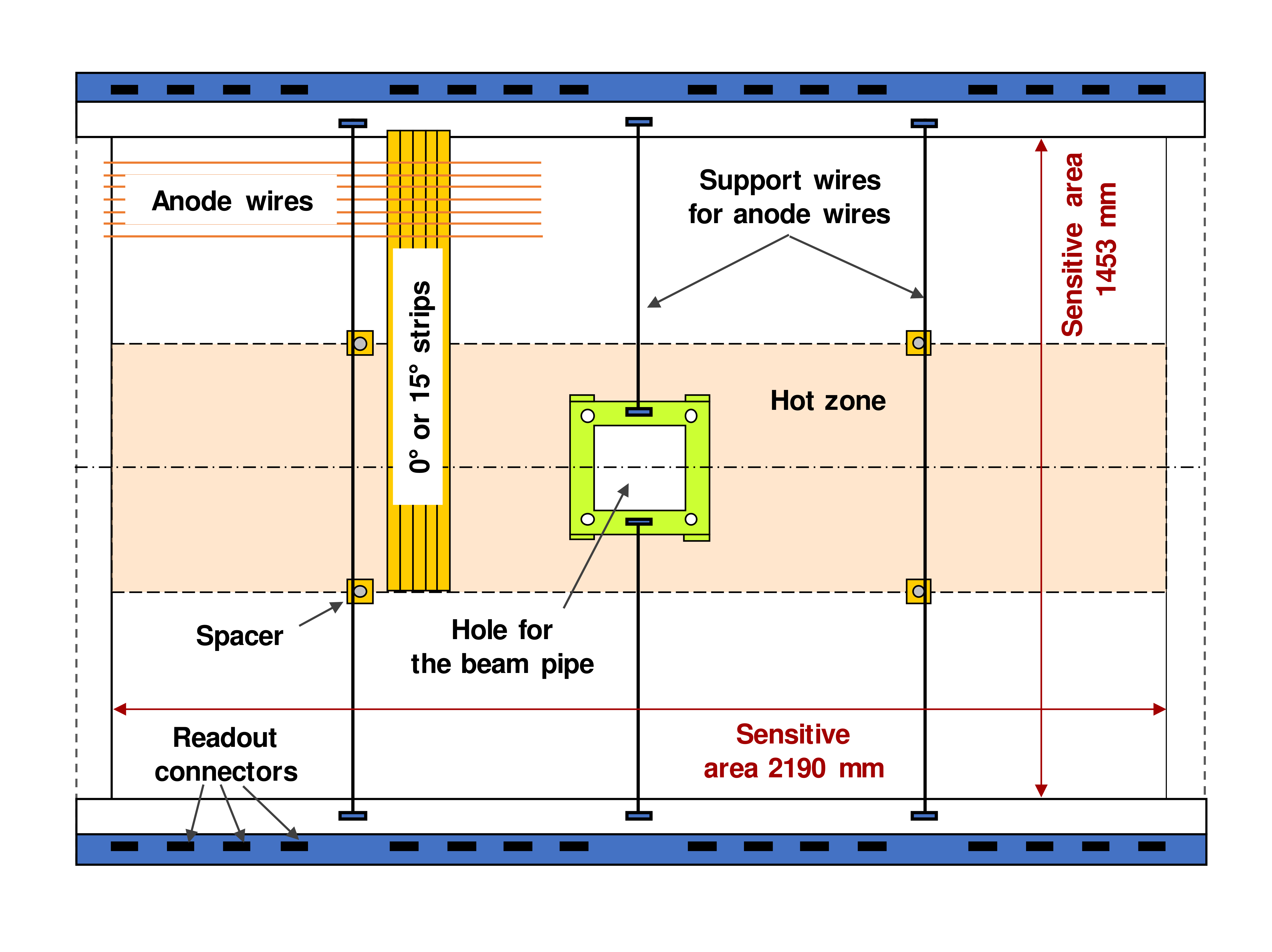}
  \label{fig:CSC_technical_draw}
\end{figure*}

The active area of the small CSC is $113\times107\,cm^2$, while the large chambers have an active area of $219\times145\,cm^2$. Both types of the CSC have similar design features. All chambers have one detection layer consisting of a plane of anode wires stretched between two cathode planes (see Figs.~\ref{fig:CSC_schematic_view},~\ref{fig:CSC_technical_draw}). The anode wires form a horizontal grid with a step of $2.5\,mm$ in \textit{Y}. They are made of gilded tungsten and have a diameter of $30\,\mu m$. In order to reduce anode wire deflection and to reinforce the flatness of the anode plane, vertical support wires are added to the mechanical design of the chambers. The support wires are made of stainless steel, have a diameter of $0.3\,mm$ and are insulated by a $0.8\,mm$ thick Teflon cladding. 

The readout of induced signals is arranged on both front and back cathode planes made of PCBs with parallel metal strips. The inclination angles of the strips with respect to the vertical axis is 0 degrees (\textit{X} coordinate) in one of the cathode planes and 15 degrees (\textit{Y} coordinate) in the other. The pitch of the \textit{X} and \textit{Y} strips is $2.5\,mm$. 

Due to the large multiplicity of charged particles in heavy ion collisions, the readout layer is divided into outer (cold) and inner (hot) zones, as shown in Fig. ~\ref{fig:CSC_technical_draw}. The size of the inner zone is $-14<Y<14\,cm$  and  $-24<Y<24\,cm$ in the small and large chambers, respectively. Each cathode plane in the small CSC is composed of two printed circuit boards, top and bottom. The cathode planes of the large chambers are assembled of eight PCBs, four in the top half of the plane and four in the bottom one.  

To enhance the structure rigidity the PCBs are glued on support honeycomb panels. Furthermore, in order to prevent chamber deformation, the distance between the two cathode planes is fixed by additional spacers. The distance between cathode and anode planes was a subject of optimization and varies from chamber to chamber in the range from 3.4 to $3.8\,mm$. A larger gap increases the number of adjacent strips with induced signal above the threshold, the width of the cluster on average spans over 6 strips, i.e. $15\,mm$. On the other hand, a smaller than $3.4\,mm$ gap results in a higher probability of electric discharge. The cathode planes are grounded, and a high voltage of about $+2.4\,kV$ is applied to the anode wires.        

The CSC front-end electronics is based on the same charge sensitive preamplifier chip VA163 as used for the GEM detectors. The multiplexed data from each FEE board are transmitted through a twisted pair flat cable to 12-bit analog-to-digital converter (ADC) modules read out by the data acquisition system. The full configuration of the Outer Tracker with six CSCs will have $\sim30100$ readout channels and is planned to be integrated into the BM@N experimental setup in the next physics run.

\begin{figure*}[tbp]
  \centering
  \caption{The gas line for the Outer Tracker. a) The layout of the mixer module: HV - on/off valve, PCV - pressure control (constant) valve, PSV - pressure safety valve, Pur. - Purifier ($H_{2}O$ and $O_{2}$), F - filter, MFC - mass flow controller, MKS 647C - power supply and readout. b) The component layout of the distributor module: FM - flowmeter (manual flow adjustment), oil bubbler - pressure and air protection.}
  \includegraphics[width=1.0\textwidth]{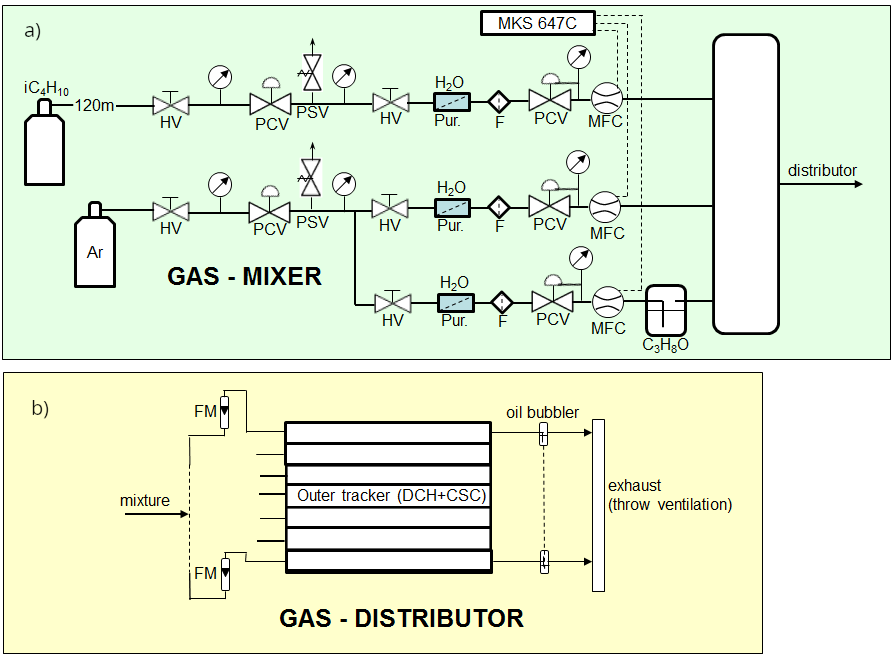}
  \label{fig:gas_outer}
\end{figure*}

\subsection{Gas system}

All chambers of the Outer Tracker were operated with the  $Ar(75\,\%) + C_{4}H_{10}(25\,\%) /  C_{3}H_{8}O (vapor) $ gas mixture. The gas system (Fig.~\ref{fig:gas_outer}) consists of two parts: 1) the mixer system, which delivers a mixture of gases in a required ratio and pressure to downstream elements; 2) the distribution system, which delivers the gas in well defined quantities to the individual detectors. The power supply and readout module MKS 647C and mass flow controllers used in the gas distribution system are produced by MKS Instruments, Inc. (Andover, Massachusetts, USA).

%\clearpage

\section{Forward Spectator Detectors} 

Several detectors, which measure the energy or charge of the projectile spectators, are located at the very end of the BM@N setup. These are the Forward Hadron Calorimeter (FHCal), the Forward Quartz Hodoscope (FQH), and the Scintillation Wall (ScWall). These detectors are used to determine the centrality of the collision and orientation of the reaction plane. Moreover, the ScWall and FQH can also be used to study the charge distributions of spectator fragments produced in nucleus-nucleus interactions.

\subsection{Forward Hadron Calorimeter}

\begin{figure*}[tbp]
  \centering
  \caption{ Forward Hadron Calorimeter: a) Schematic view of the FHCal;
            b) Photo of the FHCal installed on the movable platform (blue) at BM@N.}
\vspace*{0.3cm}
  \includegraphics[width=1.0\textwidth]{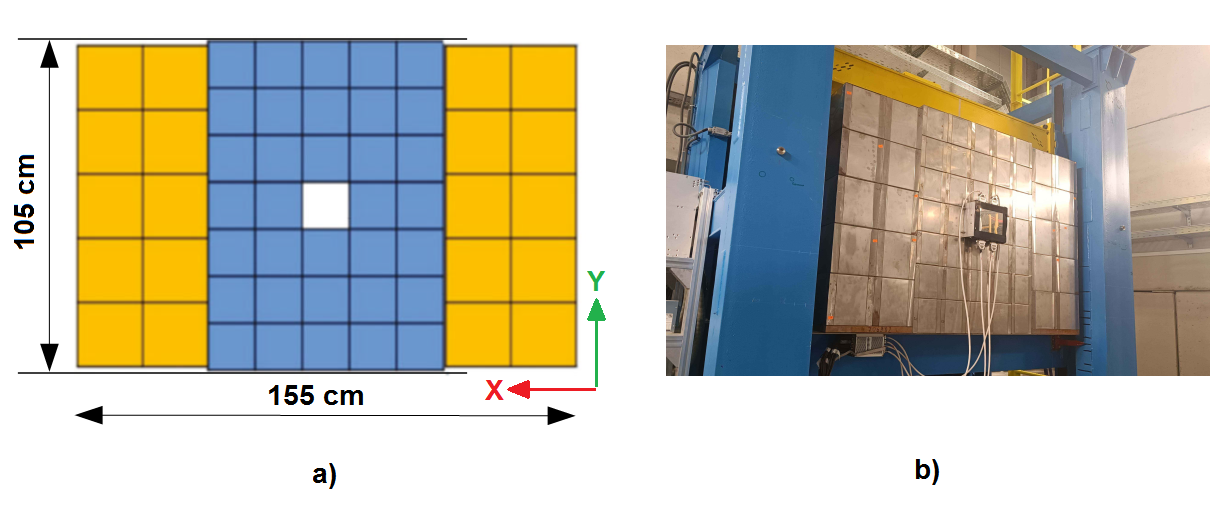}
  \label{fig:FHCal_schematic_view}
\end{figure*}

The FHCal has a granular structure in the transverse and longitudinal planes. It consists of 54 separate modules in the transverse plane (see Fig.~\ref{fig:FHCal_schematic_view}). The internal part of the FHCal consists of 34 small modules with transverse sizes of $15\times15\,cm^2$ and a length equivalent to 4.0 nuclear interaction lengths. These modules are identical to the modules of the forward hadron calorimeters of the Multi-Purpose Detector (MPD) experiment at the NICA accelerator complex~\cite{MPD_TDR}. Each of the two outer lateral parts of the calorimeter contains 10 larger modules with a transverse size of $20\times20\,cm^2$ and a length equivalent to 5.6 nuclear interaction lengths. These modules were initially constructed for the hadron calorimeter of the Compressed Baryonic Matter (CBM) experiment (FAIR, Darmstadt, Germany)~\cite{CBM_PSD_TDR} and are temporarily used in the BM@N experiment. 

Beam ions that did not interact pass to a beam dump located behind the FHCal through the hole
in the center of the calorimeter. The transverse size of the hole is $15\times15\,cm^2$. 
This design feature is dictated by the requirement to protect internal modules and the front-end electronics of the FHCal against the high radiation dose and strong activation, typical for experiments with relativistic heavy ion beams.

\begin{figure*}[tbp]
  \centering
  \caption{Forward Hadron Calorimeter: a) Scheme of a large calorimeter module, with 10 sections shown in different colors; b) 3D view of an assembled large calorimeter module.}
\vspace*{0.3cm}
  \includegraphics[width=0.95\textwidth]{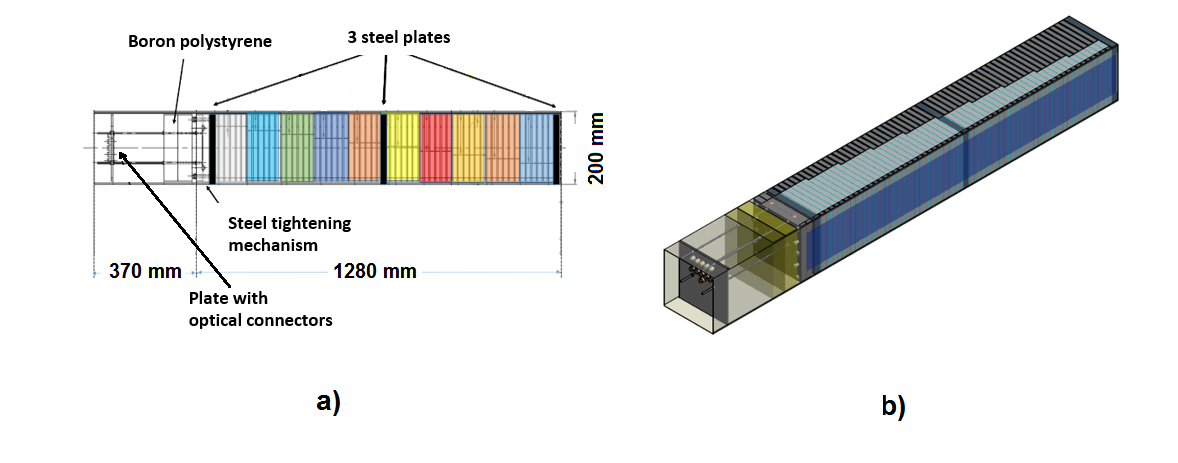}
  \label{fig:FHCal_module_view}
\end{figure*}

The FHCal modules have a sampling structure and consist of lead/scintillator layers with a sampling ratio of 4:1 (the thickness of the lead plates and scintillator tiles are $16\,mm$ and $4\,mm$, respectively) and fulfill the compensation condition ($e/h=1$) for the hadron calorimeter. The small modules have 42 lead/scintillator layers, while the large modules have 60 such layers. To get rather high rigidity of the lead plates, they are made of lead-antimony alloy. The assembly of 60 (42) alternating layers of scintillator and lead plates is bound into one package by a $0.5\,mm$ thick stainless steel band tightened using a special tensioning mechanism.  After tightening, the band is welded to additional steel plates inserted at the beginning, in the middle and at the end of the module (Fig.~\ref{fig:FHCal_module_view}). Behind the tightening mechanism, a block of boron polystyrene with a thickness of $10\,cm$ is installed in the large modules. Once the package is assembled, it is closed by a cover box made of a $0.5\,mm$ thick stainless steel sheet.

The scintillator plates are made of polystyrene-based plastic scintillator produced by Uniplast (Vladimir, Russia). The light from the scintillator plates is collected by Kuraray Y11(200) wavelength shifting optical fiber glued into a $1.2\,mm$ deep groove on the surface of the scintillation plate and transported to the end of the module. The grooves in the scintillators of the large modules are circular, while those in the scintillators of the small modules are spiral. The end of each fiber on the scintillator side is coated with reflective paint.  After the fiber is glued in the scintillator plate, the plate is wrapped in a Tyvek reflector. Outside of the scintillator plate, the fibers are placed in thin black plastic pipes to be optically shielded. When assembling the module, the plates are oriented in such a way that all optical fibers exit on the same side of the module (at the top). After that, groups of fibers from six consecutive scintillation plates are glued into individual optical connectors (7 and 10 groups in the small and large modules, respectively), which are placed on a panel mounted on the rear side of the module box (see Fig.~\ref{fig:FHCal_module_view}). 
Thus, each of the large modules has ten longitudinal sections, and each of the small modules has seven sections. The longitudinal segmentation provides high homogeneity of light collection along the modules, a large dynamic range of the calorimeter response, and makes it possible to perform detailed energy calibration of the FHCal with cosmic muons~\cite{FHCal_calib_cosmic}.

The stability of the operation of the photodetectors and the calorimeter readout chain is controlled by a system of LEDs; one LED per calorimeter module. Correspondingly, the light from each LED is split to ten (seven) fibers, which are added into the optical connectors.

The weight of a single small and large module is about $200\,kg$ and $500\,kg$, respectively. The total weight of the FHCal is about 17 tons. The calorimeter is mounted on a special platform  (Fig.~\ref{fig:FHCal_schematic_view} b), which is able to move the FHCal in \textit{X} and \textit{Y} directions.

\subsubsection{FHCal photodetectors, FEE and readout electronics}

The Hamamatsu S12572-010P MPPCs with a $3\times3\,mm^2$ sensitive area are used as photodetectors for light detection from the FHCal sections. These photodetectors have a gain of $1.35\times10^5$ and a photon detection efficiency of about $10\,\%$ at a peak sensitivity wavelength of $470\,nm$. Due to a very small pixel pitch ($10\,\mu m$), the total number of pixels is 90 000, which is important for response linearity in a wide dynamic range of the signal. The FHCal front-end electronics are composed of two separate PCBs. Ten (seven) photodetectors are installed on the first PCB directly coupled with light connectors at the end of each large (small) module. A temperature sensor is mounted near the photodetectors on an aluminum heat sink. The second PCB contains signal preamplifiers with differential ADC driver output and individually adjustable voltage regulation circuits for the photodetectors. This board also has an LED flash generation circuit with a synchronization input. All FEE boards are remotely controlled via a System Module specially designed and manufactured by HVSys (JINR, Dubna, Russia).

The total number of FHCal readout channels is 438. The digitization of signal waveforms is performed by eight ADC64s2 boards produced by AFI Electronics (JINR, Dubna, Russia). The boards have 64-channel 12-bit ADCs with a sampling rate of $62.5\,MHz$ and a memory depth of up to 1024 points per channel. The ADC64s2 are capable of time synchronization via White Rabbit network, can operate in self-triggered or externally triggered modes, and digitize signals with or without zero suppression. 

In addition to 438 signals from the individual longitudinal sections of the modules, the FEE boards provide 54 summed signals, one for each calorimeter module. In order to operate with summed signals, a custom-made 12-channel analog fan-in electronic module with individually adjustable attenuation of input signals has been designed and manufactured at JINR. The fan-ins can be used to sum up the analog outputs from various groups of the FHCal modules, if needed. One possible application of the fan-in modules is generation of a trigger signal based on energy deposition either in the whole FHCal or in its ``neutron" zone. In addition, the summed signals are used to provide trigger for cosmic calibration of the calorimeter modules.

\subsubsection{FHCal calibration with cosmic muons, energy resolution and linearity of the response}

The energy calibration of the FHCal is performed using cosmic muons. Longitudinal and transverse segmentation of the calorimeter allows reconstructing muon tracks~\cite{FHCal_calib_cosmic} and accounting for track length variation in the scintillator tiles, depending on the track orientation. 
The distribution of signal amplitudes, corrected with the known muon track length, are fitted by a Landau distribution convoluted with a Gaussian. The MPV of the fit represents one MIP (minimum ionizing particle) response of the section. It can be characterized by the number of photoelectrons. Typically, the MIP response for individual sections corresponds to 40\,--\,50 photoelectrons.

A detailed study of the linearity of the response and energy resolution for an array of 9 large modules was performed using proton beams with a kinetic energy range of 1--$9\,GeV$ at the CERN T9 and T10 beamlines~\cite{FHCal_at_T9_T10}. Good linearity and $0.54/\sqrt{E}$ energy resolution were obtained.

\begin{figure*}[tbp]
  \centering
  \caption{Forward Quartz Hodoscope: a) An FQH strip with SiPM photodetectors mounted; 
           b) Photo of the Forward Quartz Hodoscope (inside view). 1 - the PCB with SiPMs, 
	      2 - the quartz strips wrapped in reflective foils. }
\vspace*{0.3cm}
\includegraphics[width=0.9\textwidth]{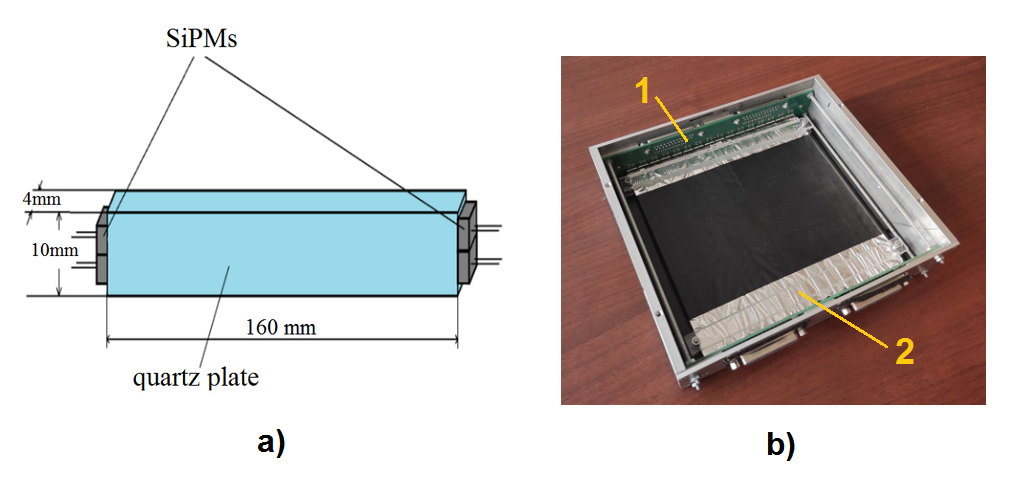}
  \label{fig:FQH_view}
\end{figure*}

\subsection{Forward Quartz Hodoscope} \label{FQH}

The FHCal beam hole is covered with the FQH beam hodoscope. The main purpose of the FQH is to measure the charge of spectator fragments, which pass the beam hole of the calorimeter. In particular, the combined FHCal and FQH response allows one to estimate the collision centrality~\cite{FHCal_FQH_centrality}. The FQH consists of 16 quartz strips, which act as Cherenkov detectors. The size of the strips is $16\times1\times0.4\,cm^3$. The light from each FQH strip is viewed by two individual silicon photomultipliers mounted on both sides of the strip (see Fig.~\ref{fig:FQH_view} a). The Hamamatsu S14160-3015PS MPPCs with a sensitive area of $3\times3\,mm^2$ and an efficiency of $32\,\%$ are used as photodetectors. The hodoscope strips with photosensors are placed inside a single light tight box (see Fig.~\ref{fig:FQH_view} b).

Four FEE boards, each of which can process eight input signals, are used in the readout of the whole FQH. The FEE boards incorporate signal amplifiers with two-gain outputs, the gains being $1\times$ and $4\times$. The low gain channel is used to cover maximum dynamic range up to the highest ion charge expected. The high gain channel is used to measure the charge of low--$Z$ fragments. Four TQDC-16 boards with a total of 64 channels are used to read out the two-gain outputs from each photodetector. Because the hodoscope is placed directly in the beamline, the charge calibration of the FQH strips is performed with ion beam.
The FQH strips were tested on the $280\,MeV$ electron beam of "Pakhra" synchrotron (LPI, Troitsk, Russia) and the light yield of about 5 photo-electrons on one MIP has been observed~\cite{FQH_tests}.

\vspace*{-0.5cm}
\begin{figure*}[tbp]
  \centering
  \caption{Scintillation Wall: a) schematic view of the ScWall; b) view of the ScWall detector with the beam hole mounted at the BM@N.}
\vspace*{0.3cm}
  \includegraphics[width=0.9\textwidth]{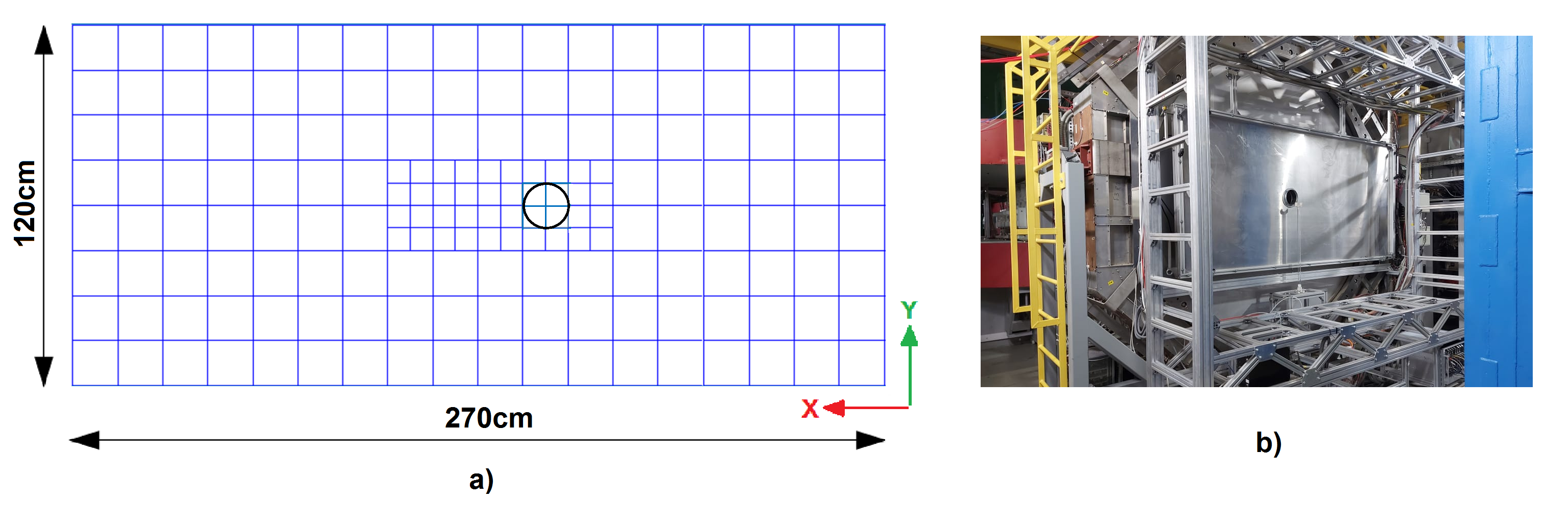}
  \label{fig:ScWall_view}
\end{figure*}

\begin{figure*}[tbp]
  \centering
  \caption{Schematic view of the ScWall components: a) large cell; b) small cell; c) assembly of a small cell with SiPM on a PCB with connectors.}
\vspace*{0.3cm}
  \includegraphics[width=0.52\textwidth]{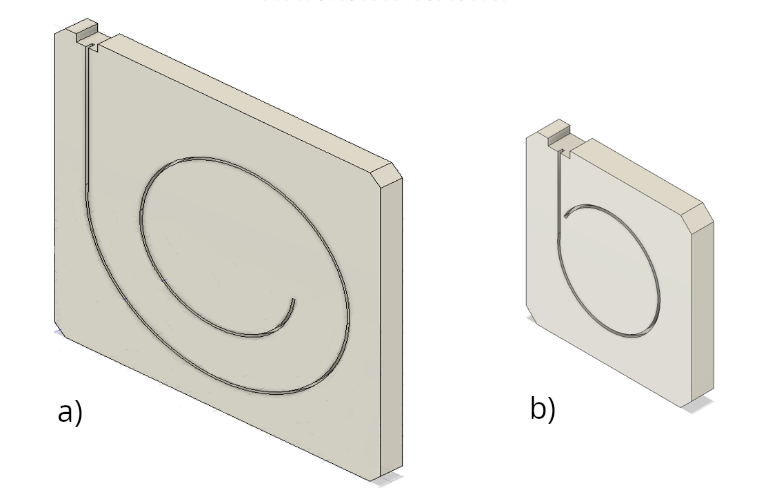}
  \includegraphics[width=0.2\textwidth]{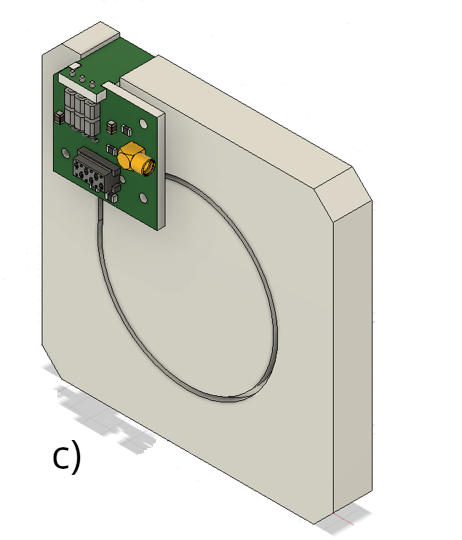}
  \label{fig:ScWall_cells}
\end{figure*}

\begin{figure*}[htbp]
  \centering
  \caption{Slow Control system for the forward detectors at BM@N.}
\vspace*{0.3cm}
  \includegraphics[width=1.0\textwidth]{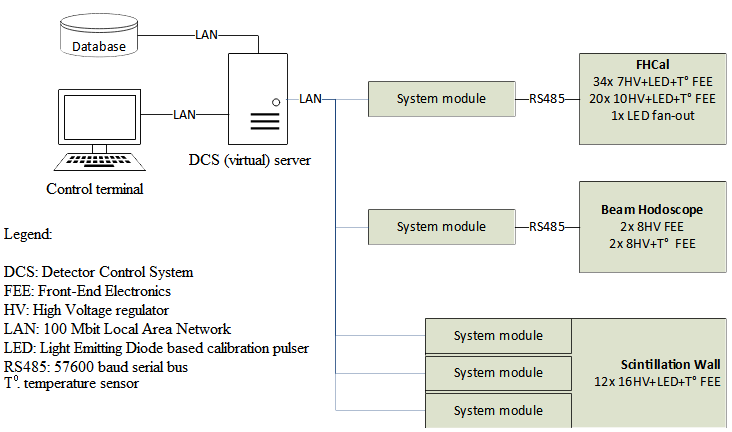}
  \label{fig:DCS_view}
\end{figure*}

\vspace*{0.3cm}

\subsection{Scintillation Wall}

The ScWall is a large area detector aimed at measuring the charged particles in the forward rapidity region. It consists of an array of scintillating plates placed in an aluminum box.
A view of the ScWall is shown in Fig.~\ref{fig:ScWall_view}. The full detector size is $270\times130\,cm^2$. The ScWall has 40 inner small ($7.5\times7.5\times1\,cm^3$) scintillator detectors (cells) and 138 large outer cells ($15\times15\times1\,cm^3$). In order to avoid radiation damage caused by the heavy ion beam, as well as to minimize background counts in other detectors, the very central part of the ScWall has a $15\times15\,cm^2$ beam hole (see Fig.~\ref{fig:ScWall_view} b). The cells are made of polistirol-based scintillators manufactured by Uniplast (Vladimir, Russia).

The light produced in the cells is collected by WLS Y11(200) S-type (Kuraray) wavelength shifting fibers  embedded in $1.5\,mm$ deep grooves (see Fig.~\ref{fig:ScWall_cells}). At the end of the fibers, the light is detected by Hamamatsu S13360-1325CS SiPMs, which have an active area of $1.3\times1.3\,mm^2$, a gain of $7\times10^5$, and a photo detection efficiency of $25\,\%$. The light yield from a minimum ionizing particle passing the large and small cells is about 32 and 55 photoelectrons, respectively~\cite{ScWall_tests}. The full area of the ScWall is divided into twelve readout zones. The readout is performed by ADC64s2 boards combined with FEE boards similar to the readout of the FHCal signals. Three ADC64s2+FEE boxes are used to digitize the signals from all the ScWall cells. Initial calibration of the ScWall channels is performed using cosmic muons, and later verified and refined in the analysis of the experimental data using hits from particles with $Z=1$ charge produced in the interactions.

\subsection{Slow Control for the forward detectors}

As light sensors, the FHCal, FQH and ScWall detector systems use SiPMs, whose amplification depends on temperature and the applied bias voltage. Therefore, a Slow Control (SC) system developed for these detectors monitors the bias voltage (HV) and measures the temperature of the electronic boards with photodetectors~\cite{DCS_forward_detectors}. If needed, the system automatically adjusts the HV based on temperature changes. The hardware part of the SC was designed and manufactured by HVSys (JINR, Dubna, Russia). A schematic view of the system is shown in Fig.~\ref{fig:DCS_view}. Multichannel HV power supply modules are operated via a microcontroller interface.  Each HVSys module has a unique IP address for communication through an individual proxy-server. The communication of the HVSys box with FEE microcontrollers is done via an RS-485 interface. All proxy-servers have connections to a GUI panel, which allows monitoring the detector status and performing temperature correction for all SiPMs. 
%The software part of the SC is written in Python3. 
In order to record actual values of HV and temperature in a general database, the SC for the forward detectors periodically relays this information to the main BM@N Slow Control System described in section~\ref{SCS}.

%\begin{figure}[ht]
%  \centering
%  \includegraphics[width=1.0\textwidth]
% {pictures/Forward_detectors/DCS_forward_detectors_panel.png}
%  \caption{The Slow Control GUI panels for forward detectors
% at BM@N: FHCal (left), ScWall (center) and FQH (right).}
%  \label{fig:DCS_panel}
%\end{figure}

%\clearpage

\section{Trigger and data acquisition}

In the 2023 Xe run, the experiment operated at $\sim0.5\,MHz$ beam intensity and most of the data were taken with a $2\,\%$ interaction length target. Such conditions correspond to an interaction rate of about $10\,kHz$, which exceeds the optimal DAQ event rate of a few $kHz$. Therefore, trigger settings were chosen to ensure high efficiency for the most central and semi-central collisions, while other types of events were added to the readout with downscaling factors.

\subsection{Trigger logic implementation}

The BM@N trigger consists of hardware and software parts. The hardware part includes 
detectors based on fast plastic scintillators described in 
section\,\ref{BeamAndTriggerDetectors}, low and high voltage power supply modules, and a programmable trigger logic unit T0U. The software part includes a graphic trigger interface and programs, which control trigger performance and beam quality.

The beam trigger (BT) is formed by the $20\,ns$ pulse coincidence from the BC1, BC2 beam counters and the absence of the pulse from the Veto counter (VC):

\[
                  BT=BC1 \times BC2 \times \overline{VC} 
\]

The minimum bias trigger (MBT), in addition to the BT requirements, sets the criterion that only 
events with pulse heights in the FD less than a preset threshold (below the beam ion peak) are considered as beam ion interactions in the space between the BC2 and FD counters, i.e. primarily in the target:

\[
                  MBT=BT  \times \overline{FD}
\]

The interaction trigger, called the Central Collision Trigger (CCT), is composed of the
minimum bias trigger and the signal from the Barrel Detector (BD) generated when the multiplicity of hits in the BD exceeds a certain threshold:

\[
                  CCT = MBT \times BD(> N) 
\]

\begin{figure}[htb]
\centering
\caption{Scheme of physics trigger generation in the T0U module.}
\includegraphics[width=0.5\textwidth]{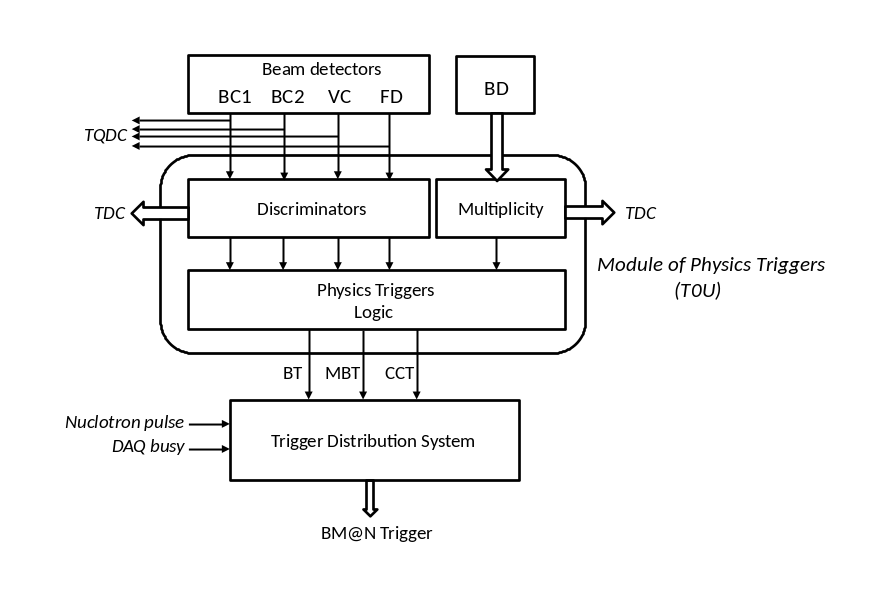}
\label{fig3}
\end{figure}

The logic of all the physics triggers mentioned above (BT, MBT, CCT) is implemented 
in a special custom-made electronic module T0U, designed to accommodate the main tasks of the BM@N trigger (Fig.\,\ref{fig3}). The T0U has a modular structure built on a motherboard that can be supplemented by mezzanine boards of four different types: four-channel discriminator input cards, five-channel FEE power supply, TTL-NIM convertor output cards, and Ethernet interface card. The T0U accepts analog signals from the BC1, BC2, VC and FD counters, as well as the LVDS pulses from the BD front-end electronics. Signal discrimination, delays and coincidence conditions are implemented using FPGA functionality.     

The trigger signals formed by the T0U are sent to the Trigger Distribution System, where they are processed with corresponding downscaling factors.

\subsection {Scalers}

\begin{figure}[htb]
  \centering
  \caption{MSC16VE module.}
\vspace*{0.3cm}
  \includegraphics[width=0.5\textwidth]{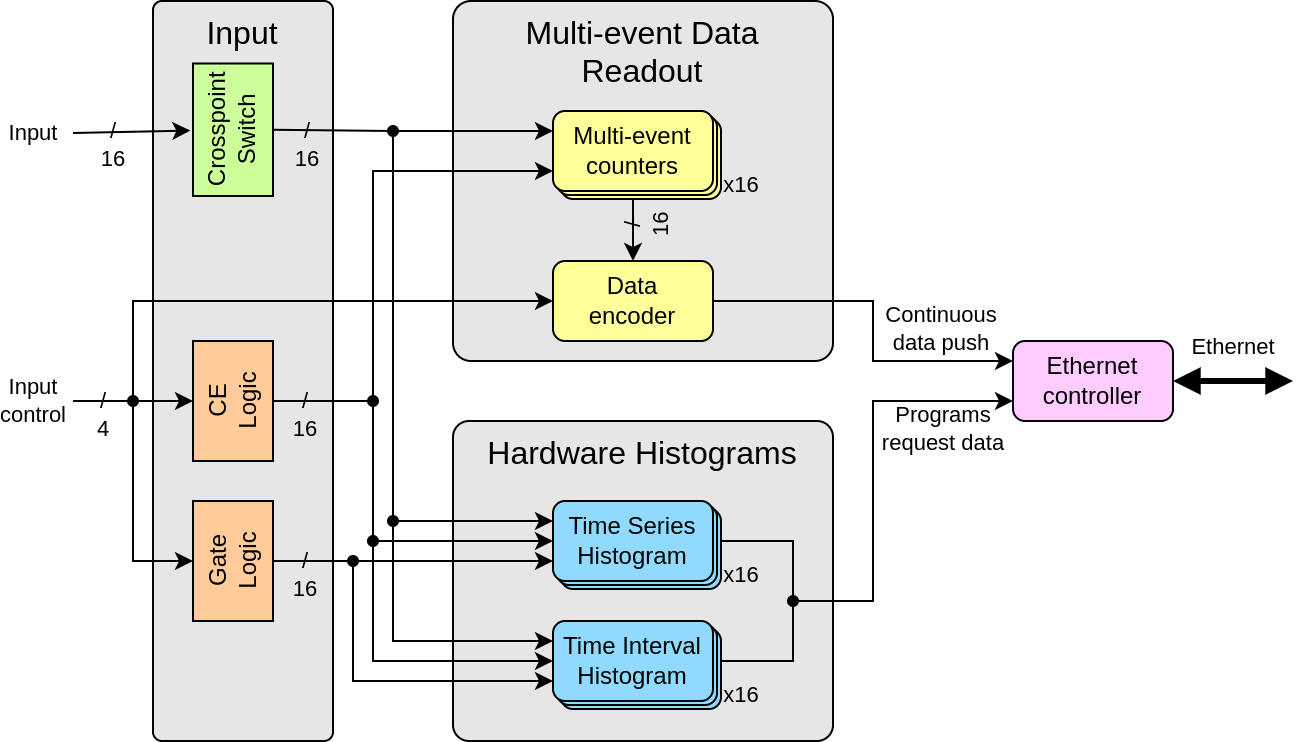}
  \label{daq0}
\end{figure}

Trigger signals generated by the T0U are sent not only to the higher level trigger modules for further processing, but also delivered to the MSC16VE 16-channel multihit scaler, which allows monitoring the trigger count rate during data taking. 
Each channel input of the MSC16VE has $50\,\Omega$ impedance and accepts pulses of $\pm2.5\,V$ range. Discrimination thresholds for input signals can be adjusted in a $\pm1\,V$ range. The module has four LVTTL count enable (CE) inputs. Data readout and module control are organized via Ethernet 1000BASE-X connection. The MSC16VE module has three main logic parts: input part, multihit data readout and hardware histograms (Fig.\,\ref{daq0}).

%\vspace*{-0.7cm}

 \begin{figure*}[htb]
  \centering
  \caption{BM@N trigger architecture.}
\vspace*{0.3cm}
  \includegraphics[width=1.0\textwidth]{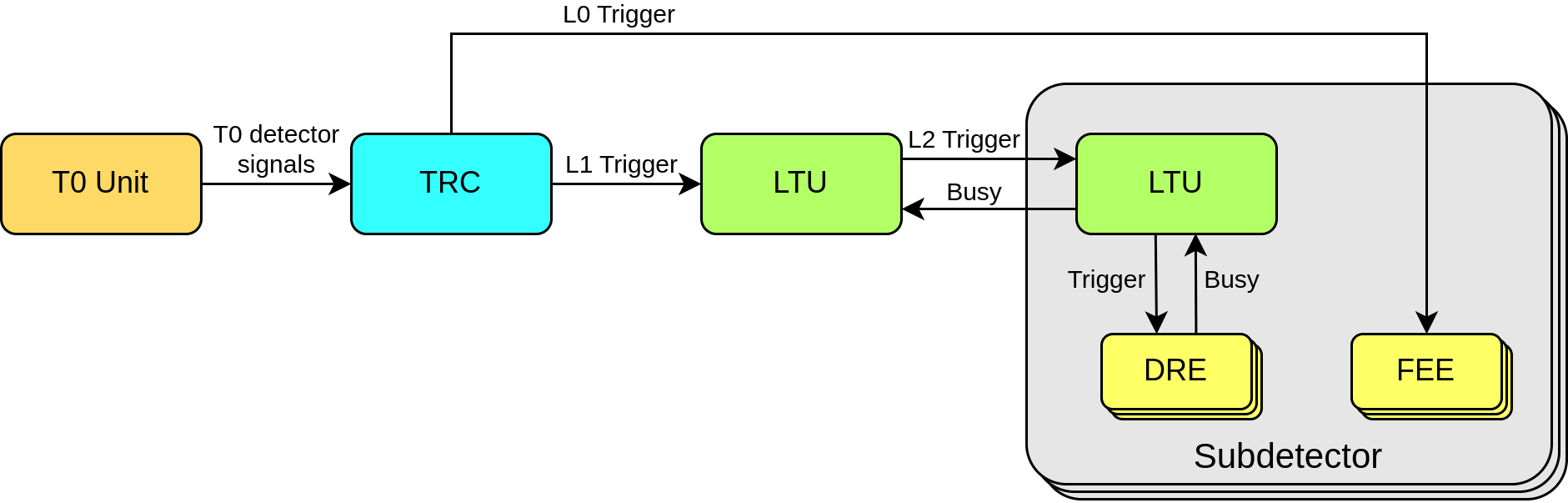}
  \label{daq3}
\end{figure*}

The input part has a crosspoint switch that allows any input channel to be processed by any 
multihit counter and histogram. CE and Gate logics have 16 independent Look-Up Tables (LUT) each. The Gate logic determines reset conditions for hardware histograms.
Multihit counters data are continuously subdivided into numbered time slices, which are pushed to a data encoder and sent further to the Ethernet. The length of time slices 
is adjustable with a minimum of $64\,ns$ and $8\,ns$ increment. 
The data encoder performs zero suppression and data packing. The hardware histograms are used for online monitoring of input counts in two possible forms: 1) count rate distribution in time, 2) time interval between two adjacent hits. Both types of hardware histograms are available for online monitoring via GUI control software.

\subsection {General architecture of the trigger distribution system}
\label{sub3.1}

The BM@N trigger distribution system can accept up to 16 input triggers and process them at three levels of decision making: L0, L1 and L2 (Fig.\,\ref{daq3}). 
All signals in the trigger distribution system are transmitted via coaxial cables in the LVTTL standard.

\begin{figure}[htb]
  \centering
  \caption{Trigger handshake chronogram.}
\vspace*{0.3cm}
  \includegraphics[width=0.5\textwidth]{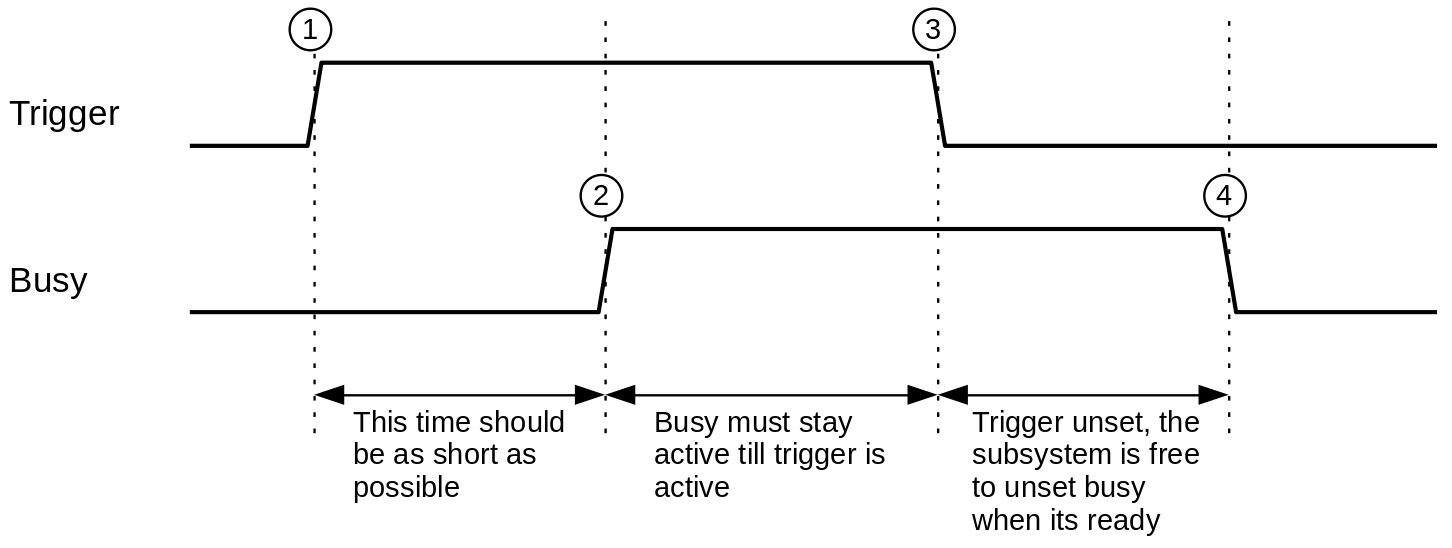}
  \label{daq1}
\end{figure}

The L0 and L1 triggers are generated by a custom-made TRigger Control (TRC) module, which receives signals representing the physics driven triggers such as BT, CCT, MBT, etc., formed by the T0U. L0 is a fast signal produced with a typical latency of $300\,ns$ and delivered to the front-end electronics of the tracking detectors (SiBT, FSD, GEM and CSC) as a trigger for sample holder circuits. L1 signals are trigger candidates derived from the physics triggers after applying downscaling factors. The formation time of the L1 trigger is adjustable and was set to $\sim1\,\mu s$ in the 2023 Xe run. In addition to the downscaling factor, each of the TRC input channels has individual settings adjustable by the user: signal delay and before/after protection time window. The before/after protection logic is used for pile-up event rejection. The output delays of L1 triggers can be set in the range from $8\,ns$ to $100\,\mu s$. More than one L1 trigger can satisfy the downscaling conditions. All of such triggers are transferred to the Logical Trigger Unit (LTU), where the L2 trigger is generated and distributed. The first out of the L1 triggers, which appears when all busy conditions are cleared,
is accepted to generate the L2 trigger. 

\begin{figure}[htb]
  \centering
  \caption{The average busy time for all subsystems.}
  \includegraphics[width=0.5\textwidth]{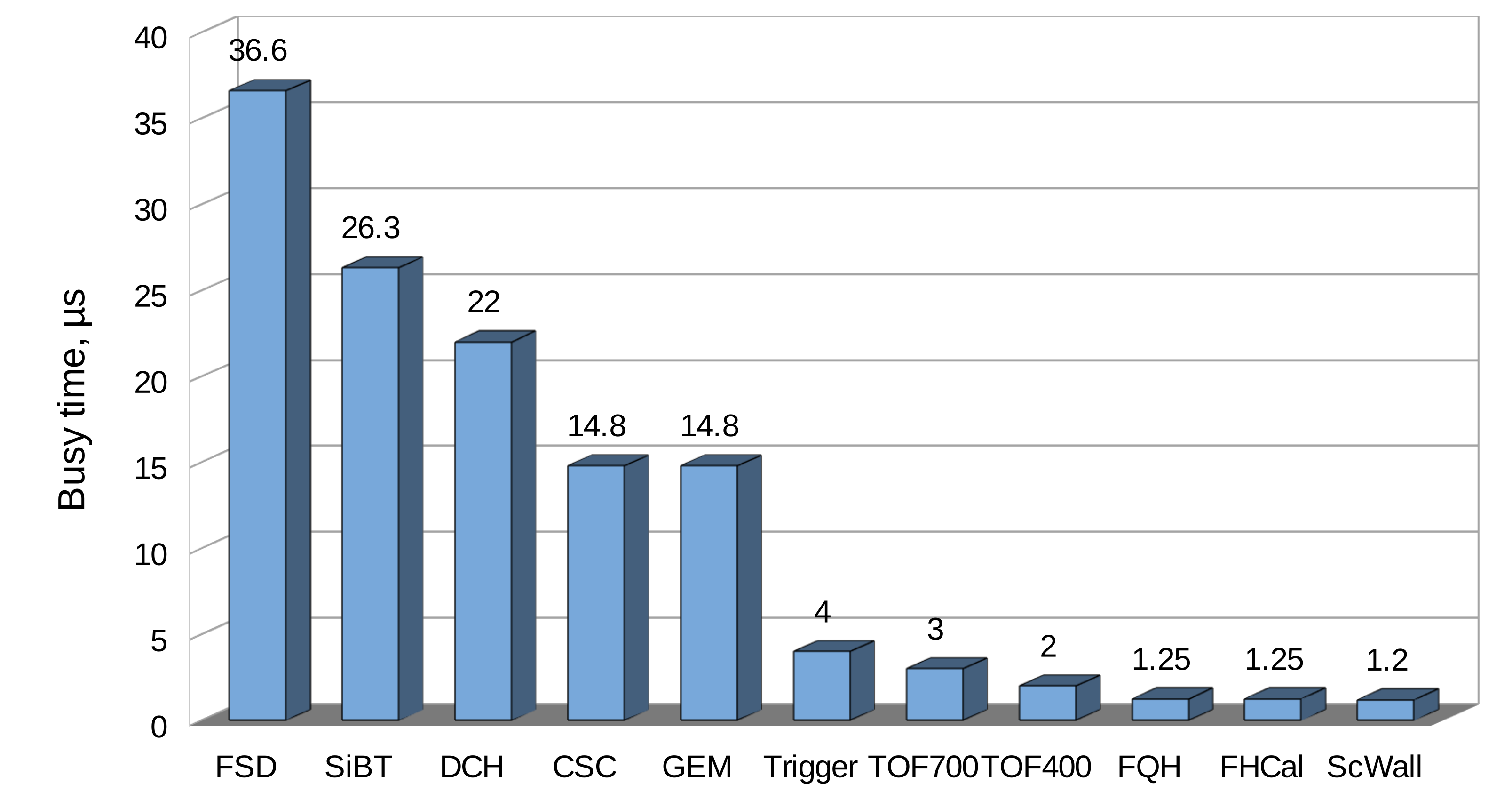}
  \label{daq2}
\end{figure}

 \begin{figure*}[!ht]
  \centering
  \caption{General architecture of the DAQ system.}
\vspace*{0.3cm}
  \includegraphics[width=0.8\textwidth]{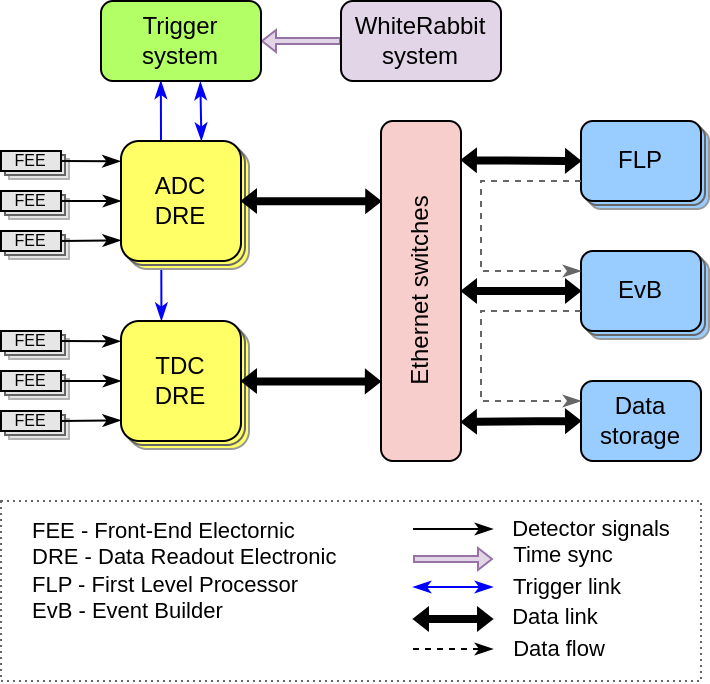}
  \label{DAQarch}
\end{figure*}

The LTU ensures the operation of the Trigger-Busy handshake algorithm and can process up to 16 busy channels. The Trigger-Busy handshake algorithm for the L2 trigger was implemented to guaranty the delivery of all triggers to the corresponding subsystems. This algorithm is shown in Fig.\,\ref{daq1}. The rising edge of the trigger signal (1) after a certain delay defines the start of the busy signal of a subsystem (2). After that the trigger signal is deasserted (3). Upon completion of data collection, the subsystem deasserts its busy signal (4). 

Busy signals can be received either from the detector readout electronics or from hierarchically lower LTU modules. The time intervals between accepted triggers and the duration of the busy signals are histogrammed in the LTU internal memory. Typical busy time of the BM@N subsystems is shown in Fig.\,\ref{daq2}. Two longest busy times are set by the readout of the silicon tracking systems, FSD and SiBT. Most of this time is required by the ADCs to process the multiplexed output of the FEE ASICs. The two systems have $128$ and $64$ input channels per chip, respectively, with multiplexer frequencies of $3.5$ and $2.5 MHz$ . The multiplexer frequency of the FSD ASICs is set to the highest level, which ensures signal transfer between the FEE and ADC without amplitude distortion.

\subsection{Detector readout data flow}

The core function of the DAQ system is the realization of data transfer from  
detectors to the storage system. It includes the data flow from the readout 
electronics to the First Level Processor (FLP) fabric, to the Event Building 
(EvB) and to the Storage System. The main DAQ components are readout electronic 
modules, a clock and time synchronization system, 
data transfer networks, data processing servers and an online storage 
system. The general DAQ architecture and data flow are illustrated in Fig.\,\ref{DAQarch}.

 \begin{figure*}[htb]
  \centering
  \caption{Data flow from a detector to the storage system.}
\vspace*{0.3cm}
  \includegraphics[width=1.0\textwidth]{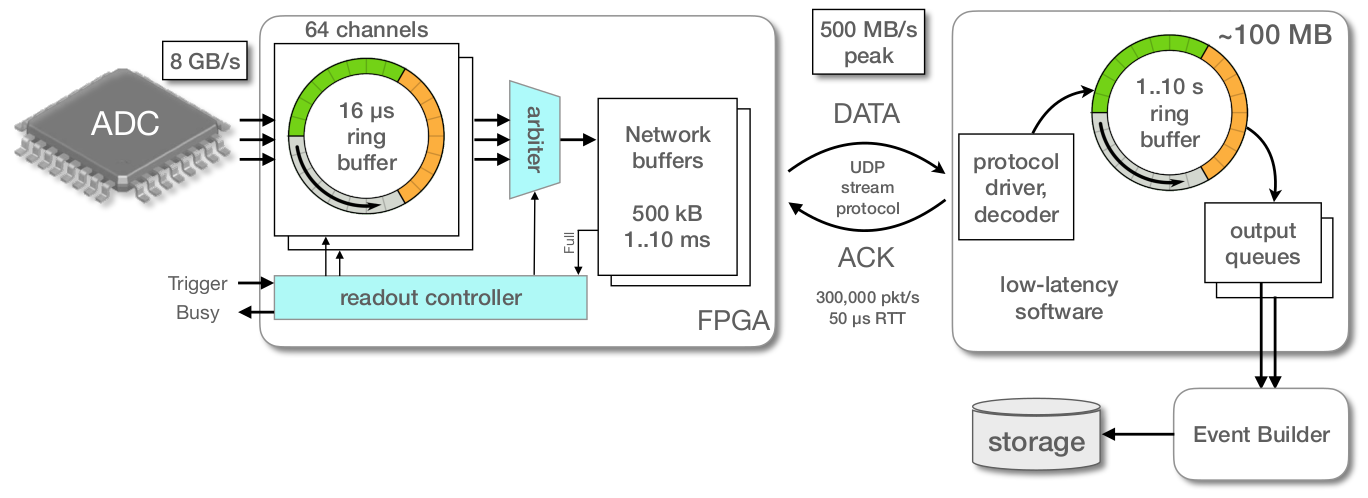}
  \label{daq001}
\end{figure*}

\subsubsection {TDC and ADC boards}

Detector Readout Electronic (DRE) boards record detector signals. 
BM@N has two main types of DRE boards grouped by function: time stamping in 
Time to Digital Converters (TDC) and amplitude sampling in Amplitude to Digital Converters (ADC).
The TQDC DRE board combines both TDC and ADC functions.

The HPTDC based TDC DRE board performs timestamping of multiple discrete signals (hits) 
with typical accuracy of $25\,ps$. Hit timestamps are kept for $51\,\mu s$ in the ring type memory. The total trigger latency should not exceed this value.
The ADC DRE board is a waveform digitizer which samples an analogue input signal at fixed time intervals. It can be run in a zero suppression mode based on baseline estimation and signal threshold value. Signal shaping can be performed in digital form with FIR 
filters. It allows reducing the number of waveform points required for digital signal representation with a minimum loss of accuracy. The ring type memory provides a possibility to read back the last $32\,\mu s$ time window of digitized waveforms. This value sets the limit on the maximum trigger latency.

\subsubsection {Timing synchronization system}
Timestamping TDCs, which are used in the readout electronics of the trigger detectors, TOF400 and TOF700 detector systems, have a time resolution of $25\,ps$, while the DCH TDCs have a $100\,ps$ resolution. These digitizer boards require precise reference clock for high quality 
measurements. They process signals using common notion of time and frequency regulated by the White Rabbit network. The time reference is provided by a GPS/GLONASS receiver and backup precision frequency reference (Rubidium clock).

The White Rabbit ensures sub-nanosecond accuracy and picosecond precision of time  synchronization for distributed systems. The DRE boards include White Rabbit Node Core and tunable crystal oscillators that are synchronized to a reference clock with a $10\,ps$ accuracy.
The WR Node Core provides a local clock with a $125\,MHz$ frequency and can set timestamps specified as TAI (International Atomic Time), which is an absolute number of seconds and nanoseconds since $01.01.1970$. Frequency dividers synchronized by a 1 PPS (pulse per second) signal are used to produce digitizer clocks: $41.667\,MHz$ for HPTDC ASICs and $62.5\,MHz$ for the waveform digitizers. 

\subsubsection {DAQ data flow}

All BM@N subdetectors, except the DCH, use Ethernet to transfer data from the readout electronics to the First Level Processors (FLP). The primary FLP task is to receive data stream in real time, buffer, validate, format and enqueue data blocks to an event building network. The FLP decouples the fast microsecond-scale synchronous data acquisition process from the slower, seconds-scale, software data processing by buffering the data in the computer RAM. As a typical example, the path of data transfer from a readout electronic module to the event building network and storage system is shown in Fig.\,\ref{daq001} for a waveform digitizer based on a 64-channel ADC. 

The electronic modules of the BM@N DAQ share a common 
communication architecture. Network connectivity is provided 
by hardware IP stack implementation in a programmable logic code 
synthesized for an onboard FPGA processor. Taking into account limited memory and logic resources of FPGA chips available, 
and the implementation complexity of the TCP protocol, a custom data 
transfer protocol MStream has been designed for data streaming 
over $1\,Gb/s$ or $10\,Gb/s$  Ethernet networks. It uses UDP over IP 
as the transport layer and implements an ordered and reliable data 
packet delivery using acknowledgments.

\begin{table*}[htb]
\begin{center}
\caption{Characteristics of the BM@N DAQ server equipment.}
\vspace*{0.3cm}
\begin{tabular}{|l|l|l|l|} \hline
Qty&Function                      &Specifications                                   & Network       \\ \hline
20& Compute node                  & Dual 18-core $3\,GHz$ CPU, $384\,GB$ RAM              & Dual $100\,Gb/s$ \\ \hline
10& NVMe storage server           & $10\times3.5\,TB$ NVMe                      & Dual $100 Gb/s$ \\ \hline
 8& HDD storage server 1          & $24\times12\,TB$ HDD, $1.8\,TB$ SSD cache      & Dual $100\,Gb/s$ \\ \hline
 4& HDD storage server 2          & $24\times18\,TB$ HDD, $3\,TB$ SSD cache        & Dual $100\,Gb/s$ \\ \hline
 6& Control server                & 4-core CPU, $64\,GB$ RAM                           & Dual $25\,Gb/s$  \\ \hline
 1& Bootstrap server              & 4-core CPU, $16\,GB$ RAM, $4\times300\,GB$ HDD & Dual $1\,Gb/s$    \\ \hline
\end{tabular}
\label{tab:DAQtable001}
\end{center}
\end{table*}

 \begin{figure*}[htb]
  \centering
  \caption{BM@N DAQ Network.}
  \includegraphics[width=1.0\textwidth]{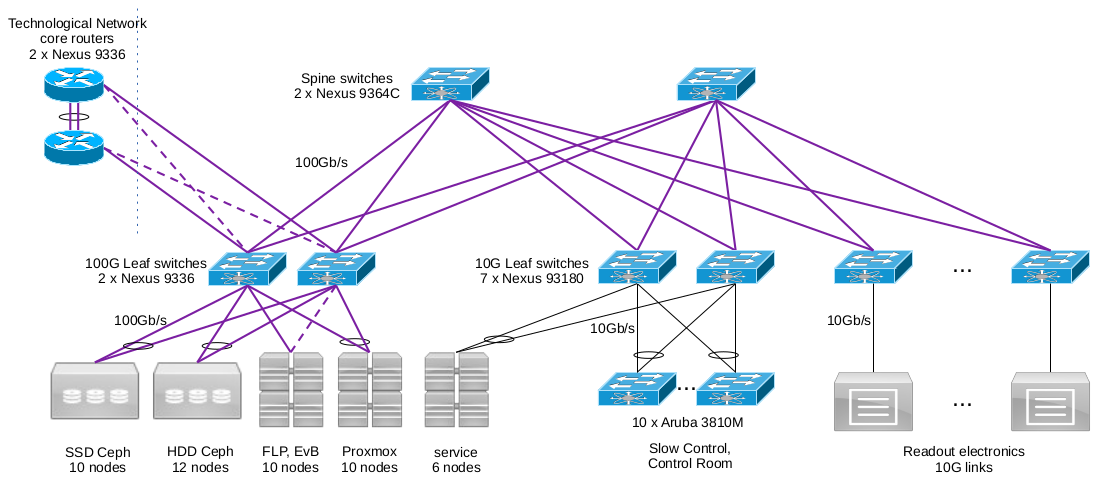}
  \label{daq002}
\end{figure*}

The FLP receives the data stream in real time using dedicated Linux servers with dual 18-core CPUs and equipped with dual $100\,Gb/s$ Ethernet adaptors. Tuning for real-time operation includes the CPU frequency and supply voltage management, network adapter interrupt coalescence mitigation and system task scheduler adjustments. The BM@N readout electronics deliver $6\,GB/s$ raw data over 200 streams in peak at a $10\,kHz$ trigger rate. A single data stream has a maximum sustained throughput of $500\,MB/s$ when using $10\,Gb/s$ Ethernet. Operation during the 2023 Xe run showed that a single manually tuned FLP server was capable of hosting 10–12 data stream receivers with minimal contribution to overall busy time. 

Software event building in BM@N is part of asynchronous processing and does not affect the readout busy time under normal operation. Event builders are cascaded in multiple layers for load distribution, and the last layer writes data files to the storage system. Event builder programs associated with data intensive subdetectors run on dedicated hardware servers, while event builders for low data rate subdetectors, as well as readout control programs, run in a KVM virtual environment. This allows efficient utilization of computer resources. The typical size of a BM@N event in the 2023 Xe run was approximately $0.6 MB$. Since the multiplexing method implemented in the readout of the tracking detectors does not provide a zero-suppression mode, the event size  is independent of the multiplicity of hits in the detectors and, thus, does not vary with the event centrality.

%=======================================
\subsection {DAQ storage system}

The DAQ server equipment is located in 4 racks of the modular data 
center (MDC). A total of 49 servers occupy 81 units of rack space. 
Table~\ref{tab:DAQtable001} shows server types and functions.

The core of the data network is a two-level Ethernet fabric with Clos 
architecture that has two switches on the spine level and multiple 
switches on the leaf level (Fig.\,\ref{daq002}). The Ethernet VPN virtualization technology is used to allow flexible traffic 
management, high availability and efficient link utilization. 
The underlay network provides connectivity between the fabric nodes. 
It is formed by leaf and spine switches connected with L3 
routed links. The network topology is managed by the OSPF dynamic routing 
protocol. The overlay network that carries user traffic is realized 
with the MP-BGP protocol at the control plane and VXLAN encapsulation 
at the data plane.

The DAQ network supports jumbo Ethernet frames up to 9000 bytes 
to maximize throughput of data transfer from readout electronics. 
The network uses Any-Source Multicast that is necessary for 
automatic discovery of readout electronics modules and software 
components of the distributed DAQ system.

Two spine and four leaf switches are located in the MDC racks. 
Other leaf switches and access switches of slow control systems of various detectors 
are located in the electronics racks in the experimental area. Two core 
routers of the DAQ technological network are located in the experimental hall,
close to the BM@N electronics. These routers provide connectivity to the outside network with a $200\,Gb/s$ bandwidth.

\begin{figure*}[tbp]
  \centering
  \caption{Main display of the Slow Control System.}
\vspace*{0.3cm}
  \includegraphics[width=1.0\textwidth]{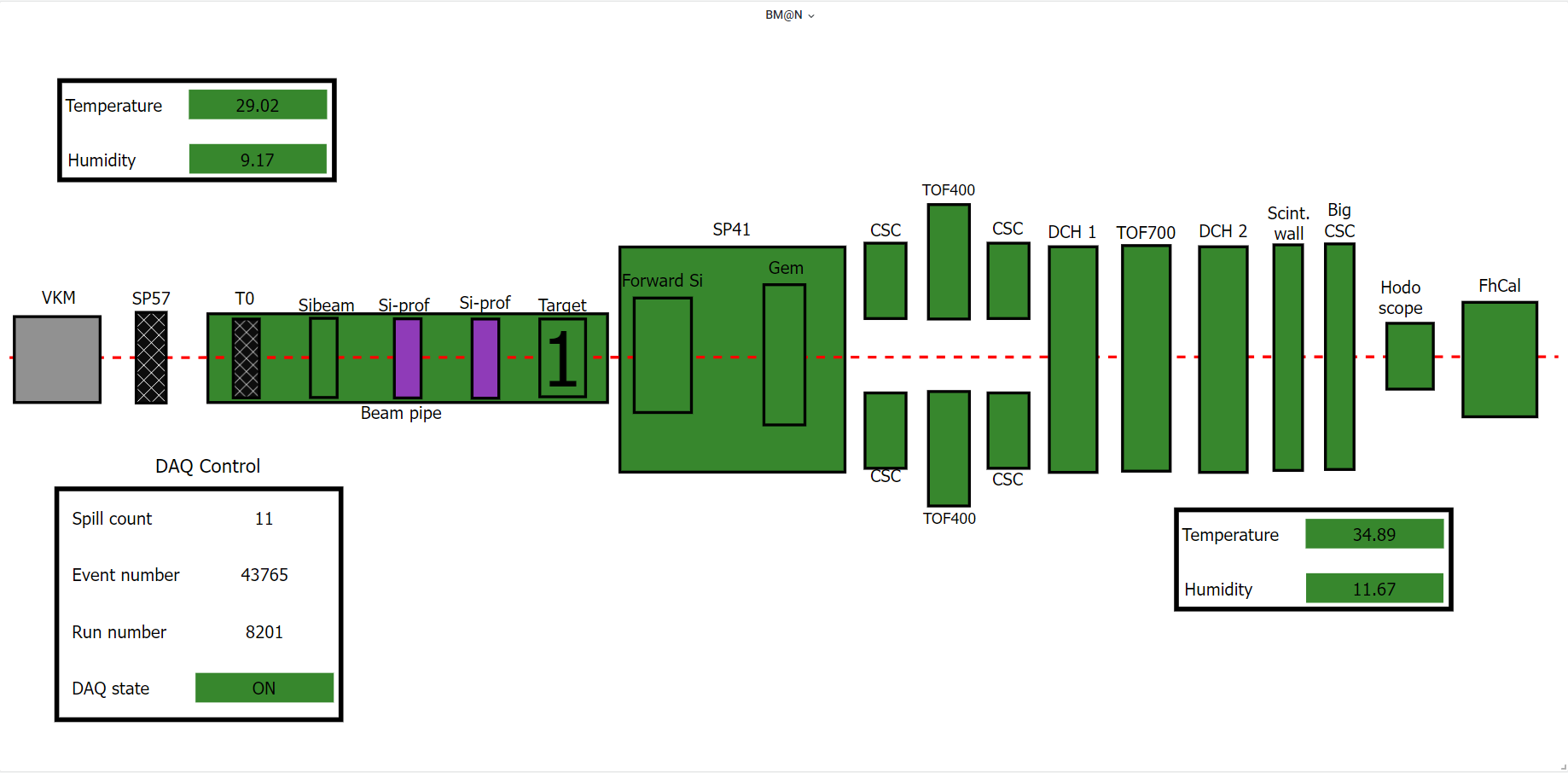}
  \label{fig:scs_bmn}
\end{figure*}

The DAQ network showed no critical problems during the BM@N data taking in
the 2023 Xe run. The Ethernet switching fabric bandwidth proved to be adequate for peak 
traffic conditions and showed no negative impact on the data taking 
performance. No significant packet drops or errors were registered 
by the monitoring system on network fabric switches that could indicate 
network saturation and packet buffer overflows. The design of the DAQ network takes into account a potential increase in both the trigger rate and event size in future experimental runs. If necessary, the fabric bandwidth can be doubled by introducing additional leaf to spine connections.

%\clearpage

\section{Slow Control System} \label{SCS}

The Slow Control System (SCS) is used to monitor the hardware status and to archive the operational conditions of the BM@N setup. It includes a user-friendly graphical interface and an alarm management system. The SCS was built around ``TANGO Controls''~\cite{Tango_Controls}, an open-source toolkit, widely used in scientific experiments. 

Slow Control data from the experiment subdetectors such as: high voltage, low voltage, magnetic field, vacuum level, gas flow and mixture, etc. are aggregated by the SCS.  These parameters are then stored in the TANGO Historical Database implemented using the PostgreSQL database with the TimescaleDB extension~\cite{PostgreSQL_DB}. The SCS is configured as a distributed cluster with backup and load balancing.

The TANGO Database, which hosts the configuration of the whole system, and the TANGO Historical Database are running on the BM@N virtual machine cluster, whereas the programs controlling and/or monitoring the hardware status of a particular subsystem can run either on a virtual cluster or on a dedicated PC for this subsystem.

The user interface for online monitoring and retrieving previously stored data is developed with Grafana~\cite{Grafana_app}, an open-source analysis and interactive visualization web application. A schematic display of the experiment hardware status and alarms (Fig.~\ref{fig:scs_bmn}) was also implemented using Grafana.

The color scheme of the SCS reflects various possible states of the subsystems: green --- normal operational conditions; red --- an alarm that indicates an unexpected malfunction; orange --- an abnormal condition for some of the parameters, but a known problem; violet --- standby mode, for example, for the beam profilometers being in the position out of the beam, as shown in Fig.\,\ref{fig:scs_bmn}; black --- no input data from the subsystem, either because it is not yet being monitored or because it is switched off; grey --- the subsystem is planned to be included in the SCS, but not fully configured.

%\clearpage

\section{Summary}
BM@N is a fixed target experiment recently put into operation at the Nuclotron/NICA complex aiming at the study of nucleus-nucleus collisions at energies up to $4.5\,AGeV$. We presented a detailed description of the BM@N spectrometer and its subsystems. The spectrometer design is driven by the requirements to handle high beam intensity and high multiplicity of produced particles typical for central and semi-central nucleus-nucleus collisions at relativistic energies. The major subsystems of the experiment include tracking detectors capable to measure the momentum of produced particles in a wide rapidity interval, time-of-flight detectors for charged particle identification, as well as forward spectator detectors designed to determine the centrality and the reaction plane in each nucleus-nucleus collision. The details of the trigger, data acquisition and control systems are also described. The BM@N setup presented in this article corresponds to the configuration used in the 2023 Xe run. During a three-week data taking period, about $0.4\times10^9$ Xe+CsI interaction events were collected at the beam energy of $3.8\,AGeV$. All the experiment subsystems operated at the expected level. A detailed evaluation of the detector performance is ongoing. Further upgrade of some of the detector subsystems is foreseen, but the described setup is close to the final configuration intended for experiments with heaviest (Au, Bi) beam ions.      

\section*{Acknowledgements}

The BM@N collaboration gratefully acknowledges the efforts of the 
staff of the accelerator division of the Laboratory of 
High Energy Physics at JINR that made the data taking possible and 
successful. We appreciate the work done by our JINR colleagues who recently passed away: V.\,Balandin, N.\,Kuzmin, A.\,Morozov, Yu.\,Petukhov, S.\,Sychkov, S.\,Vasiliev, A.\,Vishnevsky.

\bibliographystyle{alpha}

\end{document}